\documentclass[11pt]{article}
\usepackage{youngtab}
\usepackage{amscd,amsmath,amssymb,amsfonts,xspace,mathrsfs,mathtools,amsthm}
\usepackage{silence}%This removes the warnings :/
\usepackage{xr-hyper}
\usepackage{jheppub}
\usepackage[dvipsnames]{xcolor}
\usepackage{bbm}
\usepackage{graphicx}
\usepackage{tabu} 
\usepackage[color,matrix,arrow]{xy}
\usepackage{wasysym} 
\usepackage{stmaryrd}
\usepackage[shortlabels]{enumitem}
\usepackage{array,amsmath}
\usepackage{slashed}
\usepackage{tensor}
\usepackage{enumitem}
\usepackage{hyperref}
\usepackage[textsize=tiny]{todonotes}
\usepackage{cellspace, hhline}
\usepackage{cleveref}
\usepackage{lineno}
\usepackage{multirow}
\usepackage{svg}
\usepackage{booktabs, cellspace, hhline}
\usepackage{longtable}
\usepackage{comment}
\usepackage{placeins} 
\usepackage{lipsum}

\usepackage{tikz}
\usepackage{tikz-cd}
\usepackage{diagbox}
\usetikzlibrary{positioning}
\usetikzlibrary{calc}
\usetikzlibrary{decorations.pathreplacing,calligraphy}

\usepackage{xstring}
\usetikzlibrary{decorations.pathmorphing} 
\usetikzlibrary{decorations.markings} 
\usetikzlibrary{arrows} 
\usetikzlibrary{shapes} 
\usetikzlibrary{matrix} 
\usetikzlibrary{positioning} 
\usepackage[english]{babel} 
\usepackage[autostyle]{csquotes}
\usetikzlibrary{positioning,fit}
\usepackage[dvipsnames]{xcolor}
\tikzset{->-/.style={decoration={
  markings,
  mark=at position #1 with {\arrow{>}}},postaction={decorate}}}

\theoremstyle{definition}

\DeclarePairedDelimiterX\braket[2]{\langle}{\rangle}{#1\delimsize\vert\mathopen{}#2}%

\newcommand{\SBE}{\mathcal{S}_{\text{BE}}}

\newcommand{\tr}{\text{tr}\,}

\newcommand{\rk}{\text{rk}}

\newcommand{\be}{\begin{equation}} 
\newcommand{\ee}{\end{equation}} 
\newcommand{\bea}{\begin{equation} \begin{aligned}} \newcommand{\eea}{\end{aligned} \end{equation}}
\newcommand{\bes}{\begin{equation*}}
\newcommand{\ees}{\end{equation*}}

\newcommand{\nn}{\nonumber}

\newcommand{\kk}{\mathsf{k}}

\newcommand{\CC}{\mathcal{C}}

\newcommand{\CH}{\mathcal{H}}
\newcommand{\CI}{\mathcal{I}}

\newcommand{\CN}{\mathcal{N}}
\newcommand{\CO}{\mathcal{O}} 

\newcommand{\CQ}{\mathcal{Q}}
\newcommand{\CR}{\mathcal{R}}
\newcommand{\CS}{\mathcal{S}}

\newcommand{\CW}{\mathcal{W}}

\newcommand{\Tr}{\text{Tr}}

\newcommand{\SL}{\text{SL}(2,\BZ)}

\renewcommand{\t}{\widetilde }
\newcommand{\Z}{\mathbb{Z}}
\newcommand{\C}{\mathbb{C}}

\renewcommand{\SL}{{\mathscr L}}

\newcommand{\qcoh}{\mathsf{q}}
\newcommand{\qk}{q}
\newcommand{\bbC}{\mathbb{C}}

\usetikzlibrary{fit,backgrounds}
\numberwithin{equation}{section}  % make eq labels (sec.num)
\allowdisplaybreaks  % allow page breaks in displayed eqs
\numberwithin{table}{section}

\usepackage{lscape}

\usepackage[normalem]{ulem}

\usepackage{bclogo}%  \bcpanchant

 \title{On the Schubert calculus of the quantum K-theory for partial flag manifolds: a 3d A-model perspective}

\abstract{We further investigate the 3d gauged linear sigma model (GLSM)/~quantum K-theory correspondence for partial flag manifolds $X \equiv {\rm Fl}(\boldsymbol{k};n)$. This is a 3d uplift of the 2d GLSM/quantum cohomology correspondence with the 3d theory compactified on $\mathbb{R}^2\times S^1_\beta$. Recently, a set of half-BPS line operators, called Schubert line defects, were constructed that correspond to the Schubert classes in the K-theory ring of $X$. Utilizing algebro-geometric algorithms, we compute $2$-point and $3$-point correlation functions of these line operators in the 3d A-model regime of the theory. These are interpreted as genus-$0$ K-theoretic Gromov--Witten invariants, and they produce the K-theoretic Littlewood--Richardson coefficients of the quantum K-theory ring of $X$. We show how this works explicitly in examples, going beyond the existing results in the literature. Taking the small $\beta$ limit, we apply these techniques to the resulting 2d GLSM. We explicitly compute the quantum cohomology ring relations of $X$ for some cases and match with existing results in the literature in examples.
}

\preprint{CERN-TH-2026-135}

\author[a,b]{Zhihao Duan,}
\emailAdd{xduanz@gmail.com}
\affiliation[a]{Section de Mathématiques, Université de Genève, 1211 Genève 4, Switzerland}
\affiliation[b]{Theoretical Physics Department, CERN, 1211 Geneva 23, Switzerland}

\author[c]{Osama Khlaif,}
\emailAdd{osama.khlaif@phys.ens.fr}
\affiliation[c]{Philippe Meyer Institute, Physics Department, École Normale Supérieure (ENS), Université PSL, 24 rue Lhomond, F-75231 Paris, France}

\author[d,e]{Hao Zou}
\affiliation[d]{Center for Mathematics and Interdisciplinary Sciences, Fudan University, Shanghai 200433, China}
\affiliation[e]{Shanghai Institute for Mathematics and Interdisciplinary Sciences, Shanghai 200433, China}
\emailAdd{haozou@fudan.edu.cn}

\begin{document}

\maketitle

\section{Introduction}

The 3d gauge theory/quantum K-theory correspondence~\cite{Bullimore:2014awa, Jockers:2018sfl, Jockers:2019wjh,Jockers:2019lwe, Jockers:2021omw, Ueda:2019qhg, Koroteev:2017nab,Bullimore:2018jlp,Bullimore:2019qnt,Bullimore:2020nhv, Gu:2020zpg,Dedushenko:2021mds, Gu:2022yvj,Dedushenko:2023qjq, Gu:2023tcv, Gu:2023fpw, Gu:2025abc, Gu:2021yek, Gu:2021beo,Sharpe:2024ujm,Closset:2023bdr,Closset:2023izb, Huq-Kuruvilla:2024tsg, Huq-Kuruvilla:2025nlf, ahkmox,Khlaif:2025ccg} relates the quantum K-theory ring of a K\"ahler manifold $X$ with the ring of half-BPS line operators (a.k.a. the twisted chiral ring) $\mathcal{R}_{\rm 3d}$ of the 3d $\mathcal{N}=2$ gauged linear sigma model (GLSM) on $\mathbb{R}^2\times S^1_\beta$ with $X$ realized as its Higgs branch. Recently, this correspondence has been extensively studied for the case where $X$ is a partial flag manifold -- See for instance \cite{fulton2013representation,Donagi:2007hi,Ohmori:2018qza} for more introduction on the flag manifolds and their realization in physics. Recall that, for an increasing sequence of $s$ integers $0<k_1<\cdots<k_s<n$, we define the partial flag manifold Fl$(\boldsymbol{k};n)$ as:
\begin{equation}
	{\rm Fl}(\boldsymbol{k};n) := \left\{\,V_\bullet \equiv (0\subset V_1\subset V_2\subset\cdots\subset V_s\subset V_{s+1}\equiv\C^n)~\mid~\dim(V_\ell) = k_\ell\,\right\}~.
\end{equation}
Here, $\boldsymbol{k} \equiv (k_1, \cdots, k_s)$. The 3d GLSM whose target is Fl$(\boldsymbol{k};n)$ is depicted in figure~\ref{fig:parFlag GLSM} below.

In the series of works~\cite{Closset:2025akk,Closset:2026bnk} by the last two authors and other collaborators, a new set of line defects, called Schubert defects, was constructed in terms of supersymmetric quantum mechanical quiver gauge theories coupled to the 3d GLSM in the ultraviolet (UV). It was shown that, in the infrared (IR), these lines flow to the Schubert classes spanning the K-theory ring of Fl$(\boldsymbol{k};n)$. The construction builds on the earlier works \cite{Closset:2023bdr,Gu:2025tda} for the complex Grassmannian and Lagrangian Grassmannian manifolds, respectively.

For a fixed partial flag manifold Fl$(\boldsymbol{k};n)$, the set of all possible Schubert line defects $\SL_w$ forms a basis of the ring $\mathcal{R}_{\rm 3d}$. The fusion rules of these lines are of the form:
\begin{equation}\label{QK relations intro}
        \SL_w\,\star\,\SL_{w'}\,=\,{\mathcal{N}_{w\,w'}}^{w''}(q,y)\,\SL_{w''}~.
\end{equation}
They translate into ring relations of the quantum K-theory of Fl$(\boldsymbol{k};n)$ with the OPE structure constants on the RHS being the K-theoretic Littlewood--Richardson (LR) coefficients, which are determined by the genus-0 K-theoretic Gromov--Witten (GW) invariants. By the finiteness theorem for homogeneous space \cite{10.1093/imrn/rnaa108}, the OPE structure constants ${\mathcal{N}_{w\,w'}}^{w''}(q,y)$ only involve finitely many powers of $q_i$'s. The explicit computation of these coefficients was left out in \cite{Closset:2025akk,Closset:2026bnk}. 

In this work, we will utilize some algebro-geometric computational algorithms to compute them and provide explicit results in some examples. In the literature, the quantum K-theory ring and its Schubert calculus have been computed for the special cases of the Grassmannians \cite{buch-lmi,Jockers:2019lwe,Ueda:2019qhg,Gu:2020zpg,Gu:2022yvj,Closset:2023izb}, the incidence flags \cite{xu2024quantum}, and the complete flags \cite{MR4881172,MNS23,ahkmox}. Chevalley-type multiplication formulas were derived in parabolic cases, including type~$A$ and type~$C$ Grassmannians and type~$A$ two-step flag manifolds \cite{kouno2023quantumktheorychevalleyformulas}. Using our computational algorithm, we will argue that one can compute the ring relations for any partial flag manifold.

\paragraph{Presentations of the quantum K-theory ring.}
The quantum K-theory of partial flag manifolds can be presented in terms of relations between the K-theoretic Chern roots $x^{(\bullet)}_\bullet$ of the tautological vector bundles $\mathcal{S}_\bullet$\footnote{For definitions see \protect\cite{Closset:2025akk,Closset:2026bnk} and reference therein.} -- i.e., in terms of the 3d Coulomb branch parameters -- and/or the K-theoretic Chern roots $x_\bullet$ of the quotient bundles $\mathcal{Q}_\bullet$. To distinguish between the three possible cases, we will use the following terminology:
\begin{itemize}
    \item \textit{The Bethe presentation.} In this case, the quantum K-theory ring is written in terms of the variables $x^{(\bullet)}_{\bullet}$:
    \begin{equation}\label{Bethe presentation}
        {\rm QK}_T({\rm Fl}(\boldsymbol{k};n))\,\cong\,\frac{\mathbb{K}[x^{(\bullet)}]^{{\rm W}_G}}{\CI_{\rm Bethe}^{(x^{(\bullet)})}}~.
    \end{equation}
    Here,  $\mathbb{K} \equiv \Z(q,y)$ with $q$ being the exponentiation of the 3d Fayet--Iliopoulos (FI) parameter and $y$ are the 3d fugacities associated with the $SU(n)$ flavour symmetry of the 3d theory. Moreover, W$_G$ is the Weyl subgroup of the gauge symmetry, and we are quotienting by the ideal $\CI_{\rm Bethe}^{(x^{(\bullet)})}$ spanned by the BAEs. See discussion around \eqref{R3d=QK}.
    \item \textit{The Whitney presentation.} The Whitney presentation for partial flags was first conjectured in \cite{Gu:2023fpw,Gu:2023tcv}, and then later was proved in \cite{Huq-Kuruvilla:2024tsg,Gu:2025abc}. In this presentation, the quantum K-theory ring is written in terms of both sets of variables $x^{(\bullet)}_\bullet$ and $x_\bullet$ \cite[Equation~(4.2)]{Closset:2026bnk}:
    \begin{equation}\label{whiteny present}
        {\rm QK}_T({\rm Fl}(\boldsymbol{k};n))\,\cong\,\frac{\mathbb{K}[x^{(\bullet)}_\bullet\,,\,x_\bullet]^{{\rm W}_G\times{\rm W}_P}}{\mathcal{I}_{\rm Whitney}}~,
    \end{equation}
    with W$_P$ the parabolic subgroup defined in \cite[Equation~(4.3)]{Closset:2026bnk}. Here we are quotienting by the quantum ideal $\mathcal{I}_{\rm Whitney}$ that is defined by the quantum Whitney relations derived in \cite{Gu:2023tcv,Gu:2023fpw}. See also \cite[Equation~(2.36)]{Closset:2026bnk} for the notational conventions that we are following in this work.
    \item \textit{The Toda presentation.} Using the quantum Whiteny relations mentioned above, one can reduce the defining relations in \eqref{whiteny present} to be completely in terms of variables $x_\bullet$~\cite[Equation~(4.4)]{Closset:2026bnk}:
    \begin{equation}\label{toda presentation}
        {\rm QK}_T({\rm Fl}(\boldsymbol{k};n))\,\cong\,\frac{\mathbb{K}[x_\bullet]^{{\rm W}_P}}{\mathcal{I}_{\rm Toda}}~.
    \end{equation}
    Here, $\mathcal{I}_{\rm Toda}$ is the new quantum ideal spanned by relations in terms of $x_\bullet$ only. Another form of the Toda presentation for the quantum K theory of partial flags was proposed and proved in \cite[theorem~3.4]{ahkmox}. These two Toda presentations were conjectured to be isomorphic to each other in \cite{Closset:2026bnk}.
    \end{itemize}

\paragraph{Computing the quantum ring relations via Gr\"obner basis.} In this work, we will be mainly focusing on the first and last presentations discussed above. Using the computational algorithm in terms of Gr\"obner basis that was developed in \cite{Closset:2023vos, Closset:2023bdr}, we compute the quantum ring relations \eqref{QK relations intro} of QK$_T({\rm Fl}(\boldsymbol{k};n))$. We review the details of this algorithm in subsection~\ref{subsec:bethe grob algorithm} for the Bethe presentation \eqref{Bethe presentation} and in subsection~\ref{subsec:toda grob algorithm} for the Toda presentation \eqref{toda presentation}. In either case, the end result is that the quantum ring can be written as follows:
\begin{equation}
    \mathcal{R}_{\rm 3d}\,\cong\,{\rm QK}_T({\rm Fl}(\boldsymbol{k};n))\,\cong\,\frac{\mathbb{K}[\CO]}{\widetilde{\CI}_{\rm Bethe}^{(\CO)}}~.
\end{equation}
Here, we are quotienting by the ideal $\widetilde{\CI}_{\rm Bethe}^{(\CO)}$ spanned by the quantum ring relations \eqref{QK relations}:
\begin{equation}\label{Ow Ow'}
    \CO_w\,\star\,\CO_{w'}\,=\,{\mathcal{N}_{w\,w'}}^{w''}(q,y)\,\CO_{w''}~. 
\end{equation}

One can instead compute the topological metric and the structure constants:
\begin{equation}\label{gww Nwww}
    g_{w\,w'}(q,y)\,:=\,\left\langle\,\SL_w\,\SL_{w'}\right\rangle_{\mathbb{P}^1\times S_\beta^1}~, \qquad \CN_{w\,w'\,w''}(q,y)\,:=\,\left\langle\,\SL_w\,\SL_{w'}\,\SL_{w''}\right\rangle_{\mathbb{P}^1\times S_\beta^1}~,
\end{equation}
in terms of which, the K-theoretic LR coefficients are given by:
\begin{equation}
    {\mathcal{N}_{w\,w'}}^{w''}\,=\,g^{w''\,v}\,\mathcal{N}_{w\,w'\,v}~.
\end{equation}
Here, repeated indices are summed over, and $g^{w\,w'}$ is the inverse of $g_{w,w'}$.

Using the 3d A-model formalism, one can compute the correlation functions using the sum-over-Bethe-vacua formula \cite{Nekrasov:2009uh,Nekrasov:2014xaa,Closset:2016arn, Closset:2017zgf} -- See \eqref{<L...>} below. As was explained in \cite{Closset:2023vos,Closset:2023bdr}, the Gr\"obner basis algorithm can be applied here as well to work out these correlation functions more efficiently. It turns out that the correlators introduced in \eqref{gww Nwww} can be written as a trace of a product of square matrices referred to as the companion matrices:
\begin{equation}\label{compMatr top and Str intro}
\begin{split}
    &g_{w\,w'}(q,y)\,=\,\Tr\left(\mathfrak{M}_{\CH}^{-1}\,\times\,\mathfrak{M}_w^{(\boldsymbol{k};n)}\,\times\,\mathfrak{M}_{w'}^{(\boldsymbol{k};n)}\right)~,\\
    &\mathcal{N}_{w\,w'\,w''}(q,y)\,=\,\Tr\left(\mathfrak{M}_{\CH}^{-1}\,\times\,\mathfrak{M}_w^{(\boldsymbol{k};n)}\,\times\,\mathfrak{M}_{w'}^{(\boldsymbol{k};n)}\,\times\,\mathfrak{M}_{w''}^{(\boldsymbol{k};n)}\right)~.
\end{split}
\end{equation}
The different notations appearing on the RHS of the above two equations are explained in section~\ref{sec: 3d A model}. As we will discuss in more detail in that section, this can be done in both the Bethe and Toda presentations, where, undoubtedly, we get the same results matching \eqref{Ow Ow'}.

\paragraph{Dual Schubert lines and K-theoretic LR coefficients.} To directly compute the K-theoretic LR coefficients appearing in \eqref{QK relations}, one can compute 3-point correlation functions with one of the operators being the \textit{dual Schubert line defect} $\SL^{\vee \,w}$. For this, one still needs to compute the topological metric and its inverse. These lines are defined as a linear combination of $\SL_w$ as follows:
\begin{equation}
    \SL^{\vee \,w}\,:=\,g^{w\,w'}(q,y)\,\SL_{w'}~.
\end{equation}
More explicitly, in terms of these dual lines, the LR coefficients are given by:
\begin{equation}
    {\mathcal{N}_{w\,w'}}^{w''}(q,y)\,=\,\left\langle\SL_w\,\SL_{w'}\,\SL^{\vee\,w''}\right\rangle_{\mathbb{P}^1\times S^1_\beta}~.
\end{equation}

From the 3d GLSM/QK theory correspondence, the line operators $\SL^{\vee\,w}$ correspond to the ideal sheaves $[\CO^{\vee \,w}]\in {\rm K}({\rm Fl}(\boldsymbol{k};n))$ -- see \cite{brion-lec} for the definition. Moreover, by definition, these lines satisfy the following property:
\begin{equation}\left\langle\SL_w\,\SL^{\vee\,w'}\right\rangle_{\mathbb{P}^1\times S^1_\beta}\,=\,\delta_w^{w'}\,=\,\begin{cases}
        1~,\qquad &w\,=\,w'~,\\
        0~,\qquad &w\,\neq\,w'~.
    \end{cases}
\end{equation}
In subsection~\ref{subsec: dual lines} below, we revisit this concept and list all the Chern characters associated with these dual lines in some simple partial flag manifolds.

\paragraph{Plan of the paper.}
The article is structured as follows. In section~\ref{sec: 3d A model}, we give a lightning review of the 3d A-model associated with partial flag manifolds and review its connection with the quantum K-theory. Building on this, in section~\ref{sec:QK from Grobner}, we discuss the computational algorithm that we will follow to explicitly compute the quantum K-theory ring relations, including the topological metric and the structure constants,. We do this in both the Bethe and Toda presentations. We also exhibit some examples of simple partial flag manifolds.

Going to the 2d limit of the theory, in section~\ref{sec:2d A model}, we review the structure of the corresponding 2d A-model and provide some examples on how one can use the Gr\"obner basis algorithms to compute the quantum cohomology rings of partial flag manifolds.

We finish with some conclusions and comments on further directions in section~\ref{sec:conclusion}. In appendix \ref{sec:appendix}, we add some further examples on quantum K-theory ring relations for partial flag manifolds.
%%%%%%%%%%%%%%%%%%%%%%%%%%%%%%%%%%%%%%%%%%%%%%
%%%%%%%%%%%%%%%%%%%%%%%%%%%%%%%%%%%%%%%%%%%%%%
%%%%%%%%%%%%%%%%%%%%%%%%%%%%%%%%%%%%%%%%%%%%%%
%%%%%%%%%%%%%%%%%%%%%%%%%%%%%%%%%%%%%%%%%%%%%%

\section{Ingredients of the 3d A-model of partial flags} \label{sec: 3d A model}
In this section, we give a short review of some of the essential ingredients of the 3d A-model for partial flag manifolds. We will mention only the components that are directly related to our discussion in this work. We will also provide references for further details.

\begin{figure}
    \centering
    \includegraphics[width=1\linewidth]{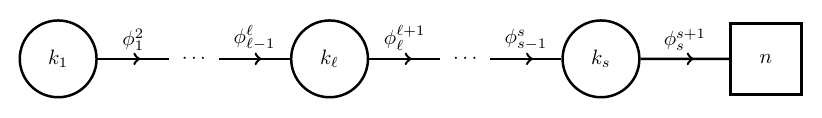}
    \caption{3d $\CN =2$ quiver gauge theory whose Higgs branch, in the geometric phase, is given by the partial flag manifold Fl$(\boldsymbol{k};n)$.}
    \label{fig:parFlag GLSM}
\end{figure}

\subsection{3d GLSMs to partial flags}
We are interested in studying the quantum K-theory ring of the partial flag manifold Fl$(\boldsymbol{k};n)$ with $\boldsymbol{k}=(k_1, \cdots, k_s)$ a vector of strictly increasing integers such that $k_s< n$. Recall that this is defined as:
\begin{equation}\label{partial-flag-defn}
	{\rm Fl}(\boldsymbol{k};n) := \left\{\,V_\bullet \equiv (0\subset V_1\subset V_2\subset\cdots\subset V_s\subset V_{s+1}\equiv\C^n)~\mid~\dim(V_\ell) = k_\ell\,\right\}~.
\end{equation}

As reviewed in \cite[section~2]{Closset:2026bnk}, from a gauge theoretic perspective, such a manifold can be realized as the Higgs branch of the 3d $\mathcal{N}=2$ quiver gauge theory described in figure \ref{fig:parFlag GLSM}. Here we have the gauge group $G = \bigtimes_{\ell=1}^s U(k_\ell)$. Between every two consecutive gauge nodes, we have a bifundamental matter multiplet $\phi_\bullet^{\bullet+1}$ and, at the last node, we have $n$ matter multiplets $\phi_{s}^{s+1}$ in the fundamental representation of $U(k_s)$.

For the (mixed) Chern--Simons (CS) levels of the theory, we will pick them to be \cite[Equations~(2.26)--(2.27)]{Closset:2026bnk}:
\begin{equation}\label{standard levels}
\begin{aligned}
    		&\kk_\ell \,=\, k_\ell\, -\, \frac{k_{\ell-1} + k_{\ell+1}}{2}~,\qquad \mathsf{l}_\ell \,=\, -1~,\qquad \forall \ell \,=\, 1,\cdots,s~, \\
            &\kk_{\ell, \ell+1} = \frac{1}{2}~,\qquad \qquad \qquad \qquad \qquad \qquad \qquad \,\forall \ell\, =\, 1,\cdots,s-1~.
\end{aligned}
\end{equation}
The first two levels are for each gauge node $U(k_\ell)_{\kk_\ell, \kk_\ell+\mathsf{l}_\ell\,k_\ell}$. Meanwhile, the last level is a mixed CS level between each two consecutive gauge nodes. For our conventions of the CS levels, see \cite[section~2.2]{Closset:2016arn}. This is the standard choice of the CS levels that guarantees that our 3d GLSM is in the geometric phase -- that is, the phase where the moduli space of vacua consists only of the Higgs branch Fl$(\boldsymbol{k};n)$ \cite{Gu:2023tcv,Gu:2023fpw,Closset:2025akk,Closset:2026bnk}.

\subsection{The 3d A-model of partial flag manifolds}
Let us now put our 3d theory on the geometry $\bbC\times S^1_\beta$ where $\beta\in \mathbb{R}_{>0}$ is the size of the circle fibre, and let us perform a topological A-twist along the complex plane $\bbC$. For more details on the 3d A-model formalism, see~\cite{Nekrasov:2009uh,Nekrasov:2014xaa,Closset:2016arn,Closset:2017zgf, Closset:2023vos}. The physical operators in the theory are half-BPS line operators wrapping the circle fibre at some point $p\in \bbC$. These line operators live in the 3d twisted chiral ring $\mathcal{R}_{\rm 3d}$ which, with the special choice of the CS levels in \eqref{standard levels}, is isomorphic to the (equivariant) quantum K-theory ring of the partial flag manifold QK$_T$(Fl($\boldsymbol{k};n$)).\footnote{The equivariance here is with respect to the $GL(n)$ isometry of the partial flag manifold. From the gauge theory perspective, this corresponds to the flavour symmetry rotating the $n$ fundamental chiral multiplets $\phi_s^{s+1}$.}

\subsubsection{The quantum ring in terms of BAEs}
To be more explicit, the defining relations of the quantum ring are given in terms of the Bethe ansatz equations (BAEs) of the 3d theory. These can be worked out starting with the effective twisted superpotential, which has the form:
% \footnote{In the first three terms we are absorbing a factor of $2\pi i \beta$ in the definition of $\widetilde{\sigma}^{(\ell)}$.}
\begin{multline}\label{eff-twisted}
    \mathcal{W} = \sum_{\ell=1}^{s}\, \left[-\,\frac{\beta}{2}\, \tr \widetilde{\sigma}^{(\ell)}\,\left(\beta\,\tr \widetilde{\sigma}^{(\ell)}\, +\, 1\right) \,+\,\frac{k_\ell\,\beta}{2}\, \sum_{a_\ell=1}^{k_\ell} \,\widetilde{\sigma}_{a_\ell}^{(\ell)}\, (\beta\,\widetilde{\sigma}_{a_\ell}^{(\ell)}\, +\, 1)\, +\, \frac{\beta\,\log \qk_\ell}{2\pi i}\,  \tr \widetilde{\sigma}^{(\ell)} \right]\\ +
    \frac{1}{(2\pi i)^2}\,\sum_{\ell=1}^{s-1} \,\sum_{a_\ell=1}^{k_\ell}\,\sum_{a_{\ell+1}=1}^{k_{\ell+1}}\, \text{Li}_2\left(x^{(\ell)}_{a_\ell}\, {x^{(\ell+1)}_{a_{\ell+1}}}^{-1}\right) \,+\, \frac{1}{(2\pi i)^2}\,\sum_{a_s=1}^{k_s} \,\sum_{i=1}^{n}\, \text{Li}_2\left(x^{(s)}_{a_s}\, y_i^{-1}\right)~.
\end{multline}
Here, $\widetilde{\sigma}^{(\ell)}_a$ are the $k_\ell$ complexified Coulomb branch parameters for each gauge group $U(k_\ell)$, and $x_a^{(\ell)}\equiv e^{-\beta\,\widetilde{\sigma}_a^{(\ell)}}$. $q_\ell$ is the exponentiation of the 3d FI parameter associated with $U(1)\subseteq U(k_\ell)$. And, $y_i$ are the flavour (equivariant) parameters associated with the $SU(n)$ flavour symmetry group.

The BAEs are defined as the critical locus of \eqref{eff-twisted}. Namely, $\exp\left(2\pi i\,\partial\mathcal{W}\right) = 1$. More explicitly, these equations are of the form:
\begin{equation}\label{BAE stand}
          (-1)^{k_\ell-1}\,\qk_\ell\, (x_{a_\ell}^{(\ell)})^{k_\ell}\, { \prod_{a_{\ell-1}=1}^{k_{\ell-1}}\,\left(1\,-\,\frac{x_{a_{\ell-1}}^{(\ell-1)}}{x_{a_\ell}^{(\ell)}}\right)}\, =\,\det x^{(\ell)}\, {\prod_{a_{\ell+1}=1}^{k_{\ell+1}}\,\left(1\,-\,\frac{x_{a_\ell}^{(\ell)}}{x_{a_{\ell+1}}^{(\ell+1)}}\right)}~,\qquad \forall \ell, a_\ell~.
\end{equation}
with the understanding that $x_b^{(s+1)}\,\equiv y_b$ for $b=1, \cdots, n$. Moreover, $\det x^{(\ell)}  = e_{k_{\ell}}(x^{(\ell)}) = x_1^{(\ell)}\,\cdots\,x_{k_{\ell}}^{(\ell)}$ is the $k_\ell$-th symmetric polynomial of the variables $x_{a_\ell}^{(\ell)}$. These equations can be rearranged into polynomial equations $\mathsf{P}_{a_\ell}^{(\ell)}=0$. In terms of these polynomials, the quantum K-theory ring of the partial flag is given as:
\begin{equation}\label{R3d=QK}
    \mathcal{R}_{\rm 3d}\,\cong\,{\rm QK}_T({\rm Fl}(\boldsymbol{k};n))\,\cong\,\frac{\mathbb{K}[x_\bullet^{(\bullet)}]^{{\rm W}_G}}{\mathcal{I}_{\rm Bethe}^{(x^{(\bullet)})}}~,
\end{equation}
where, $\mathbb{K}\equiv \mathbb{Z}(q_
\ell, y_i)$, and ${\rm W}_G \equiv \bigtimes_{\ell=1}^s S_{k_\ell}$ is the Weyl subgroup of the gauge group $G$. This takes care of the residual gauge symmetry on the Coulomb branch. Moreover, the \textit{Bethe ideal} $\mathcal{I}_{\rm Bethe}^{(x^{(\bullet)})}$ is the ideal generated by the polynomials $\mathsf{P}_{a_\ell}^{(\ell)}(x^{(\bullet)},q, y)$.

\paragraph{The set of Bethe vacua.} The solutions to the BAEs \eqref{BAE stand} give us the set of supersymmetric vacua of the 3d quiver gauge theory. More explicitly, we define the set of these vacua as:
\begin{equation}\label{SBE}
    \SBE\,:=\, \{\widehat{x}\,\equiv\,(\widehat{x}_{\bullet}^{(\bullet)})\,:\,\mathsf{P}_{a_\ell}^{(\ell)}(\widehat{x})\,=\,0~, \quad \widehat{x}^{(\ell)}_{a_\ell}\,\neq\,\widehat{x}_{b_{\ell}}^{(\ell)}~, \quad \forall a_\ell\neq b_\ell~, ~\forall\ell\}\,/\,{\rm W}_G~.
\end{equation}
Similar to what we mentioned above, here we are quotienting by the action of the residual gauge symmetry on the Coulomb branch W$_G$ to avoid overcounting of the vacuum states. The number of such vacua, i.e., the 3d flavoured Witten index \cite{Witten:1999ds}, is equal to the Euler number of the Higgs branch \cite{Witten:1981nf}:\footnote{Note that the first equality here is correct since, for partial flag manifolds, only even cohomology is non-vanishing. For cases where there are non-trivial odd cohomology classes -- for instance, for the quintic case --, it is not clear to us how this equation should be modified.}
\begin{equation}\label{|SBE|}
    \left|\SBE\right|\,=\,\chi({\rm Fl}(\boldsymbol{k};n))\,=\,\frac{n!}{\prod_{\ell=0}^{s}\,(k_{\ell+1}\,-\,k_\ell)!}~,
\end{equation}
with the convention that $k_0 \equiv 0$ and $k_{s+1}\equiv n$.

\subsubsection{Genus-$0$ correlation functions and the quantum ring structure}
Picking a basis $\{\SL_w\}_{w\in I}$ for the twisted chiral ring $\mathcal{R}_{\rm 3d}$, the ring relations -- i.e. the fusion rules of parallel lines -- can be worked out fully by computing the 2-point and 3-point functions. These are usually referred to as the topological metric and the structure constants, respectively. More explicitly, we have:
\begin{equation}\label{top met and str const}
    g_{w\,w'}(q,y)\,:=\,\left\langle\,\SL_w\,\SL_{w'}\right\rangle_{\mathbb{P}^1\times S_\beta^1}~, \qquad \CN_{w\,w'\,w''}(q,y)\,:=\,\left\langle\,\SL_w\,\SL_{w'}\,\SL_{w''}\right\rangle_{\mathbb{P}^1\times S_\beta^1}~.
\end{equation}
In terms of these two sets of correlators, the fusion rules are given by:
\begin{equation}\label{3d OPE relations}
    \SL_w\,\star\,\SL_{w'}\,=\,{\mathcal{N}_{w\,w'}}^{w''}(q,y)\,\SL_{w''}~.
\end{equation}
Here, we used the inverse of the topological metric $g^{w\,w'}(q,y)$ to raise the third index of the structure constants. Also, we are following the convention where repeated indices are summed over.

From the isomorphism \eqref{R3d=QK}, let us take the set $\{\mathcal{O}_w\}_{w\in I}$ to be the basis of the quantum K-theory ring of the partial flag corresponding to the basis of $\mathcal{R}_{\rm 3d}$ we chose above. Then, the fusion rules above are interpreted in terms of the quantum-deformed tensor product of the (equivariant) classical K-theory ring, namely:
\begin{equation}\label{QK relations}
    \CO_w\,\otimes_q\,\CO_{w'}\,=\,{\mathcal{N}_{w\,w'}}^{w''}(q,y)\,\CO_{w''}\,=\, \CO_w\,\otimes\,\CO_{w'}\,+\,q\,(\cdots)~.
\end{equation}

\paragraph{A special choice of basis.} We are interested in a particular basis of the K-theory ring of the partial flag Fl$(\boldsymbol{k};n)$. This is the one given by the Schubert classes $\mathcal{O}_w$ where $w\in {\rm W}^{(\boldsymbol{k};n)}$. For definition and further properties, see \cite[Subsection~4.1]{Closset:2025akk} for the full flag case and \cite[Subsection~3.2]{Closset:2026bnk} for more general partial flags. One point to recall here is that the indexing set ${\rm W}^{(\boldsymbol{k};n)}$ contains $S_n$ permutations $w$ that have descents at most at positions $k_1, \cdots,k_s$, and the order of this set is given by the RHS of \eqref{|SBE|}.

\paragraph{Correlation functions of Schubert line defects.} For the special choice of basis given above, the corresponding set of line operators was constructed in \cite{Closset:2025akk} for the complete flag case, and in \cite{Closset:2026bnk} for any partial flag manifold. These line operators were dubbed the Schubert line defects. Given this set of line operators, one can compute the corresponding topological metric and structure constants as defined in \eqref{top met and str const}. In the 3d A-model, and upon supersymmetric localization, these correlation functions can be computed explicitly and exactly via the following formula:\footnote{One shall note that, in addition to this Bethe-vacua formula, there is another JK residue formula \cite{Closset:2023bdr} that can be generalized to the partial flag manifolds.}
\begin{equation}\label{<L...>}
    \left\langle\SL_w\,\cdots\,\right\rangle_{\mathbb{P}^1\times S_\beta^1}\,=\,\sum_{\widehat{x}\,\in\,\SBE}\,\CH(\widehat{x},y)^{-1}\,{\rm ch}_T(\CO_w)\,\cdots~.
\end{equation}
The sum is taken over the set of Bethe vacua defined in \eqref{SBE}. To write down the explicit form of the handle-gluing operator $\CH(x,y)$, we need to recall here the form of the effective dilaton potential \cite{Nekrasov:2014xaa,Closset:2016arn,Closset:2017zgf,Closset:2023vos}:\footnote{There is also another contribution coming from the mixed gauge-$U(1)_R$ CS terms. In this discussion, we are fixing the bare CS levels such that the effective ones are trivial.}
\begin{equation}\label{omega total}
    \Omega(x,y) \,=\, \Omega_{\rm BF}(x,y)\,+\,\Omega_{\rm W-bos}(x,y)~, 
\end{equation}
where $\Omega_{\rm BF}(x,y)$ encodes the contribution of the bifundamental chiral multiplets along the quiver. More explicitly, this can be written as:
\begin{equation}\label{omega bf}
    \Omega_{\rm BF}(x,y)\,=\,\sum_{\ell=1}^{s}\,\Omega_{\rm BF}^{(\ell)}(x,y)~, \qquad \Omega_{\rm BF}^{(\ell)}(x,y)\,:=\,\frac{1}{2\pi i}\,\sum_{a=1}^{k_\ell}\,\sum_{b=1}^{k_{\ell+1}}\,\log\left(1\,-\,\frac{x_a^{(\ell)}}{x_{b}^{(\ell+1)}}\right)~,\\
    \end{equation}
 Note that here we are fixing the R-charges of all the bifundamental matter multiplets to be $0$. As for the $\Omega_{\rm W-bos}$, this comes from the $G$ W-bosons in the gauge vector multiplet. Their contribution is of the form:
\begin{equation}
      \Omega_{\rm W-bos}(x,y) \,=\,\sum_{\ell=1}^{s}\,\Omega_{\rm W-bos}^{(\ell)}(x,y)~, \qquad \Omega_{\rm W-bos}^{(\ell)}(x,y)\,:=\,-\,\frac{1}{2\pi i}\,\sum_{\substack{a,b=1\\ a\neq b}}^{k_\ell}\,\log\left(1\,-\,\frac{x_a^{(\ell)}}{x_{b}^{(\ell)}}\right)~.
\end{equation}

In terms of the two potentials $\mathcal{W}$ \eqref{eff-twisted} and $\Omega$ \eqref{omega total}, the handle-gluing operator of the theory can be worked out. This is defined as:
\begin{equation}\label{handle-gluing op}
    \mathcal{H}(x,y)\,:=\,e^{2\pi i\Omega(x,y)}\,\times\,\det({\rm Hessian}(\mathcal{W}))~, \qquad {\rm Hessian}(\mathcal{W})_{a_{\ell'}, b_{\ell}}\,:=\,
  \frac{1}{\beta^2}\,\frac{\partial^2\mathcal{W}}{\partial \sigma^{(\ell')}_{a_{\ell'}}\,\partial \sigma_{b_{\ell}}^{(\ell)}}~,
\end{equation}
for all $\ell, \ell'=1, \cdots, s$ and all corresponding $a_\ell, b_{\ell'}$. Note that the Hessian matrix of $\CW$ is of size $\rk(G)\times\rk(G)$ with $\rk(G)$ being the rank of the gauge group $G$. From \eqref{eff-twisted}, one can write down the explicit form of the elements of this matrix \cite[Equation~(2.18)]{Closset:2023vos}:
\begin{equation}
    {\rm Hessian}(\CW)_{a_{\ell'}, b_{\ell}}\,=\,\delta_{\ell, \ell'-1}\,{\rm H}^{(1)}_{a_{\ell'}, b_{\ell}}\,+\,\delta_{\ell,\ell'}\,{\rm H}^{(2)}_{a_{\ell'}, b_{\ell}}\,+\,\delta_{\ell, \ell'+1}\,{\rm H}^{(3)}_{a_{\ell'}, b_{\ell}}\,+\,\delta_{\ell,\ell'}\,(\delta_{a_{\ell'}, b_{\ell}}\,k_\ell\,-\,1)~.
\end{equation}
where, 
\begin{equation}
    \begin{split}
        &{\rm H}^{(1)}_{a_{\ell'}, b_{\ell}}(x,y)\,=\,-\,\frac{x_{b_{\ell}}^{(\ell)}/x_{a_{\ell'}}^{(\ell')}}{1\,-\,x_{b_{\ell}}^{(\ell)}/x_{a_{\ell'}}^{(\ell')}}~,\\
        &{\rm H}^{(2)}_{a_{\ell'},b_\ell}(x,y)\,=\,\delta_{b_\ell, a_{\ell'}}\,\left(\sum_{c_{\ell'+1}}^{k_{\ell'+1}}\,\frac{x_{a_{\ell'}}^{(\ell')}/x_{c_{\ell'+1}}^{(\ell'+1)}}{1\,-\,x_{a_{\ell'}}^{(\ell')}/x_{c_{\ell'+1}}^{(\ell'+1)}}\,+\,\sum_{c_{\ell'-1}}^{k_{\ell'-1}}\,\frac{x_{c_{\ell'-1}}^{(\ell'-1)}/x_{a_{\ell'}}^{(\ell')}}{1\,-\,x_{c_{\ell'-1}}^{(\ell'-1)}/x_{a_{\ell'}}^{(\ell')}}\right)~,\\
        &{\rm H}^{(3)}_{a_{\ell'},b_\ell}(x,y)\,=\,-\,\frac{x_{a_{\ell'}}^{(\ell')}/x_{b_{\ell}}^{(\ell)}}{1\,-\,x_{a_{\ell'}}^{(\ell')}/x_{b_{\ell}}^{(\ell)}}~.
    \end{split}
\end{equation}

As for the last ingredient appearing on the RHS of \eqref{<L...>}, this is the equivariant Chern character of the Schubert class $\mathcal{O}_w\in{\rm QK}_T({\rm Fl}(\boldsymbol{k};n))$ associated with $\SL_w$. Recall that:
\begin{equation}\label{ch defn}
    {\rm ch}\,:\,{\rm K}({\rm Fl}(\boldsymbol{k};n))\,\longrightarrow\,{\rm H}^\bullet({\rm Fl}(\boldsymbol{k};n))
\end{equation}
is an isomorphism. Following the discussion in \cite{Huq-Kuruvilla:2024tsg, Closset:2026bnk}, the Chern character of the Schubert class $\CO_w$ is given by the parabolic (double) Whitney polynomial $\mathfrak{W}_w^{(\boldsymbol{k};n)}(x,y)$. For definition, see \cite[Theorem~5.9]{Huq-Kuruvilla:2024tsg} and \cite[Equation~(4.9)]{Closset:2026bnk}.

%%%%%%%%%%%%%%%%%%%%%%%%%%%%%%%%%%%%%%%%%%%%%%
%%%%%%%%%%%%%%%%%%%%%%%%%%%%%%%%%%%%%%%%%%%%%%

%%%%%%%%%%%%%%%%%%%%%%%%%%%%%%%%%%%%%%%%%%%%%%
%%%%%%%%%%%%%%%%%%%%%%%%%%%%%%%%%%%%%%%%%%%%%%
%%%%%%%%%%%%%%%%%%%%%%%%%%%%%%%%%%%%%%%%%%%%%%
%%%%%%%%%%%%%%%%%%%%%%%%%%%%%%%%%%%%%%%%%%%%%%
\section{Quantum K-theory ring and Gr\"{o}bner basis algorithm}
\label{sec:QK from Grobner}

In this section, we discuss an algorithm to compute the quantum ring relations for QK$_T({\rm Fl}(\boldsymbol{k};n))$. We do this using the Gr\"obner basis algorithm and the companion matrix technique that were discussed in \cite{Closset:2023vos, Closset:2023bdr}.

\subsection{Quantum K-theory relations from the BAEs}
In this subsection, we will show how, starting with the BAEs of the theory \eqref{BAE stand}, one can directly compute the ring relations \eqref{QK relations} defining the quantum K-theory of the partial flag. We will discuss this algorithm using two different presentations of the quantum K-theory ring of the partial flag. We will refer to these presentations as the \textit{Bethe} and \textit{Toda} presentations. 

Briefly, in the Bethe presentation, the quantum ring is written in terms of the 3d Coulomb branch parameters -- i.e., the K-theoretic Chern roots of the tautological bundles on the partial flag. Meanwhile, in the Toda case, we rewrite the quantum ring in terms of the quotient bundle Chern roots. These two are connected via the quantum Whitney relations of the flag manifold, as we will review later in this subsection. For more discussion, see \cite[section~4]{Closset:2026bnk}

\subsubsection{Quantum K-theory ring in the Bethe presentation}\label{subsec:bethe grob algorithm}
Recall from \eqref{R3d=QK} that the quantum K-theory of the partial flag manifold Fl$(\boldsymbol{k};n)$ can be computed directly by computing the 3d twisted chiral ring of the corresponding 3d A-model. What we would like to do now is to perform this computation explicitly using the Gröbner basis algorithm as discussed in \cite{Closset:2023vos,Closset:2023bdr}. The details of this technique are explained in full generality in \cite[Subsection~2.3]{Closset:2023vos}. Therefore, here we will give a brief review adapted to our 3d GLSM given in figure \ref{fig:parFlag GLSM}.

Our starting point is the Bethe ideal $\CI_{\rm Bethe}^{(x^{(\bullet)})}$ that we introduced around \eqref{R3d=QK}. Recall, this ideal is generated by the polynomials $\mathsf{P}_\bullet^{(\bullet)}(x^{(\bullet)},q, y)$ that we get from the BAEs \eqref{BAE stand}. As pointed out in \eqref{SBE}, when solving for the Bethe vacua, one avoids the non-physical solutions where $\widehat{x}^{(\ell)}_{a_\ell}\,=\,\widehat{x}^{(\ell)}_{b_\ell}$ for $a_\ell\,\neq\,b_\ell$. This is implemented at the level of the Bethe ideal by augmenting the following set of polynomials to it:
\begin{equation}
    \widetilde{\mathsf{P}}_{a_\ell,b_\ell}^{(\ell)}\,:=\,\frac{\mathsf{P}_{a_\ell}^{(\ell)}\,-\,\mathsf{P}_{b_\ell}^{(\ell)}}{x_{a_\ell}^{(\ell)}\,-\,x_{b_\ell}^{(\ell)}}\,\in\,\mathbb{K}[x^{(\bullet)}]~, \qquad \forall a_\ell\,> \,b_{\ell}~,\,\forall\ell~.
\end{equation}
We will denote the new Bethe ideal after adding these new polynomials by $\widetilde{\CI}_{\rm Bethe}^{(x^{(\bullet)})}$. Another way non-physical vacua can appear is in cases where at least one of the Coulomb branch variables is equal to zero. To discard these solutions, we further introduce a new set of variables $\mathsf{z}_\ell$ and the corresponding polynomials \cite[Equation~(2.62)]{Closset:2023vos}:
\begin{equation}
    \mathsf{Z}_\ell(x^{(\bullet)},\mathsf{z})\,:=\,\mathsf{z}_\ell\,\det x^{(\ell)}\,-\,1~,\qquad \ell=1, \cdots, s~.
\end{equation}
So, at this stage, our Bethe ideal is defined as:
\begin{align}\label{tilde IBE x z}
    \widetilde{\CI}_{\rm Bethe}^{(x^{(\bullet)},\mathsf{z})}\,:=\,\left\langle\mathsf{P}_{\bullet}^{(\bullet)}~,~\widetilde{\mathsf{P}}_{\bullet,\bullet}^{(\bullet)}~,~\mathsf{Z}_\bullet\right\rangle\,\subset\,\mathbb{K}[x^{(\bullet)}, \mathsf{z}]^{{\rm W}_G}~.
\end{align}
The \textit{Bethe variety} $\mathcal{V}_{\rm BE}$ is defined as the zero set of the ideal $\widetilde{\CI}_{\rm Bethe}^{(x^{(\bullet)},\mathsf{z})}$. For our case here, this variety is zero-dimensional and its order $|\mathcal{V}_{\rm BE}|$ is none other than the number of Bethe vacua of the 3d theory \eqref{|SBE|} \cite[Equation~(2.67)]{Closset:2023vos}.

\medskip
\noindent
\textbf{Symmetrization and quantum ring relations: the final step.} Now we come to the last step of deriving the relations of QK$_T$(Fl$(\boldsymbol{k};n)$) from the Bethe ideal. This is where we take advantage of the residual Weyl symmetry W$_G$ on the Coulomb branch of the theory. As indicated in \eqref{tilde IBE x z}, this symmetry is exhibited in the elements of the Bethe ideal being symmetric in the variables $x^{(\bullet)}$; therefore, one can rewrite the basis of this ideal in terms of other symmetric polynomials \cite[Subsection~4.1]{Closset:2023bdr}. We choose these polynomials to be the parabolic double Whitney polynomials $\mathfrak{W}_w^{(\boldsymbol{k};n)}(x^{(\bullet)},y)$ mentioned around \eqref{ch defn} for each possible permutation $w\in{\rm W}^{(\boldsymbol{k};n)}$. 

Analogous to \cite[Equation~(4.5)]{Closset:2023bdr}, we implement this new basis via introducing a new set of variables $\mathcal{O}_w$ and adding the polynomials:
\begin{equation}\label{symmetrization step}
    \mathsf{W}_w^{(\boldsymbol{k};n)}(x^{(\bullet)}, y,\mathcal{O}_w)\,:=\,\mathfrak{W}_w^{(\boldsymbol{k};n)}(x^{(\bullet)},y)\,-\,\CO_w\,\in\,\mathbb{K}[x^{(\bullet)},\mathcal{O}_w]~,\qquad \forall w\,\in\,{\rm W}^{(\boldsymbol{k};n)}~,
\end{equation}
to the ideal \eqref{tilde IBE x z}. We can eliminate all $x^{(\bullet)}, \textsf{z}$ variables in the new ideal and get:\footnote{This step is easily performed in \textsc{Mathematica} using the \textit{Singular.m} package \cite{Singularm} which is an interface to \textsc{Singular} \protect\cite{DGPS}.}
\begin{equation}
    \widetilde{\mathcal{I}}_{\rm Bethe}^{(\CO)}\,:=\,\widetilde{\CI}_{\rm Bethe}^{(x^{(\bullet)}, \textsf{z}, \CO)}\mid_{\rm reduce}\,\subset\,\mathbb{K}[\CO]~.
\end{equation}
The defining relations for this ideal are none other than the quantum relations \eqref{QK relations} and the 3d twisted chiral ring $\mathcal{R}_{\rm 3d}$ \eqref{R3d=QK} is given by \cite[Equation~(4.8)]{Closset:2023bdr}:
\begin{equation}\label{eq:quotient_ring_3d}
    \mathcal{R}_{\rm 3d}\,\cong\,{\rm QK}_T({\rm Fl}(\boldsymbol{k};n))\,\cong\,\frac{\mathbb{K}[\CO]}{\widetilde{\CI}_{\rm Bethe}^{(\CO)}}~.
\end{equation}

\subsubsection{Quantum K-theory in the Whitney and Toda presentations}\label{subsec:toda grob algorithm}
The Whitney-type presentation of the quantum K-theory of partial flags was first conjectured in \cite{Gu:2023tcv}, and it was later proven in \cite{Gu:2025abc}. This presentation can be understood as the quantum deformation of the classical Whitney relations associated with the short exact sequences:
\begin{equation}\label{eq:flsec}
    0\, \longrightarrow \,\mathcal{S}_\ell \,\longrightarrow\, \mathcal{S}_{\ell+1}\, \longrightarrow\, \mathcal{Q}_{\ell+1}\, \longrightarrow 0~,
\end{equation}
for $\ell=1,\dots, s$, where $\mathcal{S}_\ell$ is the $\ell$-th tautological vector bundle on Fl$(\boldsymbol{k};n)$, and $\mathcal{Q}_{\ell+1} = \mathcal{S}_{\ell+1} / \mathcal{S}_{\ell} $ is the quotient bundle. We shall introduce $\mathcal{S}_0 = 0$ and $\mathcal{S}_{s+1} = \mathbb{C}^n$ for convenience.

For a vector bundle $E$ of rank $\mathsf{r}$ over the partial flag manifold Fl$(\boldsymbol{k};n)$, the Hirzebruch $\lambda$-class $\lambda_t(E)\in {\rm K}_T(\text{Fl}(\boldsymbol{k};n))[t]$ is defined as:
\[
    \lambda_t(E) \:=\: 1 \,+ \,t\,E\, +\, \cdots\, +\, t^{\mathsf{r}}\, \wedge^{\mathsf{r}} E ~.
\]
In terms of these classes, the quantum K-Whitney relations are given by \cite{Gu:2023tcv, Gu:2023fpw}:
\begin{equation}
    \lambda_t(\mathcal{S}_{\ell}) \otimes_q \lambda_t(\mathcal{Q}_{\ell+1}) \:=\: \lambda_t(\mathcal{S}_{\ell+1}) - t^{k_{\ell+1} - k_\ell} \frac{q_\ell}{1- q_\ell} \det(\mathcal{Q}_{\ell+1})\otimes_q \left( \lambda_t(\mathcal{S}_{\ell}) - \lambda_t(\mathcal{S}_{\ell-1})  \right)\,.
\end{equation}
Following the conventions in \cite{Closset:2026bnk}, let 
\[
x^{(\ell)}=\left\{x_{a_{\ell}}^{(\ell)}: a_\ell = 1,\dots, k_{\ell}\right\}~,\quad \text{and}\quad  \widehat{x}^{(\ell)}=\left\{x_j : j={k_{\ell-1}+1 },\dots,{k_{\ell}} \right\}
\]
denote the set of the K-theoretical Chern roots of $\mathcal{S}_{\ell}$ and the set of the K-theoretical Chern roots of $\mathcal{Q}_{\ell}$, respectively. In the special case when $\ell = s+1$, denote $y_i = x_i^{(s+1)}$ for $i=1,\dots,n$, which are the equivariant parameters. Then, the $\lambda$-classes $\lambda_t(\mathcal{S}_{\ell})$ and $\lambda_t(\mathcal{Q}_{\ell})$ can be represented by:
\begin{equation}
    \lambda_t(\mathcal{S}_{\ell})\:=\: \prod_{a_\ell=1}^{k_\ell}\,\left( 1\,+\,t\, x_{a_{\ell}}^{(\ell)} \right)~,  \qquad \lambda_t(\mathcal{Q}_{\ell})\:=\: \prod_{j=k_{\ell-1}+1}^{k_\ell}\,\left( 1\,+\,t\, x_{j} \right)~.
\end{equation}
Therefore, the quantum K-ring relations can be written in a more explicit form as follows:
\begin{multline} \label{eq:qkwhitney}
    \sum_{i+j=p} e_i(x^{(\ell)})\, e_j(\widehat{x}^{(\ell+1)}) \:=\: e_{p}(x^{(\ell+1)})  \\
     -\, \frac{q_\ell}{1\,-\,q_\ell} \,\Big(\prod_{m=k_\ell+1}^{k_{\ell+1}} x_m\Big)\, \left( e_{p-k_{\ell+1}+k_\ell}(x^{(\ell)}) \,-\,  e_{p-k_{\ell+1}+k_\ell}(x^{(\ell-1)})  \right)~,
\end{multline}
for $p = 0,1,\dots, k_{\ell+1}$, and $\ell = 0,\dots,s$. In the above, $e_i(x)$ is the $i$-th elementary symmetric polynomial. Let $\mathcal{I}_{\rm Whitney}$ be the ideal generated by the quantum Whitney relations given in \eqref{eq:qkwhitney}, then QK$_T(\text{Fl}(\boldsymbol{k};n))$ is isomorphic to the following quotient ring:
\begin{equation}\label{eq:whitneyqr}
    {\rm QK}_T(\text{Fl}(\boldsymbol{k};n)) \:\cong\: \frac{\mathbb{K}[x^{(\bullet)},x]^{{\rm W}_G\times {\rm W}_P}}{\mathcal{I}_{\rm Whitney}}~.
\end{equation}
Here, ${\rm W}_G$ is again the Weyl subgroup of the gauge group $G$, acting on tautological bundle variables $x^{(\bullet)}$, and ${\rm W}_P\equiv\bigtimes_{\ell=1}^s S_{k_\ell - k_{\ell-1}} $ is the Weyl group of the parabolic subgroup of Fl$(\boldsymbol{k};n)$, acting on the quotient bundle variables $x$. The ideal $\mathcal{I}_{\rm Whitney}$ is invariant under the Weyl group ${\rm W}_G\times {\rm W}_P$ due to the apparent invariance of \eqref{eq:qkwhitney}, therefore the quotient ring \eqref{eq:whitneyqr} is well-defined.

One can further eliminate the tautological bundle variables $x^{(\bullet)}$ in the quotient ring \eqref{eq:whitneyqr} and obtain the Toda presentation for the quantum K-ring of Fl$(\boldsymbol{k};n)$. Note that \eqref{eq:qkwhitney} can be viewed as recursive relations for $e_p(x^{(\ell)})$ as:
\begin{multline}\label{eq:recursive}
    e_p(x^{(\ell+1)}) \:= \:\sum_{i+j=p} e_i (x^{(\ell)}) e_j(\widehat{x}^{(\ell+1)}) \\
    + \frac{q_\ell}{1-q_\ell} \Big(\prod_{m=k_\ell+1}^{k_{\ell+1}} x_m\Big) \left( e_{p-k_{\ell+1}+k_\ell}(x^{(\ell)}) -  e_{p-k_{\ell+1}+k_\ell}(x^{(\ell-1)})  \right)\,,
\end{multline}
with the initial conditions that $e_0(x^{(\ell)})=1$ for all $\ell$ and $e_i(x^{(0)})=0$ for all $i>0$, and the convention that $e_i(x^{(\ell)})=0$ for all $i>k_{\ell}$ or $i<0$. By these recursive relations, $e_p(x^{(\ell)})$ can be expressed as the {\it K-theoretic parabolic quantum elementary polynomials} ${{\rm F}^{({\boldsymbol{k}};n)}}^\bullet_\bullet(x;q)$, which were first introduced in \cite{Closset:2026bnk}:
\begin{equation} \label{eq:eliminate}
    e_i(x^{(\ell)}) \:=\: {{\rm F}^{({\boldsymbol{k}};n)}}^{k_{\ell}}_i(x;q)~.
\end{equation}
See \cite[Subsection~4.4]{Closset:2026bnk} for more details on the K-theoretic parabolic quantum elementary polynomials and their generating function. 

The equations in $\mathcal{I}_{\rm Whitney}$ and the tautological bundle variables can be eliminated iteratively using the relations \eqref{eq:eliminate} until $\ell = s+1$. In the final step when $\ell = s+1$ and $k_{s+1}=n$, the remaining Whitney relations become the Toda relations:
\begin{equation}\label{eq:qktoda}
    {{\rm F}^{({\boldsymbol{k}};n)}}^{n}_j(x;q) \:=\: e_j(y) ~,
\end{equation}
for $j=1,\dots,n$. The explicit formula for ${{\rm F}^{({\boldsymbol{k}};n)}}^{n}_i(x;q)$ was conjectured in \cite[Equation~(4.33)]{Closset:2026bnk}, and we add it here for completeness:
\begin{equation}
    {{\rm F}^{({\boldsymbol{k}};n)}}^{n}_i(x;q) \:=\: \sum_{\substack{I\subseteq [n]\\|I| = i}}\, 
    \left[\,\prod_{j\in I}\,{x_j} \,\prod^s_{\substack{l=1\\(k_{l-1}, k_l] \cap \mathbb{Z} \cap I \neq \emptyset\\ (k_l,  k_{l+1}] \cap \mathbb{Z} \subset I}}\,\frac{1}{1\,-\,q_l}\,\right]~,
\end{equation}

Introducing the Toda ideal to be:
\begin{equation}\label{eq:qktodaideal}
    \mathcal{I}_{\rm Toda} \,:= \,\langle\, {{\rm F}^{({\boldsymbol{k}};n)}}^{n}_j(x;q) \,-\, e_j(y)\,,\, j\, =\,1\,,\,\dots\,,\,n \,\rangle~,
\end{equation}
then, from the above discussion, we see that we can write the quantum K-theory ring of the partial flag manifold as:
\begin{equation}\label{eq:todaqr}
    {\rm QK}_T(\text{Fl}(\boldsymbol{k};n)) \:\cong\: \frac{\mathbb{K}[x]^{{\rm W}_P}}{\mathcal{I}_{\rm Toda}}~.
\end{equation}
Note that the polynomials ${{\rm F}^{({\boldsymbol{k}};n)}}^{n}_i(x;q)$ are symmetric under the Weyl group ${\rm W}_{P}$, which can be verified directly from the recursive equation \eqref{eq:recursive}, and thus ensures the ${\rm W}_P$-invariance of $\mathcal{I}_{\rm Toda}$ and so the well-definedness of \eqref{eq:todaqr}. 

The Whitney presentation uses both the tautological variables $x^{(\ell)}$ and the quotient bundle variables $x_i$'s, while the Toda presentation only keeps the quotient variables. Dealing with fewer variables makes explicit computations more efficient, such as the Gr\"obner reductions and companion matrices carried out in this paper. For computational reasons, we prefer to work with the Toda presentation rather than with the Whitney presentation.

%%%%%%%%%%%%%%%%%%
\begin{table}[t!]
\centering
\begin{equation*}
        \begin{array}{|c||c|c|c|c|c|c|}
    \hline
        w & {(1\,2\,3)}&{(1\,3\,2)}& {(2\,1\,3)}&{(2\,3\,1)}&{(3\,1\,2)} &{(3\,2\,1)} \\
        \hline
        \hline
         \ell(w)&  0 &1 &1&2&2&3\\
        \hline
         \dim(X_w)& 3  & 2 &2&1&1&0\\
        \hline
        \end{array}
    \end{equation*}
\caption{For each possible permutation in $S_3$, we give its length and the dimension of the corresponding Schubert variety. Recall $\dim(X_w) = \dim ({\rm Fl}(3))-\ell(w)$.}
\label{tab:permutations for fl(3)}
\end{table}
%%%%%%%%%%%%%%%%%%%

\subsubsection{Some explicit examples}
Let us now provide some explicit examples for the algorithms discussed above. For the Grassmannian case, these two algorithms -- i.e., in the two different presentations of the quantum K-theory ring -- are the same. This was applied explicitly in \cite[Subsection~4.1]{Closset:2023bdr}. In this subsection, we will consider the example of Fl$(3)$. Further examples can be found in the appendices. In appendix \ref{app:QK Fl4}, we consider the complete flag Fl$(4)$. In appendix \ref{app: QK Fl134}, we consider the incidence flag Fl$(1,3;4)$ where we find matching with the partial results of \cite{xu2024quantum}. And, in appendix \ref{app: QK Fl124}, we consider the last partial flag case with $n=4$, which is Fl$(1,2;4)$. On the other hand, those examples can also be computed using schubmult package \cite{samuel2026triplequantumschubertcalculus,schubmult}.

\begin{table}[t!]
{\footnotesize
\begin{align} \nn
    \begin{split}
&\mathcal{O}_{(2\,1\,3)} \,\otimes_q\, \mathcal{O}_{(2\,1\,3)} \,=\, \left( 1\,-\, \frac{y_2}{y_1}\right)\,\mathcal{O}_{(2\,1\,3)} \,+\, \frac{y_2}{y_1} \,\mathcal{O}_{(3\,1\,2)}\, +\, q_1 \,\frac{ y_2}{y_1}\, \left(1\,-\,\mathcal{O}_{(1\,3\,2)}\right)~,\\
        &\mathcal{O}_{(2\,1\,3)} \,\otimes_q \,\mathcal{O}_{(1\,3\,2)} \,=\, \mathcal{O}_{(2\,3\,1)}\,-\,\mathcal{O}_{(3\,2\,1)}\,+\,\mathcal{O}_{(3\,1\,2)}~,\\
        &\mathcal{O}_{(2\,1\,3)} \,\otimes_q\, \mathcal{O}_{(2\,3\,1)} \,=\, \left( 1\,-\,\frac{y_2}{y_1} \right)\, \mathcal{O}_{(2\,3\,1)} \,+\, \frac{y_2}{y_1} \mathcal{O}_{(3\,2\,1)}~,\\
        &\mathcal{O}_{(2\,1\,3)}\, \otimes_q\, \mathcal{O}_{(3\,1\,2)} \,= \,\left(1\,-\,\frac{y_3}{y_1}\right)\, \mathcal{O}_{(3\,1\,2)}\, +\, q_1\,\frac{ y_3}{y_1}\, \mathcal{O}_{(1\,3\,2)}~,\\
        &\mathcal{O}_{(2\,1\,3)} \,\otimes_q\, \mathcal{O}_{(3\,2\,1)} \,=\, \left( 1\,-\,\frac{y_3}{y_1}\right)\,\mathcal{O}_{(3\,2\,1)} \,+\,q_1\, \frac{y_3}{y_1}\, \mathcal{O}_{(2\,3\,1)} \,+\,  q_1\, q_2\,\frac{ y_3}{y_1}\, \left(1\,-\,\mathcal{O}_{(2\,1\,3)}\right)~,\\
        &\mathcal{O}_{(1\,3\,2)}\, \otimes_q\, \mathcal{O}_{(1\,3\,2)} \,=\, \left(1\,-\,\frac{y_3}{y_2}\right)\,\mathcal{O}_{(1\,3\,2)} \,+\, \frac{y_3}{y_2} \left(q_2 \,(1\,-\,\mathcal{O}_{(2\,1\,3)})\,+\,\mathcal{O}_{(2\,3\,1)}\right)~,\\
        &\mathcal{O}_{(1\,3\,2)}\, \otimes_q\, \mathcal{O}_{(2\,3\,1)}\, = \,\left(1\,-\,\frac{y_3}{y_1} \right) \,\mathcal{O}_{(2\,3\,1)}\, +\, q_2\, \frac{y_3 }{y_1}\,\mathcal{O}_{(2\,1\,3)}~,\\
        &\mathcal{O}_{(1\,3\,2)} \,\otimes_q\, \mathcal{O}_{(3\,1\,2)} \,=\,\left(1\,-\,\frac{y_3}{y_2} \right)\,\mathcal{O}_{(3\,1\,2)}\, +\, \frac{y_3}{y_2} \,\mathcal{O}_{(3\,2\,1)}~,\\
        &\mathcal{O}_{(1\,3\,2)} \,\otimes_q\, \mathcal{O}_{(3\,2\,1)} \,=\,\left( 1\,-\,\frac{y_3}{y_1}\right)\,\mathcal{O}_{(3\,2\,1)} \,+\, q_1\, q_2 \,\frac{ y_3}{y_1} \,\left(1\,-\,\mathcal{O}_{(1\,3\,2)}\right)\, + \,q_2\, \frac{y_3}{y_1}\, \mathcal{O}_{(3\,1\,2)}~,\\
        &\mathcal{O}_{(2\,3\,1)} \,\otimes_q\, \mathcal{O}_{(2\,3\,1)}\, =\, \left(1\,-\,\frac{y_2}{y_1}\right) \,\left(1\,-\,\frac{y_3}{y_1}\right) \,\mathcal{O}_{(2\,3\,1)}\, +\, q_2\, \frac{y_2}{y_1}\, \mathcal{O}_{(3\,1\,2)}\,+\,q_2\,\frac{y_3}{y_1}\,\left(1\,-\, \frac{y_2}{y_1}\right) \,\mathcal{O}_{(2\,1\,3)}~,\\
        &\mathcal{O}_{(2\,3\,1)}\, \otimes_q\, \mathcal{O}_{(3\,1\,2)} \,=\, \left(1\,-\,\frac{y_3}{y_1}\right)\,\mathcal{O}_{(3\,2\,1)}\, +\,q_1\, q_2\, \frac{y_3}{y_1}~,\\
        &\mathcal{O}_{(2\,3\,1)}\, \otimes_q\, \mathcal{O}_{(3\,2\,1)} \,=\, \left( 1\,-\, \frac{y_2}{y_1}\right)\,\left(1 \,- \,\frac{y_3}{y_1} \right)\,\mathcal{O}_{(3\,2\,1)}\, +\, q_2\, \frac{y_2}{y_1}\,\left(1\,-\,\frac{y_3}{y_1}\right)\, \mathcal{O}_{(3\,1\,2)}\, +\, q_1\, q_2\,\frac{y_2\, y_3}{y_1^2}\, \mathcal{O}_{(1\,3\,2)}\, \\&\qquad \qquad\qquad \qquad+\, q_1 \,q_2\,\frac{y_3}{y_1}\, \left(1\,-\,\frac{y_2}{y_1} \right)~,\\
        &\mathcal{O}_{(3\,1\,2)} \,\otimes_q\, \mathcal{O}_{(3\,1\,2)} \,=\, \left(1\,-\,\frac{y_3}{y_1}\right)\, \left(1\,-\,\frac{y_3}{y_2}\right)\, \mathcal{O}_{(3\,1\,2)} \,+\, q_1\, \frac{y_3}{y_2} \,\mathcal{O}_{(2\,3\,1)}\,+\,q_1\,\frac{y_3}{y_1}\,\left(1\,-\,\frac{y_3}{y_2}\right) \mathcal{O}_{(1\,3\,2)}~,\\
        &\mathcal{O}_{(3\,1\,2)}\, \otimes_q \,\mathcal{O}_{(3\,2\,1)} \,=\, \left( 1\,-\,\frac{y_3}{y_1}\right)\,\left(1\,-\,\frac{y_3}{y_2}\right)\, \mathcal{O}_{(3\,2\,1)}\, +\, q_1\, \frac{y_3}{y_2}\,\left(1\,-\,\frac{y_3}{y_1} \right)\, \mathcal{O}_{(2\,3\,1)} \,+\, q_1\, q_2\, \frac{y_3^2}{y_1\, y_2} \,\mathcal{O}_{(2\,1\,3)}\\&\qquad \qquad \qquad \qquad+\, q_1 \,q_2\, \frac{y_3}{y_1}\,\left(1\, -\,\frac{y_3}{y_2} \right)~,\\
        &\mathcal{O}_{(3\,2\,1)} \,\otimes_q\, \mathcal{O}_{(3\,2\,1)}\, = \,\left[\left(1\,-\,\frac{y_2}{y_1}\right)\,\left(1\,-\,\frac{y_3}{y_1}\right)\,\left(1\,-\,\frac{y_3}{y_2}\right)\,-\, q_1\, q_2\, \frac{y_3}{y_1}\right]\,\mathcal{O}_{(3\,2\,1)} \,\\
        &\qquad \qquad \qquad\qquad +\, \left[q_1\, \left(1\,-\,\frac{y_2}{y_1}\right)\,\left(1\,-\,\frac{y_3}{y_1}\right)\, \frac{y_3}{y_2}\,+\, q_1\, q_2\, \frac{y_3}{y_1}\right]\,\mathcal{O}_{(2\,3\,1)} \,+\,q_1\,q_2\,\left(\frac{y_2}{y_1} \,-\, \frac{y_3}{y_1}\right)\,\frac{y_3}{y_1}\, \mathcal{O}_{(1\,3\,2)}\\
        &\qquad \qquad  \qquad \qquad \,+\,\left[ q_2\, \left(1\,-\,\frac{y_3}{y_1}\right)\,\left(\frac{y_2}{y_1}\,-\,\frac{y_3}{y_1}\right) \,+\, q_1 \,q_2\,\frac{y_3}{y_1} \right]\,\mathcal{O}_{(3\,1\,2)} \,+\, q_1 \,q_2\, \frac{y_3}{y_1}\,\left(\frac{y_3}{y_2}\,-\,\frac{y_3}{y_1} \right)\,\mathcal{O}_{(2\,1\,3)}\\
        &\qquad\qquad\qquad \qquad+ \,q_1 \,q_2\, \left(1\,-\,\frac{y_2}{y_1}\right)\,\left(1\,-\,\frac{y_3}{y_2}\right)\,\frac{y_3}{y_1}~.
  \end{split}
 \end{align}}
\caption{The equivariant QK product for the complete flag Fl$(3)$.  \label{tab:Fl3equivQK}}
\end{table}

\medskip
\noindent
\textbf{The quantum K-theory ring of Fl$(3)$.} For the case of the complete flag Fl$(3)$, recall that we have $6$ possible Schubert classes as listed in table \ref{tab:permutations for fl(3)}. The equivariant ring relations are given in table \ref{tab:Fl3equivQK}. Part of these relations match with the partial results of \cite[Theorem~4.5]{xu2024quantum}. Meanwhile, the remaining relations can be viewed as an extension to those of \cite{xu2024quantum}. 

In the non-equivariant limit where we send $y_i\rightarrow 1$, we get the following results:
{\small\begin{align} \label{QKFL3}
&\mathcal{O}_{(2\,1\,3)} \,\otimes_q\, \mathcal{O}_{(2\,1\,3)} \,=\, \mathcal{O}_{(3\,1\,2)}\, +\, q_1 \, \left(1\,-\,\mathcal{O}_{(1\,3\,2)}\right)~,\\
        &\mathcal{O}_{(2\,1\,3)} \,\otimes_q \,\mathcal{O}_{(1\,3\,2)} \,=\, \mathcal{O}_{(2\,3\,1)}\,-\,\mathcal{O}_{(3\,2\,1)}\,+\,\mathcal{O}_{(3\,1\,2)}~,\\
        &\mathcal{O}_{(2\,1\,3)} \,\otimes_q\, \mathcal{O}_{(2\,3\,1)} \,=\, \mathcal{O}_{(3\,2\,1)}~,\\
        &\mathcal{O}_{(2\,1\,3)}\, \otimes_q\, \mathcal{O}_{(3\,1\,2)} \,= \, q_1\, \mathcal{O}_{(1\,3\,2)}~,\\
        &\mathcal{O}_{(2\,1\,3)} \,\otimes_q\, \mathcal{O}_{(3\,2\,1)} \,=\, q_1\, \mathcal{O}_{(2\,3\,1)} \,+\,  q_1\, q_2\, \left(1\,-\,\mathcal{O}_{(2\,1\,3)}\right)~,\\
        &\mathcal{O}_{(1\,3\,2)}\, \otimes_q\, \mathcal{O}_{(1\,3\,2)} \,=\,  q_2 \,(1\,-\,\mathcal{O}_{(2\,1\,3)})\,+\,\mathcal{O}_{(2\,3\,1)}~,\\
        &\mathcal{O}_{(1\,3\,2)}\, \otimes_q\, \mathcal{O}_{(2\,3\,1)}\, = \, q_2\,\mathcal{O}_{(2\,1\,3)}~,\\
        &\mathcal{O}_{(1\,3\,2)} \,\otimes_q\, \mathcal{O}_{(3\,1\,2)} \,= \,\mathcal{O}_{(3\,2\,1)}~,\\
        &\mathcal{O}_{(1\,3\,2)} \,\otimes_q\, \mathcal{O}_{(3\,2\,1)} \,=\, q_1\, q_2  \,\left(1\,-\,\mathcal{O}_{(1\,3\,2)}\right)\, + \,q_2\, \mathcal{O}_{(3\,1\,2)}~,\\
        &\mathcal{O}_{(2\,3\,1)} \,\otimes_q\, \mathcal{O}_{(2\,3\,1)}\, =\,  q_2\,  \mathcal{O}_{(3\,1\,2)}~,\\
        &\mathcal{O}_{(2\,3\,1)}\, \otimes_q\, \mathcal{O}_{(3\,1\,2)} \,=\, q_1\, q_2~,\\
        &\mathcal{O}_{(2\,3\,1)}\, \otimes_q\, \mathcal{O}_{(3\,2\,1)} \,=\, q_1\, q_2\, \mathcal{O}_{(1\,3\,2)}~,\\
        &\mathcal{O}_{(3\,1\,2)} \,\otimes_q\, \mathcal{O}_{(3\,1\,2)} \,=\,  q_1 \,\mathcal{O}_{(2\,3\,1)}~,\\
        &\mathcal{O}_{(3\,1\,2)}\, \otimes_q \,\mathcal{O}_{(3\,2\,1)} \,=\, q_1\, q_2 \,\mathcal{O}_{(2\,1\,3)}~,\\
        &\mathcal{O}_{(3\,2\,1)} \,\otimes_q\, \mathcal{O}_{(3\,2\,1)}\, = \,-\, q_1\, q_2\,\mathcal{O}_{(3\,2\,1)} \,+\, q_1\, q_2\,\mathcal{O}_{(2\,3\,1)}\,+\, q_1 \,q_2\,\mathcal{O}_{(3\,1\,2)}~.
 \end{align}}
This matches exactly with \cite[Theorem~5.1]{xu2024quantum} and reproduces the relations of \cite[Subsection~5.1.2]{Gu:2023fpw}

%%%%%%%%%%%%%%%%%%%%%%%%%%%%%%%%%%%%%%%%%%%%%%
%%%%%%%%%%%%%%%%%%%%%%%%%%%%%%%%%%%%%%%%%%%%%%
\subsection{Quantum K-theory relations from the companion matrix}\label{QK from ComMatrix}
In this subsection, we will discuss a different direction to computing the quantum ring relations of QK$_T$(Fl($\boldsymbol{k};n$)), which involves the explicit calculation of the K-theoretic Littlewood--Richardson (LR) coefficients appearing on the RHS of \eqref{QK relations}. This involves computing the distinct components of the topological metric and structure constants \eqref{top met and str const} using the localization formula \eqref{<L...>}.

\subsubsection{Topological correlators via the companion matrix}\label{subsec: TopinComMat}
In this subsection, we discuss the computation of the 3d A-model correlation functions \eqref{<L...>} via the companion matrix method. Let us start with the general discussion of the companion matrix method applied in the Bethe presentation of the quantum ring. For further discussion about this, see \cite[Subsection~2.3]{Closset:2023vos}. Then, we will make comments on applying this method to the Toda presentation of the quantum ring.

\medskip
\noindent
\textbf{In terms of the Bethe presentation.}
Briefly, the idea is as follows. Starting with the Bethe ideal $\CI_{\rm Bethe}^{(x^{(\bullet)})}$, One can use a computer algebra system (e.g., \textsc{singular} \cite{DGPS}) to compute a Gr\"obner basis for it. In terms of this basis, one can work out a corresponding canonical basis for the quotient ring $\mathcal{R}_{\rm 3d}$ as presented on the RHS of \eqref{R3d=QK}. As discussed in the previous subsection, one discards the non-physical vacua by applying the trick in \eqref{tilde IBE x z} and by applying the symmetrization as in \eqref{symmetrization step}.\footnote{Although one can use the parabolic Whitney polynomials $\mathfrak{W}_w^{(\boldsymbol{k};n)}(x^{(\bullet)})$ for the symmetrization, here we used the elementary symmetric polynomials as in \protect\cite[Equation~(2.60)]{Closset:2023vos}.}

To any polynomial $\mathsf{Q}\in\mathbb{K}[x^{(\bullet)}]$, one can associate a companion matrix $\mathfrak{M}_{\mathsf{Q}}$ of size $|\SBE|\times|\SBE|$ as in \cite[Equation~(2.47)]{Closset:2023vos}. This assignment satisfies the following properties:
\begin{equation}
    \begin{split}
        \sum_{\widehat{x}\in\SBE}\,\mathsf{Q}(\widehat{x})\,=\,\Tr\,\mathfrak{M}_{\mathsf{Q}}~,
    \end{split}
\end{equation}
and, 
\begin{equation}
    \mathfrak{M}_{\mathsf{Q}_1+\mathsf{Q}_2}\,=\,\mathfrak{M}_{\mathsf{Q}_1}\,+\,\mathfrak{M}_{\mathsf{Q}_2}~, \quad \mathfrak{M}_{\mathsf{Q}_1\cdot\mathsf{Q}_2}\,=\,\mathfrak{M}_{\mathsf{Q}_1}\,\times\,\mathfrak{M}_{\mathsf{Q}_2}~, \quad \mathfrak{M}_{\mathsf{Q}_1/\mathsf{Q}_2}\,=\,\mathfrak{M}_{\mathsf{Q}_1}\,\times\,(\mathfrak{M}_{\mathsf{Q}_2})^{-1}~.
\end{equation}

Let us apply this to the 2 and 3-point correlation functions given in \eqref{top met and str const}. The handle-gluing operator \eqref{handle-gluing op} is a rational function in the Coulomb branch parameters. Let us denote the corresponding companion matrix by $\mathfrak{M}_{\mathcal{H}}$. As for the Schubert line defect $\SL_w$, let us denote the companion matrix for the associated parabolic Whitney polynomial by $\mathfrak{M}_{w}^{(\boldsymbol{k};n)}$. Putting these components together, we can rewrite \eqref{top met and str const} as follows:
\begin{equation}\label{compMatr top and Str}
\begin{split}
    &g_{w\,w'}(q,y)\,=\,\Tr\left(\mathfrak{M}_{\CH}^{-1}\,\times\,\mathfrak{M}_w^{(\boldsymbol{k};n)}\,\times\,\mathfrak{M}_{w'}^{(\boldsymbol{k};n)}\right)~,\\
    &\mathcal{N}_{w\,w'\,w''}(q,y)\,=\,\Tr\left(\mathfrak{M}_{\CH}^{-1}\,\times\,\mathfrak{M}_w^{(\boldsymbol{k};n)}\,\times\,\mathfrak{M}_{w'}^{(\boldsymbol{k};n)}\,\times\,\mathfrak{M}_{w''}^{(\boldsymbol{k};n)}\right)~.
\end{split}
\end{equation}

\medskip
\noindent
\textbf{In terms of Toda presentation.}
The above calculations are carried out in the Bethe presentation of the quantum K-theory of the partial flag manifold. Instead, one can also use the Toda presentation, following the discussion we had in subsection~\ref{subsec:toda grob algorithm}. The final results are independent of the presentations. Let us first recall the basics of the regular representation. 

Let $\mathbb{K}$ be a closed field and $\mathcal{A}$ be a finite-dimensional commutative $\mathbb{K}$-algebra. The companion matrix $\mathfrak{M}_Q$ associated with $Q\in\mathcal{A}$ is actually a regular representation of $\mathcal{A}$:
\[
    \mathfrak{M}\::\: \mathcal{A} \,\longrightarrow \,{\rm Rep}(\mathcal{A})~,\qquad Q \,\longmapsto \,\mathfrak{M}_{Q}~,
\]
namely, it is obtained by associating $Q$ with a $\mathbb{K}$-linear endomorphism $\mathfrak{M}_Q$ of the underlying vector space.

Now, let us further consider two isomorphic $d$-dimensional commutative $\mathbb{K}$-algebras, $\mathcal{A}_1$ and $\mathcal{A}_2$, with the isomorphism given by: 
\[
\Phi\::\: \mathcal{A}_1 \,\xlongrightarrow{\sim}\, \mathcal{A}_2~.
\]
Then, for a given $Q\in \mathcal{A}_1$, one should have the following commutative diagram:
\[ 
\begin{tikzcd}
\mathcal{A}_1 \arrow{r}{\Phi} \arrow[swap]{d}{\mathfrak{M}_{Q}} & \mathcal{A}_2 \arrow{d}{\mathfrak{M}_{\Phi(Q)}} \\%
\mathcal{A}_1 \arrow{r}{\Phi}& \mathcal{A}_2
\end{tikzcd}
\]    
With a choice of the basis for $\mathcal{A}_1$ and $\mathcal{A}_2$, 
\[
\mathcal{B}_{\mathcal{A}_1}\:=\:\{\mathfrak{e}_1\,,\,\dots\,,\,\mathfrak{e}_d\}~,\qquad \mathcal{B}_{\mathcal{A}_2}\:=\:\{\mathfrak{f}_1\,,\,\dots\,,\,\mathfrak{f}_d\}~,
\]
one can write the isomorphism in the following matrix form:
\[ 
    \Phi(\mathfrak{e}_i) \:=\: \sum_{j}\, \Phi_{ij}\, \mathfrak{f}_j~.
\]
The commutativity of the above diagram implies that $\mathfrak{M}_Q$ and $\mathfrak{M}_{\Phi(Q)}$ are related by a conjugation with $\Phi$:
\[
    \mathfrak{M}_{\Phi(Q)} \,=\, \Phi\, \mathfrak{M}_{Q}\, \Phi^{-1}~.
\]

Applying the above general results to the Bethe presentation and the Toda presentation of ${\rm QK}_T(\text{Fl}(\boldsymbol{k};n))$, we have the following equivalence:
\[
    {\rm QK}_T(\text{Fl}(\boldsymbol{k};n)) \:\cong\: \frac{\mathbb{K}[\CO]}{\widetilde{\CI}_{\rm Bethe}^{(\CO)}} \:\cong\: \frac{\mathbb{K}[x]^{{\rm W}_P}}{\mathcal{I}_{\rm Toda}}~.
\]
With this in mind, the computation for \eqref{compMatr top and Str} in these two presentations should give the same results due to the trace in their definitions. We have also verified this in the explicit examples that we will consider below.

\medskip
\noindent
\textbf{Remark.} In general, the explicit form of the companion matrix depends on the basis we choose. Consider the automorphism from $\mathcal{A}$ to itself and follow the same discussion as above, and then the companion matrices in two different bases are related by a similarity transformation. Therefore, the topological metric and the structure constants are the same.

In addition, since both the Bethe and Toda presentations are realized as quotient-ring algebras, the companion matrix depends only on the corresponding quotient classes. More precisely, with the fixed choice of the basis, $\{[\mathfrak{e}_1],\dots, [\mathfrak{e}_d]\}$, the explicit form of the companion matrix $\mathfrak{M}_{[Q]}$ is independent of the representatives for both the basis elements $[\mathfrak{e}_i]$ and the element $[Q]$. That is, the companion matrix $\mathfrak{M}_{[Q]}$ is unchanged with $[\mathfrak{e}^\prime_i] = [\mathfrak{e}_i]$ or $[Q^\prime]=[Q]$. 

\subsubsection{Explicit examples}
In \cite[Subsection~4.2]{Closset:2023bdr}, the companion matrix method was applied to compute the topological correlators in the Grassmannian manifold case. Let us here apply it to the complete flag Fl$(3)$ model. For the topological metric, we find the following form:\footnote{Here we are following the ordering in table \protect\ref{tab:permutations for fl(3)} from left to right.}
\begin{equation}\label{Fl3 top metric}
g_{w\,w'}(q,y)\, =\, \frac{1}{(1\,-\,q_1)(1\,-\,q_2)}\,
\begin{pmatrix}
 1 & 1 & 1 & 1 & 1 & 1 \\
 1 & 1 & 1 & A & 1 & A \\
 1 & 1 & 1 & 1 & B & B \\
 1 & A & 1 & D & C & E \\
 1 & 1 & B & C & F & G \\
 1 & A & B & E & G & H \\
\end{pmatrix}~,
\end{equation}
where, for compactness, we introduce the following variables: 
\begin{align*}
 A &\,\equiv\ \left(1\,-\,\frac{y_3}{y_1}\right)\,+\,\frac{y_3}{y_1}\,q_2~, \qquad
 B \,\equiv\, \left(1\,-\,\frac{y_3}{y_1}\right)\,+\,\frac{y_3}{y_1}\,q_1~, \qquad
 C \,\equiv\, \left(1\,-\,\frac{y_3}{y_1}\right)\,+\,\frac{y_3}{y_1}\,q_1\,q_2~, \\
 D &\,\equiv\,  \left(1\,-\,\frac{y_2}{y_1}\,-\,\frac{y_3}{y_1}\,+\,\frac{y_2\,y_3}{y_1^2}\right)\,+\,\left(\frac{y_2}{y_1}\,+\,\frac{y_3}{y_1}\,-\,\frac{y_2\,y_3}{y_1^2}\right)\,q_2~, \\
 E &\,\equiv\, \left(1\,-\,\frac{y_2}{y_1}\,-\,\frac{y_3}{y_1}\,+\,\frac{y_2\,y_3}{y_1^2}\right)\,
+\,\left(\frac{y_2}{y_1}\,-\,\frac{y_2\,y_3}{y_1^2}\right)\,q_2
\,+\,\frac{y_3}{y_1}\,q_1\,q_2~, \\
 F &\,\equiv\, \left(1\,-\,\frac{y_3}{y_1}\,-\,\frac{y_3}{y_2}\,+\,\frac{y_3^2}{y_1\,y_2}\right)\,
+\,\left(\frac{y_3}{y_1}\,+\,\frac{y_3}{y_2}\,-\,\frac{y_3^2}{y_1\,y_2}\right)\,q_1~, \\
 G &\,\equiv\, \left(1\,-\,\frac{y_3}{y_1}\,-\,\frac{y_3}{y_2}\,+\,\frac{y_3^2}{y_1\,y_2}\right)
\,+\,\left(\frac{y_3}{y_2}\,-\,\frac{y_3^2}{y_1\,y_2}\right)\,q_1\,
+\,\frac{y_3}{y_1}\,q_1\,q_2~, \\
 H &\,\equiv\, \left(1\,-\,\frac{y_3}{y_2}\,-\,\frac{y_2}{y_1}\,+\,\frac{y_3^2}{y_1\,y_2}\,+\,\frac{y_3}{y_1}\,-\,\frac{y_3^2}{y_1^2}\right)\,
+\,\left(\frac{y_3}{y_2}\,-\,\frac{y_3}{y_1}\,-\,\frac{y_3^2}{y_1\,y_2}\,+\,\frac{y_3^2}{y_1^2}\right)\,q_1\,\\
&\qquad+\,\left(\frac{y_2}{y_1}\,-\,\frac{y_3}{y_1}\,+\,\frac{y_3^2}{y_1^2}\right)\,q_2\,
+\,\left(2\,\frac{y_3}{y_1}\,-\,\frac{y_3^2}{y_1^2}\right)\,q_1\,q_2~.
\end{align*}

Given the above matrix, one can straightforwardly compute its inverse. We find:
\begin{equation}\label{Fl3 top metric inverse}
g^{w\,w'}(q,y) \,=\,
\begin{pmatrix}
a & b & c & -d & -e & k \\
b & f & -g & d & k & -k \\
c & -g & h & k & e & -k \\
-d & d & k & -k & -k & k \\
-e & k & e & -k & -k & k \\
k & -k & -k & k & k & -k
\end{pmatrix}~,
\end{equation}
where the matrix elements are defined as:
\begin{equation}
    \begin{aligned}
a &\equiv \left(\frac{y_1^2}{y_2\, y_3}\,-\,\frac{y_1^2}{y_3^2}\,-\,\frac{y_1}{y_2}\,+\,\frac{y_1\,y_2}{y_3^2}\,-\,\frac{y_2}{y_3}\,+\,1\right) \,+\, \left(\frac{y_1}{y_3}\,-\,\frac{y_1\,y_2 }{y_3^2}\,+\,\frac{y_2}{y_3}\,-\,1\right)\,q_1~,\\
&+\, \left(-\,\frac{y_1^2}{y_2 \,y_3}\,+\,\frac{y_1}{y_2}\,+\,\frac{y_1}{y_3}\,-\,1\right)\,q_2\, +\,\left(1\,-\,2\,\frac{y_1}{y_3}\right)\,q_1\,q_2~,\\
b &\equiv\, \left(\frac{y_1^2}{y_3^2}\,-\,\frac{y_1}{y_3}\,-\,\frac{y_1\,y_2}{y_3^2}\,+\,\frac{y_2}{y_3}\right)\, +\, \left(\frac{y_1\, y_2}{y_3^2}\,-\,\frac{y_2}{y_3}\right)\,q_1\, +\, \frac{y_1}{y_3}\,q_1\,q_2~,\\
c &\equiv\, \left(\frac{y_1}{y_2}\,-\,\frac{y_1}{y_3}\, -\,\frac{y_1^2}{y_2\,
   y_3}\,+\,\frac{y_1^2}{y_3^2}\right)\, +\, \left(\frac{y_1^2}{y_2\, y_3}\,-\,\frac{y_1}{y_2}\right)\,q_2\, +\, \frac{y_1}{y_3}\,q_1\,q_2~,\\
d &\equiv\, \left(\frac{y_1^2}{y_3^2}\,-\,\frac{y_1}{y_3}\right)\, +\,\frac{y_1}{y_3}\,q_1~, \qquad
e \equiv\, \left(\frac{y_1^2}{y_3^2}\,-\,\frac{y_1}{y_3}\right)\, +\,\frac{y_1}{y_3}\,q_2~, \\
f &\equiv\, \left(\frac{y_1}{y_3} \,-\, \frac{y_2}{y_3} \,+\, \frac{y_1\,y_2}{y_3^2} \,-\, \frac{y_1^2}{y_3^2}\right) \,+\, \left(\frac{y_2}{y_3} \,-\, \frac{y_1}{y_3} \,-\, \frac{y_1\,y_2}{y_3^2}\right)\,q_1~,\\
g &\equiv\, \left(\frac{y_1^2}{y_3^2}\,-\,\frac{y_1}{y_3}\right) \,+\,\frac{y_1}{y_3}\,q_1\,q_2~, \qquad k \equiv\, \frac{y_1^2}{y_3^2}~,\\
h &\equiv\, \left(\frac{y_1^2}{y_2\,
   y_3}\,-\,\frac{y_1^2}{y_3^2}\,-\,\frac{y_1}{y_2}\,+\,\frac{y_1}{y_3}\right) \,+\, \left(-\,\frac{y_1^2}{y_2\, y_3}\,+\,\frac{y_1}{y_2}\,-\,\frac{y_1}{y_3}\right)\,q_2~.\\
\end{aligned}
\end{equation}

In the non-equivariant limit, the topological metric \eqref{Fl3 top metric} takes the following form:
\begin{equation}
g_{w\,w'}(q) \,=\, \frac{1}{(1\,-\,q_1)(1\,-\,q_2)}\,
\begin{pmatrix}
1 & 1 & 1 & 1 & 1 & 1 \\
 1 & 1 & 1 & q_2 & 1 & q_2 \\
 1 & 1 & 1 & 1 & q_1 & q_1 \\
 1 & q_2 & 1 & q_2 & q_1\, q_2 & q_1\, q_2 \\
 1 & 1 & q_1 & q_1 \,q_2 & q_1 & q_1\, q_2 \\
 1 & q_2 & q_1 & q_1\, q_2 & q_1\, q_2 & q_1\, q_2 \\
\end{pmatrix}~.
\end{equation}

Meanwhile, the inverse of the topological metric \eqref{Fl3 top metric inverse} has the following non-equivariant form:
\begin{equation}\label{inv_top_met_fl3}
   g^{w\,w'}(q) \,=\, \left(
\begin{array}{cccccc}
 -\,q_1\, q_2 & q_1 \,q_2\, & q_1 \,q_2 & -\,q_1 & -\,q_2 & 1 \\
 q_1 \,q_2 & -\,q_1 & -\,q_1\, q_2 & q_1 & 1 & -\,1 \\
 q_1\, q_2 & -\,q_1\, q_2 & -\,q_2 & 1 & q_2 & -\,1 \\
 -\,q_1 & q_1 & 1 & -\,1 & -\,1 & 1 \\
 -\,q_2 & 1 & q_2 & -\,1 & -\,1 & 1 \\
 1 & -\,1 & -\,1 & 1 & 1 & -\,1 \\
\end{array}
\right)~.
\end{equation}

One can do the same to compute the 3-point correlation functions. Although we are not adding the details here since the final formulas are long and complicated, we have checked that the K-theoretic LR coefficients ${\mathcal{N}_{w\,w'}}^{w''}(q,y)$ match with those appearing in the quantum ring relations in table \ref{tab:Fl3equivQK}.

\subsubsection{Some comments on the dual Schubert classes}\label{subsec: dual lines}
The computation of the LR coefficients \eqref{QK relations} can, in fact, be sped up using the \textit{dual Schubert line defects} which we will denote by $\SL^{\vee\,w}\in\CR_{\rm 3d}$. These are half-BPS line operators defined such that:
\begin{equation}
    \left\langle\SL_w\,\SL^{\vee\,w'}\right\rangle_{\mathbb{P}^1\times S^1_\beta}\,=\,\delta_w^{w'}\,=\,\begin{cases}
        1~,\qquad &w\,=\,w'~,\\
        0~,\qquad &w\,\neq\,w'~,
    \end{cases}
\end{equation}
for any $w,w'\in {\rm W}^{(\boldsymbol{k};n)}$.

In the IR, the line operator $\SL^{\vee\,w}$ flows to the dual Schubert class $[\mathcal{O}^{\vee\,w}]\in {\rm QK}_T(X)$. The latter are represented by what we will refer to as the \textit{dual Whitney polynomials} -- or \textit{dual quantum Grothendieck polynomials}, depending on the presentation we are using for the quantum K-theory ring. To our knowledge, a compact formula for these polynomials is not known in the literature. Therefore, what we will do here is to follow the standard procedure in computing these characters in some examples and compare with what is known in the literature.

Knowing the explicit form of the topological metric \eqref{top met and str const}, one can expand the dual lines $\SL^{\vee\,w}$ in terms of the Schubert lines $\SL_w$ as follows:
\begin{equation}\label{dual schubert}
    \SL^{\vee\,w}\,=\,g^{w\,w'}(q,y)\,\SL_{w'}~, \qquad \forall\,w\,\in\,{\rm W}^{(\boldsymbol{k};n)}~,
\end{equation}
where, as usual, the repeated indices are summed over. 

\paragraph{Aside remark.} To avoid confusion with the terminology used in \cite{brion-lec,buch2018,xu2024quantum,MihalceaLecture}, here the dual Schubert lines should be dubbed the {\it quantum dual Schubert lines}, and the associated Schubert classes the {\it quantum dual Schubert classes} \footnote{See also the recent work \cite{2024arXiv240702703S} for a mathematical discussion on related aspects.}. As demonstrated in the specific examples of Gr$(2,4)$ and Fl$(3)$ below, the quantum dual Schubert classes will give the (classical) dual Schubert classes in the limit of $q_i\rightarrow 0$. We conjecture that this is true for general cases and leave the proof as future work.

\vspace{1em}

Having the explicit form of $\SL^{\vee \,w}$, one can directly compute the LR coefficients directly without the need to follow the procedure highlighted below \eqref{3d OPE relations}. Namely, we have:
\begin{equation}
    {\mathcal{N}_{w\,w'}}^{w''}(q,y)\,=\,\left\langle\SL_w\,\SL_{w'}\,\SL^{\vee\,w''}\right\rangle_{\mathbb{P}^1\times S^1_\beta}~.
\end{equation}
For the examples we are discussing below, we did check that these 3-point functions match the corresponding coefficients appearing in the ring structures discussed in the previous subsection.

\begin{table}[t]
\centering
\begin{equation*}
        \begin{array}{|c||c|c|c|c|c|c|}
    \hline
        w & {(1\,2\,3\,4)}&{(1\,3\,2\,4)}& {(1\,4\,2\,3)}&{(2\,3\,1\,4)}&{(2\,4\,1\,3)} &{(3\,4\,1\,2)} \\
        \hline
        \hline
         \ell(w)&  0 & 1&2&2&3&4\\
        \hline
         \dim(X_w)& 4 & 3&2&2&1&0\\
        \hline
        \end{array}
    \end{equation*}
\caption{For each possible permutation in W$^{(2;4)}$, we give its length and the dimension of the corresponding Schubert variety. Recall $\dim(X_w) = \dim ({\rm Gr}(2,4))-\ell(w)$.}
\label{tab:permutations for Gr(2,4)}
\end{table}

\medskip
\noindent
\textbf{Example: the dual Schubert classes in QK$_T$(Gr$(2,4)$).} This example has been considered already in \cite[Equation~(4.39)]{Closset:2023bdr}, but here we review in the light of the above discussion.\footnote{Also, there was a missing non-trivial case in \protect\cite[Equation~(4.39)]{Closset:2023bdr} which is the dual class of the trivial Schubert line.} In this case, we have $6$ possible permutations as listed in table \ref{tab:permutations for Gr(2,4)}. The K-theoretic parabolic double Whitney polynomials reduce to the double Grothendieck polynomials listed in \cite[Equations~(3.72)--(3.73)]{Closset:2023bdr}. Moreover, the topological metric is given in \cite[Equation~(4.31)]{Closset:2023bdr} and its inverse is given as:
\begin{equation}
   g^{w\,w'}(q,y)\,=\, \begin{pmatrix}
    \alpha & \beta & \gamma & \delta & \epsilon & \zeta \\
    \beta & \kappa &\,\theta\,-\,\gamma &\,\theta\,-\,\delta & \,\eta\,-\,\epsilon & \,-\,\zeta \\
    \gamma & \,\theta\,-\,\gamma & \eta & -\,\theta & -\,\eta & 0 \\
    \delta & \,\theta\,-\,\delta & -\,\theta & \eta & -\,\eta & 0 \\
    \epsilon &  \,\eta\,-\,\epsilon& -\,\eta & -\,\eta & \eta & 0 \\
    \zeta & \,-\,\zeta & 0 & 0 & 0 & 0 \\
\end{pmatrix}~,
\end{equation}
where:
{\small\begin{align*}
        & \alpha \:\equiv\: -\, q \,\left(1\,-\, \frac{y_1\,y_2}{y_3\,y_4}\right) \,+\, \left(1 \,-\, \frac{y_1}{y_3} \right)\,\left(1\, -\, \frac{y_1}{y_4} \right)\,\left(1\, - \,\frac{y_2}{y_3} \right)\,\left(1\, - \,\frac{y_2}{y_4} \right)~, \\
        & \beta \:\equiv\: -\, q \,\frac{y_1 \,y_2}{y_3 \,y_4} \,+\, \frac{y_2}{y_3}\,\left(1\, -\, \frac{y_1}{y_3} \right)\, \left( 1\,-\, \frac{y_1}{y_4}\right)\,\left( 1\,-\, \frac{y_2}{y_4} \right) ~,\\
        &\gamma\:\equiv \: \frac{y_2}{y_4}\,\left(1\,-\,\frac{y_1}{y_3}\right)\,\left(1\,-\,\frac{y_1}{y_4}\right)~,\quad \quad \delta\:\equiv\: \frac{y_1}{y_3}\, \left(1\,-\,\frac{y_1}{y_4}\right)\,\left(1\,-\,\frac{y_2}{y_4}\right)~,\\
        & \epsilon\:\equiv\: \frac{y_1}{y_4}\,\left( 1\,-\,\frac{y_1\,y_2}{y_3\,y_4}\right)~,\quad \quad \zeta\:\equiv \: \frac{y_1\, y_2}{y_3\, y_4}~,\quad \quad \eta\:\equiv \: \frac{y_1^2}{y_4^2}~,\quad \quad \theta\:\equiv \: \frac{y_1}{y_4}\,\left( 1\,-\,\frac{y_1}{y_4}\right)~,\\
        & \kappa\:\equiv \: q\, \frac{y_1 \,y_2}{y_3 \,y_4} \,+\, \left(1\,-\,\frac{y_1}{y_4}\right) \,\left( 1\,-\, \frac{y_1}{y_4}\, -\, \left(1\,-\,\frac{y_1}{y_3}\right)\,\left(1\,+\,\frac{y_2}{y_3}\right) \,+\, \frac{y_2}{y_4}\, \left(1\,-\,\frac{y_1}{y_3}\right)\,\left(1\,+\,\frac{y_2}{y_3}\right)\right)~.\\
    \end{align*}}
In the non-equivariant limit, this reads:
\begin{equation} \label{eq:g24invmetric}
   g^{w\,w'}(q)\,=\, \begin{pmatrix}
    0 & -\,q& 0 & 0 & 0 & 0 \\
    -\,q& q & 0 & 0 & 1 & -\,1 \\
    0 & 0 & 1 & 0 & -\,1 & 0 \\
    0 & 0 & 0 & 1 & -\,1 & 0 \\
    0 & 1 & -\,1 & -\,1 & 1 & 0 \\
    1 & -\,1 & 0 & 0 & 0 & 0 \\
\end{pmatrix}~.
\end{equation}
Here, in writing the topological metric and its inverse, we are ordering the columns of the matrix following table~\eqref{tab:permutations for Gr(2,4)}.

From \eqref{dual schubert} and \eqref{eq:g24invmetric}, we find the following form for the dual Schubert classes $\CO^{\vee\,w}\in {\rm QK}_T({\rm Gr}(2,4))$:
\begin{equation}
    \begin{aligned}
        &\CO^{\vee\,(1\,2\,3\,4)} \,=\, - \,q \,\CO_{(1\,3\,2\,4)} \,+\, \CO_{(3\,4\,1\,2)}~, \\
        &\CO^{\vee\,(1\,3\,2\,4)} \,=\, - \,q\, +\, q\, \CO_{(1\,3\,2\,4)} \,+\, \CO_{(2\,4\,1\,3)}\, -\, \CO_{(3\,4\,1\,2)}~, \\
        &\CO^{\vee\,(1\,4\,2\,3)} \,=\, \CO_{(1\,4\,2\,3)} \,-\, \CO_{(2\,4\,1\,3)}~, \\
        &\CO^{\vee\,(2\,3\,1\,4)} \,=\, \CO_{(2\,3\,1\,4)} \,-\, \CO_{(2\,4\,1\,3)}~, \\
        &\CO^{\vee\,(2\,4\,1\,3)} \,=\, \CO_{(1\,3\,2\,4)} \,-\, \CO_{(1\,4\,2\,3)} \,-\,  \CO_{(2\,3\,1\,4)}\, +\, \CO_{(2\,4\,1\,3)}~, \\
        &\CO^{\vee\,(3\,4\,1\,2)} \,=\, 1 \,-\, \CO_{(1\,3\,2\,4)}~.
    \end{aligned}
\end{equation}
One can immediately check that, the above relations reproduce the results in \cite[Example~1.3]{2024arXiv240702703S}, and in the classical limit, these relations reproduce results of \cite[Proposition~4.3.2]{brion-lec} specializing to Gr$(2,4)$.

In terms of the Chern characters, the (quantum) dual Schubert classes can be written as:

\begin{align*}
     &\widetilde{\mathfrak{G}}^{\vee\,{(3\,4\,1\,2)}}(x,y) \,=\, \frac{s_2}{y_3\,y_4}~,\\
     &\widetilde{\mathfrak{G}}^{\vee\,{(2\,4\,1\,3)}}(x,y) \,=\, \frac{s_2}{y_2 \,y_4}\,-\,\frac{s_2^2}{y_2\, y_3\, y_4^2}~,\\
     &\widetilde{\mathfrak{G}}^{\vee\,{(2\,3\,1\,4)}}(x,y) \,=\,\frac{s_2}{y_2 \,y_3}\,-\,\frac{s_1\, s_2}{y_2\, y_3\, y_4} \,+\,\frac{s_2^2}{y_2\, y_3\, y_4^2}~,\\
     &\widetilde{\mathfrak{G}}^{\vee\,{(1\,4\,2\,3)}}(x,y) \,=\,\frac{s_1}{y_4}\,-\,\frac{s_2}{y_2 \,y_4}\,-\,\frac{s_2}{y_3\, y_4}\,-\,\frac{s_2}{y_4^2}\,+\, \frac{s_2^2}{y_2\, y_3\, y_4^2}~,\\
     %%%%%%%%%%%%%%%%%%%%%%
     &\widetilde{\mathfrak{G}}^{\vee\,{(1\,3\,2\,4)}}(x,y) \,=\,\frac{s_1}{y_3}\,-\,\frac{s_2}{y_2\, y_3}\,-\,\frac{s_2}{y_3^2}\, -\,\frac{q \,s_2}{y_3\, y_4}\,-\,\frac{s_1^2}{y_3 \,y_4}\,+\,\frac{ s_1\,s_2}{y_2\, y_3\, y_4}\,+\,\frac{s_1\,s_2 }{y_3^2\,
   y_4}\\
   &\qquad \qquad \qquad\quad+\,\frac{ s_1\,s_2}{y_3\, y_4^2}\,-\,\frac{s_2^2}{y_2\, y_3\,
   y_4^2}\,-\,\frac{s_2^2}{y_3^2\, y_4^2}~,\\
   %%%%%%%%%%%%%%%%%%%%%%%%%%%%%
    & \widetilde{\mathfrak{G}}^{\vee\,(1\,2\,3\,4)}(x,y) = 1\,-\,q\,-\,\frac{s_1}{y_4}\,-\,\frac{s_1}{y_3}\,+\,\frac{s_2}{y_3^2}\,+\,\frac{s_2}{y_4^2}\,+\,\frac{q\, s_2}{y_3\, y_4}\,+\,\frac{s_1^2}{y_3\, y_4}\,-\,\frac{s_1\, s_2}{y_3^2\,
   y_4}\\
   &\qquad \qquad\qquad \quad-\,\frac{s_1\, s_2}{y_3\, y_4^2}\,+\,\frac{s_2^2}{y_3^2\,
   y_4^2}~.\\
\end{align*}
For compactness, here, we introduced $s_1\equiv x_1+x_2$, and, $s_2\equiv x_1x_2$. In the non-equivariant limit, these polynomials take the form:
\begin{equation}
\begin{aligned}
     &\widetilde{\mathfrak{G}}^{\vee\,{(3\,4\,1\,2)}}(x) \,=\, s_2~,\\
     &\widetilde{\mathfrak{G}}^{\vee\,{(2\,4\,1\,3)}}(x) \,=\, s_2\, -\,s_2^2~,\\
     &\widetilde{\mathfrak{G}}^{\vee\,{(2\,3\,1\,4)}}(x) \,=\, s_2\,-\,s_1 s_2\,+\,s_2^2~,\\
     &\widetilde{\mathfrak{G}}^{\vee\,{(1\,4\,2\,3)}}(x) \,=\, s_1\,-\,3\, s_2\,+\,s_2^2~,\\
     &\widetilde{\mathfrak{G}}^{\vee\,{(1\,3\,2\,4)}}(x) \,=\,s_1\, -\,2\, s_2\,-\,q\, s_2\,-\,s_1^2\,+\,3\, s_1\,s_2 \,-\,2\, s_2^2~,\\
     &\widetilde{\mathfrak{G}}^{\vee\,{(1\,2\,3\,4)}}(x) \,=\, 1\,-\,q\,-\,2\, s_1\,+\,2\, s_2\,+\,q\, s_2\,+\,s_1^2\,-\,2\, s_2\, s_1\,+\,s_2^2~.\\
\end{aligned}
\end{equation}

One can eliminate the explicit dependence of the above polynomials on the quantum parameter $q$ by using the BAEs \eqref{BAE stand} of the model:
\begin{equation}
\begin{split}
   &x_a\,\prod_{i=1}^4\,(x_b\,-\,y_i)\, +\,q\,x_b\,\det y \,=\,0~,\qquad a\,\neq\, b\,\in\{1,2\}~.\\
\end{split}
\end{equation}
Here, $\det y\equiv y_1\,y_2\,y_3\,y_4$. Performing this reduction, we indeed obtain the polynomials listed in \cite[Equation~(4.39)]{Closset:2023bdr} along with:
\begin{equation}
\begin{split}
         \mathfrak{G}^{\vee\,{(1\,2\,3\,4)}}(x,y) \,&=\, \frac{s_2}{y_1\, y_2}\,-\,\frac{s_1\, s_2}{y_1\,
   y_2\, y_3}\,-\,\frac{s_1 \,s_2}{y_1\, y_2 \,y_4}\,+\,\frac{s_2^2}{y_1\, y_2\,
   y_3^2}\,+\,\frac{s_2^2}{y_1\, y_2\, y_4^2}\,+\,\frac{s_1^2\, s_2}{y_1\, y_2\, y_3\, y_4}\,\\
   &-\,\frac{s_1\, s_2^2}{y_1\, y_2\, y_3\, y_4^2}\,-\,\frac{s_1 \,s_2^2}{y_1\, y_2\, y_3^2\, y_4}\,+\,\frac{s_2^3}{y_1\, y_2\, y_3^2\, y_4^2}~.
\end{split}
\end{equation}
In the non-equivariant limit, this reduces to:
\begin{equation}
\begin{split}
         \mathfrak{G}^{\vee\,{(1\,2\,3\,4)}}(x) &\,=\, {s_2}\,-\,2\,{s_1\, s_2}\,+\,{s_2^2}\,+\,{s_2^2}\,+\,{s_1^2\, s_2}\,-\,2\,{s_1\, s_2^2}\,+\,{s_2^3}\\
         &\,=\, s_2 (1-s_1 +s_2)^2\,,
\end{split}
\end{equation}
which implies that 
\begin{equation}
    \CO^{\vee\,(1\,2\,3\,4)} \:=\: (1\, -\, \CO_{(1\,3\,2\,4)})\, \CO_{(3\,4\,1\,2)} \:=\: \det \CS \, \CO_{(3\,4\,1\,2)}~.
\end{equation}
Similarly, one can obtain the remaining non-equivariant relations as follows:
\begin{equation}
    \begin{aligned}
        & \CO^{\vee\,(1\,3\,2\,4)} \:=\: \det \CS \, \CO_{(2\,4\,1\,3)}~,\\
        & \CO^{\vee\,(1\,4\,2\,3)} \:=\: \det \CS \, \CO_{(1\,4\,2\,3)}~,\\
        & \CO^{\vee\,(2\,3\,1\,4)} \:=\: \det \CS \, \CO_{(2\,3\,1\,4)}~,\\
        & \CO^{\vee\,(2\,4\,1\,3)} \:=\: \det \CS \, \CO_{(1\,3\,2\,4)}~,\\
        & \CO^{\vee\,(3\,4\,1\,2)} \:=\: \det \CS ~.
    \end{aligned}
\end{equation}
This matches the definition of \cite{buch-lmi}. See also \cite[Equation~(4.36)]{Closset:2023bdr} for the equivariant version.\footnote{Note that there is a minus sign typo in the definition of the dual Grothendieck polynomials \protect\cite[Equation~(4.37)]{Closset:2023bdr} where the components of the dual partition $\lambda^\vee$ should be $n_f-N_c-\lambda_a$ rather than $N_c-n_f-\lambda_a$.}

\paragraph{Example: the dual Schubert classes in QK(Fl$(3)$).} Let us now look at the complete flag Fl$(3)$. Recall that in this case, we have $6$ Schubert classes which are indexed by the permutations in table \ref{tab:permutations for fl(3)}. Following the definition \eqref{dual schubert} and using the inverse of the topological metric in \eqref{inv_top_met_fl3}, we find that the dual Schubert classes $\CO^{\vee\,w}\in {\rm QK}({\rm Fl}(3))$ can be written as:
{\small\begin{equation}\label{dual_sheaves_fl3}
    \begin{split}
        &\CO^{\vee\,(1\,2\,3)}\,=\,-\,q_1\,q_2\,+\,q_1\,q_2\,\CO_{(1\,3\,2)}\,+\,q_1\,q_2\,\CO_{(2\,1\,3)}\,-\,q_1\,\CO_{(2\,3\,1)}\,-\,q_2\,\CO_{(3\,1\,2)}\,+\,\CO_{(3\,2\,1)}~,\\
        &\CO^{\vee\,(1\,3\,2)}\,=\,q_1\,q_2\,-\,q_1\,\CO_{(1\,3\,2)}\,-\,q_1\,q_2\,\CO_{(2\,1\,3)}\,+\,q_1\,\CO_{(2\,3\,1)}\,+\,\CO_{(3\,1\,2)}\,-\,\CO_{(3\,2\,1)}~,\\
        &\CO^{\vee\,(2\,1\,3)}\,=\,q_1\,q_2\,-\,q_1\,q_2\,\CO_{(1\,3\,2)}\,-\,q_2\,\CO_{(2\,1\,3)}\,+\,\CO_{(2\,3\,1)}\,+\,q_2\,\CO_{(3\,1\,2)}\,-\,\CO_{(3\,2\,1)}~,\\
        &\CO^{\vee\,(2\,3\,1)}\,=\,-\,q_1\,+\,q_1\,\CO_{(1\,3\,2)}\,+\,\CO_{(2\,1\,3)}\,-\,\CO_{(2\,3\,1)}\,-\,\CO_{(3\,1\,2)}\,+\,\CO_{(3\,2\,1)}~,\\
        &\CO^{\vee\,(3\,1\,2)}\,=\,-\,q_2\,+\,\CO_{(1\,3\,2)}\,+\,q_2\,\CO_{(2\,1\,3)}\,-\,\CO_{(2\,3\,1)}\,-\,\CO_{(3\,1\,2)}\,+\,\CO_{(3\,2\,1)}~,\\
        &\CO^{\vee\,(3\,2\,1)}\,=\,1\,-\,\CO_{(1\,3\,2)}\,-\,\CO_{(2\,1\,3)}\,+\,\CO_{(2\,3\,1)}\,+\,\CO_{(3\,1\,2)}\,-\,\CO_{(3\,2\,1)}~.
    \end{split}
\end{equation}}

In the classical limit -- i.e., looking at the $q^0$ terms -- the above expressions reproduce \cite[Proposition~4.3.2]{brion-lec}, \cite[Lemma~3.5]{buch2018}, and \cite[Equation~(4.14)]{xu2024quantum}.\footnote{Keep in mind that in those references, they are writing the dual Schubert classes in terms of the opposite Schubert classes -- let us denote them by $\CO^w$ -- rather than the Schubert classes as we are doing here. To match our expressions with theirs in the classical limit, one needs to use the fact $\CO^w = \CO_{w_0w}$ for the complete flag manifolds with $w_0$ being the longest permutation.} Therefore, our expressions above can be viewed as the quantum version of those earlier results.

Reducing the dual sheaves \eqref{dual_sheaves_fl3} in the quantum K-theory ring \eqref{QKFL3} -- this is equivalent to reducing the corresponding Chern characters in the Bethe ideal defined in terms of \eqref{BAE stand} --, we find that they can be written as follows:
\begin{equation}\label{dual schubert Fl3}
\begin{split}
    &\CO^{\vee\,(1\,2\,3)}\,=\,\det\mathcal{S}_1\,\det\mathcal{S}_2\,\left(\CO_{(1\,3\,2)}\,+\,\CO_{(2\,3\,1)}\right)\,\CO_{(3\,1\,2)}~,\\
    &\CO^{\vee\,(1\,3\,2)}\,=\,\det\mathcal{S}_1\,\det\,\mathcal{S}_2\,\CO_{(3\,1\,2)}~,\\
    &\CO^{\vee\,(2\,1\,3)}\,=\,\det\mathcal{S}_1\,\det\,\mathcal{S}_2\,\,\CO_{(2\,3\,1)}~,\\
    &\CO^{\vee\,(2\,3\,1)}\,=\,\det\mathcal{S}_1\,\det\,\mathcal{S}_2\,\left(\CO_{(3\,1\,2)}\,+\,\CO_{(2\,1\,3)}\right)~,\\
    &\CO^{\vee\,(3\,1\,2)}\,=\,\det\mathcal{S}_1\,\det\,\mathcal{S}_2\,\left(\CO_{(1\,3\,2)}\,+\,\CO_{(2\,3\,1)}\right)~,\\
    &\CO^{\vee\,(3\,2\,1)}\,=\,\det\mathcal{S}_1\,\det\,\mathcal{S}_2~.
    \end{split}
\end{equation}

Here, we used the relations:
\begin{equation}
    \det\mathcal{S}_1\,=\,1\,-\,\CO_{(2\,1\,3)}~, \qquad \det\mathcal{S}_2\,=\,1\,-\,\CO_{(1\,3\,2)}~.
\end{equation}
These two relations can be seen directly from \cite[Equation~(4.39)]{Closset:2025akk} and \cite[Equation~(4.40)]{Closset:2025akk}, respectively. Or, one can write down the short exact sequences involving these line bundles, analogous to \cite[Equation~(3.16)]{Closset:2023bdr} for the Grassmannian case.

As further examples, in appendices \ref{app: QK Fl134} and \ref{app: QK Fl124}, we also work out the topological metrics and dual Schubert classes for Fl$(1,3;4)$ and Fl$(1,2;4)$, respectively.

%%%%%%%%%%%%%%%%%%%%%%%%%%%%%%%%%%%%%%%%%%%%%%
%%%%%%%%%%%%%%%%%%%%%%%%%%%%%%%%%%%%%%%%%%%%%%
%%%%%%%%%%%%%%%%%%%%%%%%%%%%%%%%%%%%%%%%%%%%%%
%%%%%%%%%%%%%%%%%%%%%%%%%%%%%%%%%%%%%%%%%%%%%%
\section{Computing the quantum cohomology of partial flags}\label{sec:2d A model}
In this section, we study the 2d A-model of the partial flag manifold Fl$(\boldsymbol{k};n)$. We apply the Gr\"obner basis algorithm mentioned earlier to compute the corresponding quantum cohomology ring relations. Moreover, we use the companion matrix techniques to compute the 2-point and 3-point GW invariants of the partial flag.

\medskip
\noindent
\textbf{Small $\beta$ limit of the 3d GLSM.}The 2d GLSM, and hence the 2d A-model, can be obtained by taking the small-radius limit of the 3d GLSM that we discussed in the previous sections. For the Fl$(\boldsymbol{k};n)$ GLSM, this limit is performed as follows:\footnote{Similarly, the K-theoretic Chern roots of the quotient bundles reduce as:
\protect\begin{equation}\label{x 2d limit}
        x_i\, \equiv\, e^{-\beta {\sigma_i}}\, \rightarrow \,1\,-\, \beta\, {\sigma_i}\, +\, \cdots~.
\end{equation}}
% with $\sigma_\bullet$ being the Chern roots of $\mathcal{Q}_\bullet$.}
\be\label{xy 2d limit}
x^{(\bullet)}\, =\, e^{-\beta \widetilde{\sigma}^{(\bullet)}}\, \rightarrow \,1\,-\, \beta\, \widetilde{\sigma}^{(\bullet)}\, +\, \cdots~,\qquad\quad
y \,=\, e^{-\beta m}\, \rightarrow \,1\,-\, \beta\, m\, +\, \cdots~,
\ee
for all K-theoretic Chern roots. Moreover, the 3d quantum parameter $q_\ell$ gets expanded as:
\begin{equation}\label{qk to qcoh}
    \qk_\ell\,\equiv \,(-\beta)^{k_{\ell+1}-k_{\ell-1}}\,\qcoh_\ell ~,
\end{equation}
with $\qcoh_\ell$ being the quantum cohomology parameter corresponding to the exponentiation of the 2d complexified FI parameter. 

\subsection{2d A-model for the partial flag manifold}
We start with recalling the ingredients of the 2d A-model for Fl$(\boldsymbol{k};n)$. Recall that we have a quiver gauge theory as shown in figure \ref{fig:parFlag GLSM} viewed as a 2d theory rather than a 3d one. As is the case for 3d, the full dynamics of the theory are controlled by the 2d effective twisted superpotential $\widetilde{\CW}(\zeta, {\t \sigma},m)$ and the 2d dilaton potential $\widetilde{\Omega}({\t\sigma},m)$. More explicitly, the effective twisted superpotential has the following form \cite[Equation~(2.2)]{Closset:2017zgf} -- see also \cite{Witten:1993yc,Witten:1993xi, Nekrasov:2009uh}:
\begin{equation}
    \widetilde{\CW}(\zeta, {\t \sigma},m)\,=\,\sum_{\ell=1}^s\,\zeta^{\rm eff}_\ell\,\tr({\t\sigma}^{(\ell)})\,+\,\sum_{\ell=1}^{s}\,{\t\CW}_{\rm BF}^{(\ell)}({\t \sigma, m})~,
\end{equation}
where the effective complexified FI term is of the form \cite[Equation~(4.70)]{Gu:2023tcv}:
\begin{equation}
    \zeta_\ell^{\rm eff}\,=\,\zeta_\ell\,+\,\frac12\,(k_\ell\,-\,1)~,
\end{equation}
where the second term comes from the reduction of the 3d gauge CS terms in \eqref{eff-twisted}.\footnote{And this matches the one in \protect\cite[Equation~(3.19)]{Closset:2015rna} coming from the contribution of the 2d W-bosons.} The contribution of each one of the bifundamental multiplets is given by:
\begin{equation}\label{2d eff W}
    {\t\CW}^{(\ell)}_{\rm BF}({\t \sigma}, m)\,=\,-\frac{1}{2\pi i}\,\sum_{a=1}^{k_\ell}\,\sum_{b=1}^{k_{\ell+1}}\,\left({\t \sigma}_a^{(\ell)}\,-\,{\t \sigma}_b^{(\ell+1)}\right)\,\left[\log\left({\t \sigma}_a^{(\ell)}\,-\,{\t \sigma}_b^{(\ell+1)}\right)\,-\,1\right]~.
\end{equation}

As for the dilaton potential $\widetilde{\Omega}({\t\sigma},m)$, this is given explicitly by \cite[Subsection~4.6]{Witten:1993xi}, \cite[Equation~(3.31)]{Nekrasov:2014xaa}, \cite[Equation~(2.9)]{Closset:2017zgf}:
\begin{equation}
    \widetilde{\Omega}(\widetilde{\sigma}, m)\,=\,\sum_{\ell=1}^{s}\left(\widetilde{\Omega}_{\rm BF}^{(\ell)}({\t\sigma},m)\,+\,\widetilde{\Omega}^{(\ell)}_{\rm W-bos}({\t\sigma})\right)~,
\end{equation}
where, the contribution of the chiral multiplets -- keeping in mind that, as in \eqref{omega bf}, we are taking their R-charges to be $0$ -- is of the form:
\begin{equation}
    \widetilde{\Omega}^{(\ell)}_{\rm BF}({\t\sigma},m)\,:=\,\frac{1}{2\pi i}\,\sum_{a=1}^{k_\ell}\,\sum_{b=1}^{k_{\ell+1}}\,\log\left({\t \sigma}_a^{(\ell)}\,-\,{\t \sigma}_b^{(\ell+1)}\right)~.
\end{equation}
Meanwhile, the contribution of the W-bosons is given by:
\begin{equation}
    \widetilde{\Omega}^{(\ell)}_{\rm W-bos}({\t\sigma})\,=\,-\frac{1}{2\pi i}\,\sum_{\alpha\in\mathfrak{g}_\ell}\,\log\, \alpha({\t\sigma}^{(\ell)})\,=\,-\frac{1}{\pi i}\,\log\,\Delta({\t\sigma}^{(\ell)})\,+\,{\rm const}~.
\end{equation}
where, $\Delta(\t\sigma^{(\ell)})\,=\,\prod_{a<b}({\t\sigma}_a^{(\ell)}-{\t\sigma}_b^{(\ell)})$.\footnote{As for the constant term, it is:
\protect\begin{equation}
    {\rm const}\,\equiv\,-\frac{1}{\pi i}\,\log (-1)^{\binom{k_\ell}{2}}\,=\,-\,\frac{1}{2}\,\binom{k_\ell}{2}~.
\end{equation}}

\paragraph{2d BAEs and quantum cohomology.} From the explicit form of the effective twisted superpotential \eqref{2d eff W}, we can derive the following set of BAEs of our partial flag model \cite[page 29]{Gu:2023tcv}:
\begin{equation}\label{2d BAE}
   (-1)^{k_\ell-1}\, \mathsf{q}_\ell\,\prod_{b=1}^{k_{\ell-1}}\,\left({\t\sigma}_b^{(\ell-1)}\,-\,{\t\sigma}_a^{(\ell)}\right)\,=\,\prod_{c=1}^{k_{\ell+1}}\,\left({\t\sigma}_a^{(\ell)}\,-\,{\t\sigma}_c^{(\ell+1)}\right)~,
\end{equation}
for $\ell=1, \cdots, s$, and, $a=1,\cdots, k_\ell$. Here, the 2d quantum parameter $\mathsf{q}_\ell$ was introduced below \eqref{qk to qcoh}.

\subsubsection{Genus-\texorpdfstring{$0$}{0} GW invariants from the 2d A-model}
 
 For any physical point operator $\mathscr{O}$ in the 2d twisted chiral ring $\mathcal{R}_{\rm 2d}$, the correlation function can be computed explicitly via the sum-over-Bethe-vacua formula:
\begin{equation}\label{<O>}
    \langle\mathscr{O}\rangle_{\mathbb{P}^1}\,=\,\sum_{\widehat{\t \sigma}\in \SBE^{\rm 2d}}\,{\widetilde{\mathcal{H}}}(\widehat{\widetilde{\sigma}}, m)^{-1}\,\mathscr{O}(\widehat{\t\sigma},m)~. 
\end{equation}
Here, the polynomial $\mathscr{O}({\t\sigma},m)$ is the equivariant differential form corresponding to the point operator $\mathscr{O}$ via the isomorphism \cite{Witten:1993xi,Donagi:2007hi}:
\begin{equation}\label{R2d=QH}
    \CR_{\rm 2d}\,\cong\,{\rm QH}_T({\rm Fl}(\boldsymbol{k};n))\,\cong\,\frac{\widetilde{\mathbb{K}}[\widetilde{\sigma}^{(\bullet)}]^{{\rm W}_G}}{\mathfrak{I}_{\rm Bethe}}~.
\end{equation}
Here, $\widetilde{\mathbb{K}}\,\equiv\,\mathbb{Z}[\qcoh, m]$ with $m$ being the 2d complex twisted masses associated with the flavor symmetry $SU(n)$. Moreover,  ${\mathfrak{I}}_{\rm Bethe}\subset \widetilde{\mathbb{K}}[\widetilde{\sigma}^{(\ell)}]$ is the ideal defined by the 2d BAEs \eqref{2d BAE}.

As for the 2d handle-gluing operator appearing on the RHS of \eqref{<O>}, this is defined as~\cite[Equation~(4.3)]{Nekrasov:2014xaa}:
\begin{equation}
    {\t\CH}({\t\sigma},m)\,=\,e^{2\pi i \,{\t\Omega}({\t\sigma},m)}\, \det\left(-\,{\rm Hessian}({\t\CW}({\t\sigma},m))\right)~.
\end{equation}
From the explicit form of the effective twisted superpotential in \eqref{2d eff W}, we have that the Hessian is given by:
\begin{equation}
\begin{split}
        {\rm Hessian}(\t\CW)_{a_\ell, b_{\ell'}}\,&=\,2\pi i\,\frac{\partial^2{\t\CW}}{\partial{\t\sigma}_a^{(\ell)}\,\partial{\t\sigma_b^{(\ell')}}}\,\\
        &=\,\frac{\delta_{\ell-1,\ell'}}{{\t\sigma}_b^{(\ell')}-{\t\sigma}_a^{(\ell)}}\,+\,\frac{\delta_{\ell+1,\ell'}}{{\t\sigma}_a^{(\ell)}\,-\,{\t\sigma}_b^{(\ell')}}-\,\delta_{a,b}\,\delta_{\ell,\ell'}\,\sum_{c=1}^{k_{\ell-1}}\frac{1}{{\t\sigma}_{c}^{(\ell-1)}\,-\,{\t\sigma}_a^{(\ell)}}\\
        &-\,\delta_{a,b}\,\delta_{\ell,\ell'}\,\sum_{c=1}^{k_{\ell+1}}\frac{1}{{\t\sigma}_a^{(\ell)}\,-\,{\t\sigma}_c^{(\ell+1)}}~.
\end{split}
\end{equation}

\medskip
\noindent
\textbf{Topological metric and structure constants.} Using the localization formula \eqref{<O>}, one can explicitly compute $2$-point and $3$-point genus-$0$ GW invariants for the partial flag manifold Fl$(\boldsymbol{k};n)$:
\begin{equation}\label{2d top metric and str const}
    \eta_{w\,w'}(\qcoh,m)\,:=\,\left\langle\mathscr{O}_w\,\mathscr{O}_{w'}\right\rangle_{\mathbb{P}^1}~, \qquad \CC_{w\,w'\,w''}(\qcoh,m)\,:=\,\left\langle\mathscr{O}_w\,\mathscr{O}_{w'}\,\mathscr{O}_{w''}\right\rangle_{\mathbb{P}^1}~.
\end{equation}
The (equivariant) quantum cohomology ring of Fl$(\boldsymbol{k};n)$ has a special basis in terms of the Schubert classes $[\Omega_w]$ for any $w\in {\rm W}^{(\boldsymbol{k};n)}$. These are represented by the cohomological parabolic (double) Whitney polynomials $\mathfrak{Z}_w^{(\boldsymbol{k};n)}(\widetilde{\sigma}^{(\bullet)},m)$ defined in \cite[Equation~(4.7)]{Closset:2026bnk}. 
% In particular, for the special cases of the Grassmannian and complete flag manifolds, these polynomials reduce to the double Schubert polynomials $\mathfrak{S}_w(\widetilde{\sigma}^{(\bullet)})$ and their quantum deformation $\mathfrak{S}_w(\widetilde{\sigma}^{(\bullet)})$ for the complete flag. See \cite[appendix A.1]{Closset:2025akk} for definition and examples.
Moreover, in the correlation functions appearing in \eqref{2d top metric and str const}, we denote by $\mathscr{O}_w$ the corresponding Schubert point defects introduced in \cite{Closset:2025akk} for complete flag manifolds, and in \cite{Closset:2026bnk} for any partial flag manifold.

In terms of the topological metric and the structure constants computed in \eqref{2d top metric and str const}, the ring relations of QH$_T$(Fl$(\boldsymbol{k};n)$) are of the form:
\begin{equation}\label{QH relations gen}
    \Omega_w\,\wedge_\qcoh\,\Omega_{w'}\,=\,{\mathcal{C}_{w\,w'}}^{w''}(\qcoh, m)\,\Omega_{w''}~, \qquad {\mathcal{C}_{w\,w'}}^{w''}\,=\,\eta^{w''\,v}\,\mathcal{C}_{w\,w'\,v}~.
\end{equation}
Here, repeated indices are summed over. In the next subsection, we will discuss how, using the companion matrix method, one can compute the correlation functions \eqref{2d top metric and str const} and, hence, compute the full ring structures for any partial flag manifold.

%%%%%%%%%%%%%%%%%%%%%%%%%%%%%%%%%%%%%%%%%%%%%%
%%%%%%%%%%%%%%%%%%%%%%%%%%%%%%%%%%%%%%%%%%%%%%
\subsection{Quantum cohomology from 2d Bethe ideal}

Using the results that we reviewed in the previous subsection, one can repeat the discussion we had in section~\ref{sec:QK from Grobner} to compute the quantum cohomology ring for any partial flag manifold. One difference in the 2d case is that, at the symmetrization step \eqref{symmetrization step}, one uses instead the cohomological parabolic (double) Whitney polynomials defined in \cite[Equation~(4.7)]{Closset:2026bnk}. Namely, we introduce a new set of variables $\Omega_w$ defined via:
\begin{equation}\label{symmetrization step 2d}
    \widetilde{~\mathsf{W}}_w^{(\boldsymbol{k};n)}(\widetilde{\sigma}^{(\bullet)}, m,\Omega_w)\,:=\,\mathfrak{Z}_w^{(\boldsymbol{k};n)}(\widetilde{\sigma}^{(\bullet)},m)\,-\,\Omega_w\,\in\,\widetilde{\mathbb{K}}[\widetilde{\sigma}^{(\bullet)},\Omega_w]~, \qquad w\,\in\,{\rm W}^{(\boldsymbol{k};n)} ~.
\end{equation} 
Reducing the quantum cohomology ring \eqref{R2d=QH} upon using these new variables, we end up with the explicit quantum cohomology ring relations in terms of the Schubert classes $\Omega_w$ given in \eqref{QH relations gen}.

Additionally, one can apply the companion matrix algorithm discussed in subsection~\ref{QK from ComMatrix} to compute the $2$-point and $3$-point functions introduced in \eqref{2d top metric and str const} in a similar fashion to \eqref{compMatr top and Str}. In what follows, we will give some explicit results for some simple partial flag manifolds and match our results with the literature. Here, we will add the details for the examples of the projective space $\mathbb{P}^{n-1}$, the Grassmannian manifold Gr$(2,4)$, and the complete flag manifold Fl$(3)$. On the other hand, their quantum cohomology rings can also be computed using schubmult package \cite{samuel2026triplequantumschubertcalculus,schubmult}.

\medskip
\noindent
\textbf{Schubert classes in the 2d limit.} Let us also comment here on what happens to the Schubert classes of the quantum K-theory ring when we take the small-$\beta$ limit introduced in \eqref{xy 2d limit}--\eqref{qk to qcoh}. As explained in \cite[Equation~(4.10)]{Closset:2026bnk}, the parabolic Whitney polynomial $\mathfrak{W}_w^{(\boldsymbol{k};n)}(x^{(\bullet)},y)$ reduces to its cohomological counterpart $\mathfrak{Z}^{(\boldsymbol{k};n)}_{w}(\widetilde{\sigma}^{(\bullet)},m)$. Therefore, one can schematically write: 
\begin{equation}\label{Ow to Omegaw}
    \CO_w\,=\,\beta^{\ell(w)}\,\Omega_w\,+\,\cdots~,
\end{equation}
Note here that $\ell(w)$ is in fact the cohomological-grading of the class $\Omega_w$.

\medskip
\noindent
{\bf Comments on the Whitney and Toda presentation.}
In analogy to the 3d case, one can obtain the Whitney and Toda presentation. The Whitney presentation for the partial flag Fl$(\boldsymbol{k};n)$ is again obtained from the quantum deformation of the short exact sequences associated with each step \eqref{eq:flsec}:
\[
    c(\CS_\ell)\,\wedge_{\qcoh}\,c(\CQ_{\ell+1}) \:=\:c(\CS_{\ell+1}) \,+\, (-1)^{k_\ell - k_{\ell-1}} \,\qcoh_{\ell}\, c(\CS_{\ell-1})~,
\]
with the degree of $\qcoh_{\ell}$ being $(k_{\ell+1} - k_{\ell-1})$, as pointed out in \eqref{qk to qcoh}. This relation can be expanded in terms of Chern roots of the tautological bundles $\CS_{\ell}$ and quotient bundles $\CQ_{\ell}$:
\[
\widetilde{\sigma}^{(\ell)} = \{\widetilde{\sigma}_1^{(\ell)},\dots, \widetilde{\sigma}_{k_{\ell}}^{(\ell)}\}\, \quad \text{and}\quad  \widehat{\sigma}^{(\ell)} = \{ \sigma_{k_{\ell-1}+1},\dots,\sigma_{k_{\ell}}\}\,, 
\]
where $\widetilde{\sigma}^{(n)}$ corresponds to equivariant parameters -- that is, the complex twisted masses $m_i$. Performing this expansion, one can obtain:
\[
\sum_{i+j=k}\, e_i(\widetilde{\sigma}^{(\ell)})\,e_j(\widehat{\sigma}^{(\ell+1)})\:=\: e_{k}(\widetilde{\sigma}^{(\ell+1)}) \,+\, (-1)^{k_\ell-k_{\ell-1}}\,\qcoh_\ell\, e_{k+k_{\ell-1}-k_{\ell+1}}(\widetilde{\sigma}^{(\ell-1)})~.
\]
See \cite[Subsection~4.6]{Gu:2023tcv} and references therein for more details. Following the same variable elimination procedure as discussed in Subsection~\ref{subsec:toda grob algorithm} above, the Whitney presentation reduces to the Toda presentation for QH$_T$(Fl$(\boldsymbol{k};n)$), with the Toda ideal relations generated by
\begin{equation}\label{toda qcoh rels}
      {{\rm E}^{(\boldsymbol{k};n)}}^{n}_{j} (\sigma;\qcoh) \,= \,e_j(m)~.
\end{equation}
  
Here, ${{\rm E}^{(\boldsymbol{k};n)}}^{\bullet}_{\bullet}(\sigma;\qcoh)$ are the parabolic quantum elementary polynomials, and $e_{\bullet}(m)$ the elementary symmetry polynomials in the equivariant parameters. See \cite{LamShimozono,Astashkevich:1993ks}, and \cite[Subsection~4.3]{Closset:2026bnk} for more details on the parabolic quantum elementary polynomials. The Whitney and Toda presentations for QH$_T$(Fl$(\boldsymbol{k};n)$) can also be directly obtained by performing the 2d reduction by \eqref{x 2d limit}--\eqref{qk to qcoh}, see more discussions from this approach in \cite{Gu:2023tcv,Closset:2026bnk}.

Our discussion in this subsection has been in the Bethe presentation of the quantum cohomology ring. One can instead discuss what happens in these two other presentations. For instance, the quantum cohomology ring in the Toda presentation takes the following form:
\begin{equation}
    {\rm QH}_T({\rm Fl}(\boldsymbol{k};n))\,\cong\,\frac{\widetilde{\mathbb{K}}[\sigma]^{{\rm W}_P}}{\mathfrak{I}_{\rm Toda}}~,
\end{equation}
with $\mathfrak{I}_{\rm Toda}\subset \widetilde{\mathbb{K}}[\sigma]^{{\rm W}_P}$ being the ideal generated by the relations \eqref{toda qcoh rels}. Although one can compute the explicit form of the quantum ring relations in this presentation, in the rest of this section, we will only consider the Bethe presentation. We expect the results in the other presentation to be the same as we argued in the quantum K-theory ring scenario in section~\ref{sec: 3d A model}.
%%%%%%%%%%%%%%%%%%%%%%%%%%%%%%%%%%%%%%%%%%%%%%
%%%%%%%%%%%%%%%%%%%%%%%%%%%%%%%%%%%%%%%%%%%%%%
%%%%%%%%%%%%%%%%%%%%%%%%%%%%%%%%%%%%%%%%%%%%%%
\subsubsection{For projective space \texorpdfstring{$\mathbb{P}^{n-1}$}{Pn1}}
Let us start with the simplest cases, which are the projective spaces $\mathbb{P}^{n-1}$. For any $n$, we find that the topological metric in the non-equivariant limit takes the following simple form:
\begin{equation}\label{top metric 2d Pn-1}
    \eta_{w\,w'}(\qcoh)\,=\,\delta_{{\rm ord}(w)+{\rm ord}(w'),\,n}~,
\end{equation}
here, ord$(w)$ stands for the location of the permutation $w$ in the following set:
\begin{equation}
    {\rm W}^{(1;n)}\,=\,\{(1\,2\,\cdots)\,,\,(2\,1\,\cdots)\,,\,(n\,1\,\cdots)\}~.
\end{equation}

\medskip
\noindent
\textbf{Quantum cohomology ring of $\mathbb{P}^1$.} For the case with $n=2$, we get the following non-trivial quantum ring relation:
\begin{equation}
    \Omega_{(2\,1)}\,\wedge_\qcoh\,\Omega_{(2\,1)}\, =\, \qcoh\,-\,(m_1\,-\,m_2)\, \Omega_{(2\,1)}~,
\end{equation}
which, in the non-equivariant limit -- i.e., the limit $m_i\rightarrow 0$ --, reduces to $ \Omega_{(2\,1)}^2 = \qcoh$. Note that these are relations that one gets in the small-$\beta$ limit defined in \eqref{xy 2d limit}--\eqref{qk to qcoh} and \eqref{Ow to Omegaw} applied to quantum K-theory relations in \cite[Equation~(4.15)]{Closset:2023bdr}.

\medskip
\noindent
\textbf{Quantum cohomology ring for $\mathbb{P}^2$.} For the case with $n=3$, we get the following equivariant ring relations:
\begin{equation}
    \begin{split}
        &\Omega_{(2\,1\,3)}\,\wedge_\qcoh\,\Omega_{(2\,1\,3)}\,=\, -\,(m_1\,-\,m_2 )\,\Omega_{(2\,1\,3)}\,+\,\Omega_{(3\,1\,2)}~,\\
        &\Omega_{(2\,1\,3)}\,\wedge_\qcoh\, \Omega_{(3\,1\,2)}\,=\,-\,(m_1\,-\,m_3)\, \Omega_{(3\,1\,2)}\,+\,\qcoh~,\\
        &\Omega_{(3\,1\,2)}\,\wedge_\qcoh\,\Omega_{(3\,1\,2)}\,=\,(m_1\,-\,m_3)\,(m_2\,-\,m_3) \,\Omega_{(3\,1\,2)}\,+\,\qcoh
   \,\Omega_{(2\,1\,3)}\,-\,(m_2 \,-\,m_3)\, \qcoh~.
    \end{split}
\end{equation}
Indeed, these are the 2d limit of ring relations of QK$_T(\mathbb{P}^2)$ given in \cite[Equation~(4.16)]{Closset:2023bdr}.
In the non-equivariant limit, these relations simplify to:
\begin{equation}
    \begin{split}
        \Omega_{(2\,1\,3)}\,\wedge_\qcoh\,\Omega_{(2\,1\,3)}\,=\, \Omega_{(3\,1\,2)}~,\quad\Omega_{(2\,1\,3)}\,\wedge_\qcoh\, \Omega_{(3\,1\,2)}\,=\,\qcoh~,\quad\Omega_{(3\,1\,2)}\,\wedge_\qcoh\,\Omega_{(3\,1\,2)}\,=\,\qcoh
   \,\Omega_{(2\,1\,3)}~.
    \end{split}
\end{equation}
%%%%%%%%%%%%%%%%%%%%%%%%%%%%%%%%%%%%%%%%%%%%%%
\subsubsection{For Grassmannian Gr\texorpdfstring{$(2,4)$}{24}}
Let us look at the Grassmannian Gr$(2,4)$. Recall that, in this case, we have $6$ Schubert classes spanning the quantum cohomology ring. Using the companion matrix algorithm, we find that, in the non-equivariant limit, the topological metric has the following simple form:
\begin{equation}\label{eta gr}
    \eta_{w\,w'}\,=\,\eta^{w\,w'}\,=\,
   \left(
\begin{array}{cccccc}
 0 & 0 & 0 & 0 & 0 & 1 \\
 0 & 0 & 0 & 0 & 1 & 0 \\
 0 & 0 & 1 & 0 & 0 & 0 \\
 0 & 0 & 0 & 1 & 0 & 0 \\
 0 & 1 & 0 & 0 & 0 & 0 \\
 1 & 0 & 0 & 0 & 0 & 0 \\
\end{array}
\right)~.
\end{equation}
Here, we are following the ordering of the Schubert classes given in table \ref{tab:permutations for Gr(2,4)}. Moreover, using the Gr\"obner basis algorithm we discussed above, we find the relations exhibited in table \ref{tab:Gr24equivQH} defining QH$^\bullet_T({\rm Gr}(2,4))$. These match with the earlier results of Bertram \cite{bertram1997quantumschubertcalculus}.\footnote{This can also be computed by the Maple-based program \texttt{Equivariant Schubert Calculator} developed by A. Buch. See \protect\texttt{https://sites.math.rutgers.edu/$\sim$asbuch/equivcalc/}.} Moreover, they match with 2d limit defined in \eqref{xy 2d limit}--\eqref{qk to qcoh} and \eqref{Ow to Omegaw} applied to the quantum K-theory relations given in~\cite[Table 2]{Closset:2023bdr}.

%%%%%%%%%%%%%%%%%%%%%%%%%%%%%%%%%%%%%%%%%%%%%%
\begin{table}[t!]
{\small
\begin{align*}
    \begin{split}
&\Omega_{(1\,3\,2\,4)}\,\wedge_\qcoh\,\Omega_{(1\,3\,2\,4)}\,=\, \Omega_{(2\,3\,1\,4)}\,-\,\left(m_2\,-\,m_3\right)\, \Omega_{(1\,3\,2\,4)}\,+\,\,\Omega_{(1\,4\,2\,3)}~,\\
&\Omega_{(1\,3\,2\,4)}
\,\wedge_\qcoh\,\Omega_{(1\,4\,2\,3)}\,=\,\Omega_{(2\,4\,1\,3)}
\,-\,\left(m_2\,-\,m_4\right)\,\Omega_{(1\,4\,2\,3)}~,\\
&\Omega_{(1\,4\,2\,3)}\,\wedge_\qcoh\,\Omega_{(1\,4\,2\,3)}
\,=\,
\Omega_{(3\,4\,1\,2)}
\,-\,\left(m_3\,-\,m_4\right)\,\Omega_{(2\,4\,1\,3)}\,+\,\left(m_2\,-\,m_4\right)\,\left(m_3\,-\,m_4\right)\,\Omega_{(1\,4\,2\,3)}~,\\
&\Omega_{(1\,3\,2\,4)}\,\wedge_\qcoh\,\Omega_{(2\,3\,1\,4)}
\,=\,
\Omega_{(2\,4\,1\,3)}
\,-\,\left(m_1\,-\,m_3\right)\,\Omega_{(2\,3\,1\,4)}~,\\
&\Omega_{(1\,4\,2\,3)}
\,\wedge_\qcoh\,
\Omega_{(2\,3\,1\,4)}
\,=\,
\qcoh
\,-\,\left(m_1\,-\,m_4\right)\,\Omega_{(2\,4\,1\,3)}~,\\
&\Omega_{(2\,3\,1\,4)}
\,\wedge_\qcoh\,
\Omega_{(2\,3\,1\,4)}\,=\,
\Omega_{(3\,4\,1\,2)}
\,-\,\left(m_1\,-\,m_2\right)\,\Omega_{(2\,4\,1\,3)}
\,+\,\left(m_1\,-\,m_2\right)\,\left(m_1\,-\,m_3\right)\,\Omega_{(2\,3\,1\,4)}~,\\
&\Omega_{(1\,3\,2\,4)}
\,\wedge_\qcoh\,
\Omega_{(2\,4\,1\,3)}
\,=\,
\qcoh
\,+\,\Omega_{(3\,4\,1\,2)}
\,-\,\left(m_1\,-\,m_4\right)\,\Omega_{(2\,4\,1\,3)}~,\\
&\Omega_{(1\,4\,2\,3)}
\,\wedge_\qcoh\,
\Omega_{(2\,4\,1\,3)}
\,=\,
-\,\qcoh\,(m_3\,-\,m_4)\,
+\,\qcoh\,\Omega_{(1\,3\,2\,4)}
\,+\,\left(m_1\,-\,m_4\right)\,\left(m_3\,-\,m_4\right)\,\Omega_{(2\,4\,1\,3)}\\
&\qquad \qquad \qquad \qquad ~~~~~\,\,-\,\left(m_1\,-\,m_4\right)\,\Omega_{(3\,4\,1\,2)}~,\\
&\Omega_{(2\,3\,1\,4)}
\,\wedge_\qcoh\,
\Omega_{(2\,4\,1\,3)}\,=\,
-\,\qcoh\,(m_1\,-\,m_2)\,
+\,\qcoh\,\Omega_{(1\,3\,2\,4)}
\,+\,\left(m_1\,-\,m_2\right)\,\left(m_1\,-\,m_4\right)\,\Omega_{(2\,4\,1\,3)}\\
&\qquad \qquad \qquad \qquad ~~~~~\,\,-\,\left(m_1\,-\,m_4\right)\,\Omega_{(3\,4\,1\,2)}~,\\
&\Omega_{(2\,4\,1\,3)}\,\wedge_\qcoh\, \Omega_{(2\,4\,1\,3)}
\,=\,
\qcoh\, (m_1\,-\,m_2)\,(m_3\,-\,m_4)\,
\,-\, \qcoh \,(m_1\,-\,m_4)\,\Omega_{(1\,3\,2\,4)}
\,+\, \qcoh\, \Omega_{(1\,4\,2\,3)}\\
&\qquad \qquad \qquad \qquad ~~~~~\,\,+\, \qcoh\, \Omega_{(2\,3\,1\,4)}
\,-\, (m_1\,-\,m_2)\,(m_1\,-\,m_4)\,(m_3\,-\,m_4)\,\Omega_{(2\,4\,1\,3)}\\
&\qquad \qquad \qquad \qquad ~~~~~\,\,+\, (m_1\,-\,m_4)^2\, \Omega_{(3\,4\,1\,2)}~,\\
&\Omega_{(1\,3\,2\,4)} \,\wedge_\qcoh\, \Omega_{(3\,4\,1\,2)}
\,=\,
- \,(m_1\,+\,m_2\,-\,m_3\,-\,m_4)\,\Omega_{(3\,4\,1\,2)}
\,+\, \qcoh\,\Omega_{(1\,3\,2\,4)}~,\\
&\Omega_{(1\,4\,2\,3)}\, \wedge_\qcoh\, \Omega_{(3\,4\,1\,2)}
\,=\,
\,-\, \qcoh\, (m_2\,-\,m_4)\,\Omega_{(1\,3\,2\,4)}
\,+\, \qcoh\, \Omega_{(2\,3\,1\,4)}
\,+\, (m_1\,-\,m_4)\,(m_2\,-\,m_4)\,\Omega_{(3\,4\,1\,2)}~,\\
&\Omega_{(2\,3\,1\,4)}\, \wedge_\qcoh\, \Omega_{(3\,4\,1\,2)}
\,=\,
\,-\, \qcoh\, (m_1\,-\,m_3)\,\Omega_{(1\,3\,2\,4)}
\,+\, \qcoh\, \Omega_{(1\,4\,2\,3)}
\,+\, (m_1\,-\,m_3)\,(m_1\,-\,m_4)\,\Omega_{(3\,4\,1\,2)}~,\\
&\Omega_{(2\,4\,1\,3)} \,\wedge_\qcoh\, \Omega_{(3\,4\,1\,2)}
\,=\,
\qcoh\, (m_1\,-\,m_3)\,(m_2\,-\,m_4)\,\Omega_{(1\,3\,2\,4)}
\,-\, \qcoh\, (m_2\,-\,m_4)\,\Omega_{(1\,4\,2\,3)}\,+\, \qcoh\, \Omega_{(2\,4\,1\,3)}\\
&\qquad \qquad \qquad \qquad ~~~~~\,\,-\, \qcoh\, (m_1\,-\,m_3)\,\Omega_{(2\,3\,1\,4)}
\,-\, (m_1\,-\,m_3)\,(m_1\,-\,m_4)\,(m_2\,-\,m_4)\,\Omega_{(3\,4\,1\,2)}~,\\
&\Omega_{(3\,4\,1\,2)}\, \wedge_\qcoh\, \Omega_{(3\,4\,1\,2)}
\,=\,
\qcoh^2\,
-\, \qcoh\, (m_1\,-\,m_3)\,(m_2\,-\,m_3)\,(m_2\,-\,m_4)\,\Omega_{(1\,3\,2\,4)}\\
&\qquad \qquad \qquad \qquad ~~~~~\,\,+\, \qcoh\, (m_2\,-\,m_3)\,(m_2\,-\,m_4)\,\Omega_{(1\,4\,2\,3)}
\,+\, \qcoh\, (m_1\,-\,m_3)\,(m_2\,-\,m_3)\,\Omega_{(2\,3\,1\,4)}\\
&\qquad \qquad \qquad \qquad ~~~~~\,\,-\, \qcoh\, (m_1\,+\,m_2\,-\,m_3\,-\,m_4)\,\Omega_{(2\,4\,1\,3)}\\
&\qquad \qquad \qquad \qquad ~~~~~\,\,+\, (m_1\,-\,m_3)\,(m_2\,-\,m_3)\,(m_1\,-\,m_4)\,(m_2\,-\,m_4)\,\Omega_{(3\,4\,1\,2)}~.
\end{split}
\end{align*}}
\caption{The equivariant QH product for the Grassmannian manifold Gr$(2,4)$.  \label{tab:Gr24equivQH}}
\end{table}

In the non-equivariant limit -- i.e., taking $m_i \rightarrow 0$ -- these relations simplify drastically as follows:

\[
\begin{minipage}{0.48\textwidth}
\begin{align*}
\Omega_{(1\,3\,2\,4)} \,\wedge_\qcoh\, \Omega_{(1\,3\,2\,4)} 
&= \Omega_{(2\,3\,1\,4)} \,+\, \Omega_{(1\,4\,2\,3)}~, \\
\Omega_{(1\,3\,2\,4)} \,\wedge_\qcoh\, \Omega_{(1\,4\,2\,3)} 
&= \Omega_{(2\,4\,1\,3)}~, \\
\Omega_{(1\,4\,2\,3)} \,\wedge_\qcoh\, \Omega_{(1\,4\,2\,3)} 
&= \Omega_{(3\,4\,1\,2)}~, \\
\Omega_{(1\,3\,2\,4)} \,\wedge_\qcoh\, \Omega_{(2\,3\,1\,4)} 
&= \Omega_{(2\,4\,1\,3)}~, \\
\Omega_{(1\,4\,2\,3)} \,\wedge_\qcoh\, \Omega_{(2\,3\,1\,4)} 
&= \qcoh~, \\
\Omega_{(2\,3\,1\,4)} \,\wedge_\qcoh\, \Omega_{(2\,3\,1\,4)} 
&= \Omega_{(3\,4\,1\,2)}~, \\
\Omega_{(1\,3\,2\,4)} \,\wedge_\qcoh\, \Omega_{(2\,4\,1\,3)} 
&= \qcoh \,+\, \Omega_{(3\,4\,1\,2)}~, \\
\Omega_{(1\,4\,2\,3)} \,\wedge_\qcoh\, \Omega_{(2\,4\,1\,3)} 
&= \qcoh \,\Omega_{(1\,3\,2\,4)}~,
\end{align*}
\end{minipage}
\hfill
\begin{minipage}{0.65\textwidth}
\begin{align*}
\Omega_{(2\,3\,1\,4)} \,\wedge_\qcoh\, \Omega_{(2\,4\,1\,3)} 
&= \qcoh \,\Omega_{(1\,3\,2\,4)}~, \\
\Omega_{(2\,4\,1\,3)} \,\wedge_\qcoh\, \Omega_{(2\,4\,1\,3)} 
&= \qcoh \,\Omega_{(1\,4\,2\,3)} \,+\, \qcoh \,\Omega_{(2\,3\,1\,4)}~, \\
\Omega_{(1\,3\,2\,4)} \,\wedge_\qcoh\, \Omega_{(3\,4\,1\,2)} 
&= \qcoh \,\Omega_{(1\,3\,2\,4)}~, \\
\Omega_{(1\,4\,2\,3)} \,\wedge_\qcoh\, \Omega_{(3\,4\,1\,2)} 
&= \qcoh \,\Omega_{(2\,3\,1\,4)}~, \\
\Omega_{(2\,3\,1\,4)} \,\wedge_\qcoh\, \Omega_{(3\,4\,1\,2)} 
&= \qcoh \,\Omega_{(1\,4\,2\,3)}~, \\
\Omega_{(2\,4\,1\,3)} \,\wedge_\qcoh\, \Omega_{(3\,4\,1\,2)} 
&= \qcoh \,\Omega_{(2\,4\,1\,3)}~, \\
\Omega_{(3\,4\,1\,2)} \,\wedge_\qcoh\, \Omega_{(3\,4\,1\,2)} 
&= \qcoh^2~.
\end{align*}
\end{minipage}
\]

\subsubsection{For the complete flag Fl\texorpdfstring{$(3)$}{3}}
Let us now look at Fl$(3)$ case. In this case, recall that we have $6$ Schubert classes. The ring relations of QH$_T({\rm Fl}(3))$ are exhibited in table \ref{tab:Fl3equivQH}. These also can be reduced in the 2d limit -- as defined in \eqref{xy 2d limit}--\eqref{qk to qcoh} and \eqref{Ow to Omegaw} -- of the quantum K-theory ring relations given in table~\ref{tab:Fl3equivQK}.

 Moreover, in the non-equivariant limit, we get the following relations:
\[
{\small\begin{minipage}{0.47\textwidth}
\begin{align*}
\Omega_{(1\,3\,2)} \,\wedge_\qcoh\, \Omega_{(1\,3\,2)} 
&= \qcoh_2 \,+\, \Omega_{(2\,3\,1)}~, \\
\Omega_{(1\,3\,2)} \,\wedge_\qcoh\, \Omega_{(2\,1\,3)} 
&= \Omega_{(2\,3\,1)} \,+\, \Omega_{(3\,1\,2)}~, \\
\Omega_{(2\,1\,3)} \,\wedge_\qcoh\, \Omega_{(2\,1\,3)} 
&= \qcoh_1 \,+\, \Omega_{(3\,1\,2)}~, \\
\Omega_{(1\,3\,2)} \,\wedge_\qcoh\, \Omega_{(2\,3\,1)} 
&= \qcoh_2 \,\Omega_{(2\,1\,3)}~, \\
\Omega_{(2\,1\,3)} \,\wedge_\qcoh\, \Omega_{(2\,3\,1)} 
&= \Omega_{(3\,2\,1)}~, \\
\Omega_{(2\,3\,1)} \,\wedge_\qcoh\, \Omega_{(2\,3\,1)} 
&= \qcoh_2 \,\Omega_{(3\,1\,2)}~, \\
\Omega_{(1\,3\,2)} \,\wedge_\qcoh\, \Omega_{(3\,1\,2)} 
&= \Omega_{(3\,2\,1)}~, \\
\Omega_{(2\,1\,3)} \,\wedge_\qcoh\, \Omega_{(3\,1\,2)} 
&= \qcoh_1 \,\Omega_{(1\,3\,2)}~,
\end{align*}
\end{minipage}
\hfill
\begin{minipage}{0.55\textwidth}
\begin{align*}
\Omega_{(2\,3\,1)} \,\wedge_\qcoh\, \Omega_{(3\,1\,2)} 
&= \qcoh_1\, \qcoh_2~, \\
\Omega_{(3\,1\,2)} \,\wedge_\qcoh\, \Omega_{(3\,1\,2)} 
&= \qcoh_1 \,\Omega_{(2\,3\,1)}~, \\
\Omega_{(1\,3\,2)} \,\wedge_\qcoh\, \Omega_{(3\,2\,1)} 
&= \qcoh_1\, \qcoh_2 \,+\, \qcoh_2 \,\Omega_{(3\,1\,2)}~, \\
\Omega_{(2\,1\,3)} \,\wedge_\qcoh\, \Omega_{(3\,2\,1)} 
&= \qcoh_1\, \qcoh_2 \,+\, \qcoh_1 \,\Omega_{(2\,3\,1)}~, \\
\Omega_{(2\,3\,1)} \,\wedge_\qcoh\, \Omega_{(3\,2\,1)} 
&= \qcoh_1\, \qcoh_2 \,\Omega_{(1\,3\,2)}~, \\
\Omega_{(3\,1\,2)} \,\wedge_\qcoh\, \Omega_{(3\,2\,1)} 
&= \qcoh_1\, \qcoh_2 \,\Omega_{(2\,1\,3)}~, \\
\Omega_{(3\,2\,1)} \,\wedge_\qcoh\, \Omega_{(3\,2\,1)} 
&= \qcoh_1\, \qcoh_2 \,\Omega_{(2\,3\,1)} \,+\, \qcoh_1\, \qcoh_2 \,\Omega_{(3\,1\,2)}~.
\end{align*}
\end{minipage}}
\]

% \OK{Does this match sth in the literature?}

\begin{table}[t!]
{\small
\begin{align*}
    \begin{split}
&\Omega_{(1\,3\,2)}\,\wedge_\qcoh\,\Omega_{(1\,3\,2)}\,=\, \qcoh_2 + \Omega_{(2\,3\,1)}\,-\,\left(m_2\,-\,m_3\right)\, \Omega_{(1\,3\,2)}~,\\
&\Omega_{(1\,3\,2)}
\,\wedge_\qcoh\,\Omega_{(2\,1\,3)}\,=\,\Omega_{(2\,3\,1)}
\,+\,\Omega_{(3\,1\,2)}~,\\
&\Omega_{(1\,3\,2)}\,\wedge_\qcoh\,\Omega_{(2\,3\,1)} \,=\,\qcoh_2\,\Omega_{(2\,1\,3)}
\,-\,\left(m_1\,-\,m_3\right)\,\Omega_{(2\,3\,1)}~,\\
&\Omega_{(1\,3\,2)}\,\wedge_\qcoh\,\Omega_{(3\,1\,2)} \,=\,\Omega_{(3\,2\,1)}
\,-\,\left(m_2\,-\,m_3\right)\,\Omega_{(3\,1\,2)}~,\\
&\Omega_{(1\,3\,2)}\,\wedge_\qcoh\,\Omega_{(3\,2\,1)} \,=\,\qcoh_1\,\qcoh_2\,+\,\qcoh_2\,
\Omega_{(3\,1\,2)}
\,-\,\left(m_1\,-\,m_3\right)\,\Omega_{(3\,2\,1)}~,\\
&\Omega_{(2\,1\,3)}\,\wedge_\qcoh\,\Omega_{(2\,1\,3)} \,=\,\qcoh_1 \,+\,
\Omega_{(3\,1\,2)}
\,-\,\left(m_1\,-\,m_2\right)\,\Omega_{(2\,1\,3)}~,\\
&\Omega_{(2\,1\,3)}\,\wedge_\qcoh\,\Omega_{(2\,3\,1)} \,=\,
\Omega_{(3\,2\,1)}
\,-\,\left(m_1\,-\,m_2\right)\,\Omega_{(2\,3\,1)}~,\\
&\Omega_{(2\,1\,3)}\,\wedge_\qcoh\,\Omega_{(3\,1\,2)} \,=\,
\qcoh_1\,\Omega_{(1\,3\,2)}
\,-\,\left(m_1\,-\,m_3\right)\,\Omega_{(3\,1\,2)}~,\\
&\Omega_{(2\,1\,3)}\,\wedge_\qcoh\,\Omega_{(3\,2\,1)} \,=\,\qcoh_1\,\qcoh_2\,+\,\qcoh_1\,
\Omega_{(2\,3\,1)}
\,-\,\left(m_1\,-\,m_3\right)\,\Omega_{(3\,2\,1)}~,\\
&\Omega_{(2\,3\,1)}\,\wedge_\qcoh\,\Omega_{(3\,1\,2)} \,=\,\qcoh_1\,\qcoh_2
\,-\,\left(m_1\,-\,m_3\right)\,\Omega_{(3\,2\,1)}~,\\
&\Omega_{(2\,3\,1)}\,\wedge_\qcoh\,\Omega_{(2\,3\,1)} \,=\,\qcoh_2\,\Omega_{(3\,1\,2)}
\,-\,\qcoh_2\,\left(m_1\,-\,m_2\right)\,\Omega_{(2\,1\,3)}\,+\,\left(m_1\,-\,m_2\right)\,\left(m_1\,-\,m_3\right)\,\Omega_{(2\,3\,1)}~,\\
&\Omega_{(2\,3\,1)}\,\wedge_\qcoh\,\Omega_{(3\,2\,1)} \,=\,-\,\qcoh_1\,\qcoh_2\,\left(m_1\,-\,m_2\right)\,+\,\qcoh_1\,\qcoh_2\,\Omega_{(1\,3\,2)}
\,-\,\qcoh_2\,\left(m_1\,-\,m_3\right)\,\Omega_{(3\,1\,2)}\\
&\qquad \qquad \qquad \qquad \quad \,+\,\left(m_1\,-\,m_2\right)\,\left(m_1\,-\,m_3\right)\,\Omega_{(3\,2\,1)}~,\\
&\Omega_{(3\,1\,2)}\,\wedge_\qcoh\,\Omega_{(3\,1\,2)} \,=\,\qcoh_1\,\Omega_{(2\,3\,1)}
\,-\,\qcoh_1\,\left(m_2\,-\,m_3\right)\,\Omega_{(1\,3\,2)} \,+\, \left(m_1\,-m_3\right)\,\left(m_2\,-\,m_3\right)\,\Omega_{(3\,1\,2)}\\
&\Omega_{(3\,1\,2)}\,\wedge_\qcoh\,\Omega_{(3\,2\,1)} \,=\,-\,\qcoh_1\,\qcoh_2\,\left(m_2\,-\,m_3\right)\,+\,\qcoh_1\,\qcoh_2\,\Omega_{(2\,1\,3)}
\,-\,\qcoh_1\,\left(m_1\,-\,m_3\right)\,\Omega_{(2\,3\,1)}\\
&\qquad \qquad \qquad \qquad \quad \,+\,\left(m_1\,-\,m_3\right)\,\left(m_2\,-\,m_3\right)\,\Omega_{(3\,2\,1)}~,\\
&\Omega_{(3\,2\,1)}\,\wedge_\qcoh\,\Omega_{(3\,2\,1)} \,=\,\qcoh_1\,\qcoh_2\,\left(m_1\,-\,m_2\right)\,\left(m_2\,-\,m_3\right)\,-\,\qcoh_1\,\qcoh_2\,\left(m_1\,-\,m_2\right)\Omega_{(2\,1\,3)}\\
&\qquad \qquad \qquad \qquad \quad-\,\qcoh_1\,\qcoh_2\,\left(m_2\,-\,m_3\right)\,\Omega_{(1\,3\,2)}
 \,+\,\Big(\qcoh_1\,\left(m_1\,-\,m_2\right)\,\left(m_1\,-\,m_3\right)\,+\,\qcoh_1\,\qcoh_2\Big)\,\Omega_{(2\,3\,1)}\\
&\qquad \qquad \qquad \qquad \quad+\,\Big(\qcoh_2\,\left(m_1\,-\,m_3\right)\,\left(m_2\,-\,m_3\right)\,+\,\qcoh_1\,\qcoh_2\Big)\,\Omega_{(3\,1\,2)}\\
&\qquad \qquad \qquad \qquad \quad -\,\left(m_1\,-\,m_2\right)\,\left(m_1\,-\,m_3\right)\,\left(m_2\,-\,m_3\right)\,\Omega_{(3\,2\,1)}~.
\end{split}
\end{align*}}
\caption{The equivariant QH ring relations for the complete flag Fl$(3)$.  \label{tab:Fl3equivQH}}
\end{table}
%%%%%%%%%%%%%%%%%%%%%%%%%%%%%%%%%%%%%%%%%%%%%%
\subsection{Comments on the dual Schubert polynomials}
 Let us make some comments on the dual Schubert polynomials. Similar to the computations in subsection~\ref{subsec: dual lines}, we can expand the dual Schubert classes in terms of the Schubert classes using the inverse of the topological metric:
\begin{equation}
    \Omega^{\vee\,w} \,=\, \eta^{w\, w^\prime}\, \Omega_{w'}~, \qquad \forall\,w\,\in\,{\rm W}^{(\boldsymbol{k};n)}~.
\end{equation}

\paragraph{Dual Schubert classes in QH(Gr$(2,4)$).} Using the inverse of the topological metric \eqref{eta gr} for Grassmannian Gr$(2,4)$, we find the following form for the dual Schubert classes in QH$_T({\rm Gr}(2,4))$:
\begin{equation}
 {\Omega}^{\vee\,w}\,=\,\begin{cases}
        {\Omega}_w~,\qquad &w\,=\,(1\,4\,2\,3)\,,\,(2\,3\,1\,4)~,\\
        {\Omega}_{w\widetilde{w}_0}~, \qquad &\text{otherwise}~,
    \end{cases} 
\end{equation}
with $\widetilde{w}_0\equiv (3\,4\,1\,2)$ being the longest permutation in W$^{(2;4)}$.

\paragraph{Dual Schubert classes in QH(Fl$(3)$).} For the complete flag Fl$(3)$, we find the topological metric to be anti-diagonal, i.e., 
\begin{equation}
    \eta_{w\,w'}\,=\,\eta^{w\,w'}\,=\,\delta_{w',w_0w}\,=\,\begin{cases}
        1~, \qquad &w'\,=\,w_0\,w~, \\
        0~, \qquad &w'\,\neq\,w_0\,w~.
    \end{cases}
\end{equation}
Therefore, we find that the dual Schubert class in the quantum cohomology ring is given by:
\begin{equation}
    \Omega^{\vee\,w}\,=\,\Omega_{w_0w}~, \qquad \forall\,w\,\in\,S_3~.
\end{equation}

%%%%%%%%%%%%%%%%%%%%%%%%%%%%%%%%%%%%%%%%%%%%%%
%%%%%%%%%%%%%%%%%%%%%%%%%%%%%%%%%%%%%%%%%%%%%%
%%%%%%%%%%%%%%%%%%%%%%%%%%%%%%%%%%%%%%%%%%%%%%
%%%%%%%%%%%%%%%%%%%%%%%%%%%%%%%%%%%%%%%%%%%%%%
\section{Conclusions and further discussion}\label{sec:conclusion}
In this work, we explicitly computed the quantum K-theory ring relations for some partial flag manifolds in terms of the Schubert classes using the Gröbner basis algorithm. Given sufficient computational resources, the algorithm can be used to compute the ring relations for any other partial flag manifold. We also applied the same algorithm to compute the quantum cohomology rings using the 2d A-model formalism and matched our results with the literature.

Moreover, we made some comments on the dual Schubert classes $\CO^{w\,\vee}$, which are important in computing the K-theoretic LR coefficients of the quantum ring relations. What would be interesting for us in the future is to have a more systematic formula for computing the Chern characters of those sheaves and to work out a supersymmetric quantum mechanical realization of the corresponding half-BPS line operators in the twisted chiral ring, similar to \cite{Closset:2023bdr, Gu:2025tda, Closset:2025akk, Closset:2026bnk}.

Throughout the computation of the quantum ring relations, we fixed the 3d (mixed) gauge CS levels to the standard form \eqref{standard levels}. However, it is known that some alternative choices also lead to the pure geometric phase of the 3d GLSM defined in figure \ref{fig:parFlag GLSM} \cite{Jockers:2019lwe,Jockers:2021omw,Closset:2023jiq,Closset:2023bdr,Huq-Kuruvilla:2025nlf}. Although we do not show it explicitly in this work, it is straightforward to apply our algorithm in the Bethe presentation to compute the quantum rings of the partial flags in the presence of these level structures. All one needs to do is to compute the new form of the BAEs \eqref{BAE stand} with the different choice of the CS levels and then follow the same discussion that we had in subsections \ref{subsec:bethe grob algorithm} and \ref{subsec: TopinComMat}. This has already been done for the Grassmannian manifold case in \cite{Closset:2023bdr}.

Recently, it was studied in \cite{Gu:2024mqk} that there is a correspondence between the quantum K-theory and the twisted quantum cohomology of Grassmannians, demonstrated by ring presentations and correlators. A similar correspondence between quantum K-theory with another level choice and quantum cohomology was observed eariler in \cite{Kapustin:2013hpk}. Then, a natural future direction is to generalize this type of correspondence to the general partial flags by applying the algorithms in this paper.

Finally, let us also point out a small puzzle. If one needs to compute the ring relations of the quantum K-theory ring with level structures via the Toda presentation -- see subsection~\ref{subsec:toda grob algorithm}, one needs first to derive the modified form of the quantum Whitney relations of \cite{Gu:2023tcv,Gu:2023fpw}. In \cite{Closset:2025akk,Closset:2026bnk}, the way the quantum Grothendieck polynomials were derived is by applying the standard quantum Whitney relations to the parabolic Whitney polynomials. Following this procedure in the presence of the level structures, it seems that the form of the parabolic quantum Grothendieck polynomials should also be modified. We will leave this question for future work as well.

\section*{Acknowledgments}
We would like to thank Cyril Closset, Wei Gu, Leonardo Mihalcea, Eric Sharpe, Yaoxiong Wen, and Weihong Xu for valuable discussions and comments on an earlier draft. We thank Matt Samuel for independently checking the displayed quantum K-theory product tables using his schubmult software and for providing corrections and additional consistency checks. O.~Khlaif and H.~Zou would like to thank Cyril Closset, Wei Gu, Eric Sharpe, and Hao Zhang for collaboration on an earlier series of works which motivated this project. O.~Khlaif would like to thank the Department of Mathematics at the University of Geneva for the hospitality during which part of this project was done. Z.~Duan is supported by the Swiss National Science Foundation (SNSF) through a Postdoctoral Fellowship (TMPFP2\_234009). O.~Khlaif is a Junior Research Fellow at the Philippe Meyer Institute. H.~Zou is partially supported by the National Natural Science Foundation of China (NSFC) with Grant No.~12405083 and~12475005 and the Shanghai Pujiang Program with Grant No.~24PJA119.

%%%%%%%%%%%%%%%%%%%%%%%%%%%%%%%%%%%%%%%%%%%%%%
%%%%%%%%%%%%%%%%%%%%%%%%%%%%%%%%%%%%%%%%%%%%%%
%%%%%%%%%%%%%%%%%%%%%%%%%%%%%%%%%%%%%%%%%%%%%%
%%%%%%%%%%%%%%%%%%%%%%%%%%%%%%%%%%%%%%%%%%%%%%
\newpage
\appendix
\section{Further examples of quantum K-theory rings}\label{sec:appendix}
In this appendix, we add the details of the quantum K-theory ring relations for other examples of partial flag manifolds. 
%%%%%%%%%%%%%%%%%%%%%%%%%%%%%%%%%%%%%%%%%%%%%%
%%%%%%%%%%%%%%%%%%%%%%%%%%%%%%%%%%%%%%%%%%%%%%
\subsection{The complete flag Fl\texorpdfstring{$(4)$}{4}}\label{app:QK Fl4}
In the case of the complete flag Fl$(4)$, we have $4!=24$ Schubert classes in the basis. Hence, there are $276$ independent quantum ring relations that need to be computed. For the obvious reason, we will only include $10$ randomly selected relations:\footnote{It proved helpful in computing these relations to use the computer software \textsc{JULIA}.} 

{\small\begin{align*}
    &\mathcal{O}_{(1\,2\,4\,3)} \,\otimes_q\, \mathcal{O}_{(1\,3\,2\,4)}\, =\, \mathcal{O}_{(1\,3\,4\,2)}\, +\, \mathcal{O}_{(1\,4\,2\,3)} \,-\, \mathcal{O}_{(1\,4\,3\,2)}~,\\
      &\mathcal{O}_{(1\,2\,4\,3)} \,\otimes_q\, \mathcal{O}_{(3\,2\,4\,1)}\, =\, \left(1\,-\,\frac{y_4}{y_1}\right)\,\mathcal{O}_{(3\,2\,4\,1)}\, +\, q_3\,\frac{y_4}{y_1}\,\mathcal{O}_{(3\,2\,1\,4)}~,\\
      &\mathcal{O}_{(1\,2\,4\,3)} \,\otimes_q\, \mathcal{O}_{(2\,3\,4\,1)}\, =\, \left(1\,-\,\frac{y_4}{y_1}\right)\,\mathcal{O}_{(2\,3\,4\,1)}\, +\, q_3\,\frac{y_4}{y_1}\,\mathcal{O}_{(2\,3\,1\,4)}~,\\
        &\mathcal{O}_{(2\,3\,4\,1)} \,\otimes_q\, \mathcal{O}_{(3\,2\,1\,4)} \,=\, \frac{y_3\, y_2}{y_1^2}\;\mathcal{O}_{(4\,3\,2\,1)} \,+\, \left(1\,-\,\frac{y_3}{y_1}\right)\,\frac{y_2}{y_1}\mathcal{O}_{(3\,4\,2\,1)}~\\
        &\qquad \qquad \qquad\qquad \qquad +\, \left(1\,-\,\frac{y_2}{y_1}\right)\,\frac{y_3}{y_1}\mathcal{O}_{(4\,2\,3\,1)} \,+\, \left(1\,-\,\frac{y_2}{y_1}\right)\,\left(1\,-\,\frac{y_3}{y_1}\right)\mathcal{O}_{(3\,2\,4\,1)} ~,\\
        &\mathcal{O}_{(1\,2\,4\,3)} \,\otimes_q\, \mathcal{O}_{(1\,2\,4\,3)}\,=\, \frac{y_4}{y_3}\,\mathcal{O}_{(1\,3\,4\,2)}\, -\, q_3\,\frac{y_4}{y_3}\,\mathcal{O}_{(1\,3\,2\,4)}\, +\,q_3\, \frac{y_4}{y_3}\, +\,\left( 1\,-\,\frac{y_4}{y_3}\right)\, \mathcal{O}_{(1\,2\,4\,3)}~,\\
      &\mathcal{O}_{(1\,2\,4\,3)} \,\otimes_q\, \mathcal{O}_{(1\,3\,4\,2)} \,=\, \frac{y_4}{y_2}\,\mathcal{O}_{(2\,3\,4\,1)}\, +\,q_3 \,\frac{y_4}{y_2}\,\mathcal{O}_{(1\,3\,2\,4)} \,-\,q_3\, \frac{y_4}{y_2}\,\mathcal{O}_{(2\,3\,1\,4)}\,+\, \left(1\,-\,\frac{y_4}{y_2}\right)\, \mathcal{O}_{(1\,3\,4\,2)}~,\\
      &\mathcal{O}_{(1\,3\,2\,4)} \,\otimes_q\, \mathcal{O}_{(1\,3\,2\,4)} \,=\, \frac{y_3}{y_2}\,\mathcal{O}_{(1\,4\,2\,3)}\, +\,\frac{y_3}{y_2}\,\mathcal{O}_{(2\,3\,1\,4)} \,-\,\frac{y_3}{y_2}\,\mathcal{O}_{(2\,4\,1\,3)}\,+\,q_2\, \frac{y_3}{y_2}\,-\,q_2\,\frac{y_3}{y_2}\,\mathcal{O}_{(1\,2\,4\,3)} ~,\\
      &\qquad \qquad \qquad\qquad \qquad -\,q_2\,\frac{y_3}{y_2}\,\mathcal{O}_{(2\,1\,3\,4)}\,+\,q_2\,\frac{y_3}{y_2}\,\mathcal{O}_{(2\,1\,4\,3)} \,+\,\left(1\,-\,\frac{y_3}{y_2}\right)\,\mathcal{O}_{(1\,3\,2\,4)}~,\\
        &\mathcal{O}_{(1\,3\,2\,4)} \,\otimes_q\, \mathcal{O}_{(2\,1\,4\,3)} \,=\,\mathcal{O}_{(2\,3\,4\,1)}\, +\,\mathcal{O}_{(2\,4\,1\,3)} \,-\,\mathcal{O}_{(2\,4\,3\,1)}\,+\,\mathcal{O}_{(3\,1\,4\,2)} \,-\,\mathcal{O}_{(3\,2\,4\,1)} \,-\,\mathcal{O}_{(3\,4\,1\,2)}\\
    &\qquad \qquad \qquad\qquad \qquad +\,\mathcal{O}_{(3\,4\,2\,1)} +\,\mathcal{O}_{(4\,1\,2\,3)}\,-\,\mathcal{O}_{(4\,1\,3\,2)}\,-\,\mathcal{O}_{(4\,2\,1\,3)}\,+\,\mathcal{O}_{(4\,2\,3\,1)}\,+\,\mathcal{O}_{(4\,3\,1\,2)}\\
    &\qquad \qquad \qquad\qquad \qquad-\,\mathcal{O}_{(4\,3\,2\,1)}~,\\
      &\mathcal{O}_{(4\,1\,2\,3)} \,\otimes_q\, \mathcal{O}_{(4\,2\,3\,1)} \,=\, q_1\, q_2 \,q_3 \,\frac{y_4^2}{y_3\,y_1}\,\mathcal{O}_{(3\,1\,2\,4)} \,+\, \left(1\,-\,\frac{y_4}{y_1}\right)\,\left(1\,-\,\frac{y_4}{y_2}\right)\,\left(1\,-\,\frac{y_4}{y_3}\right)\,\mathcal{O}_{(4\,2\,3\,1)}\\
      &\qquad \qquad \qquad\qquad \qquad +\, q_1\,q_2\,q_3\,\left(1\,-\,\frac{y_4}{y_3}\right)\,\frac{y_4^2}{y_1\,y_2}\mathcal{O}_{(2\,1\,3\,4)} \,+\, q_1\,q_2\,q_3\,\left(1\,-\,\frac{y_4}{y_2}\right)\,\left(1\,-\,\frac{y_4}{y_3}\right)\,\frac{y_4}{y_1}\\ 
      &\qquad \qquad \qquad\qquad \qquad+\, q_1\,\left(1\,-\,\frac{y_4}{y_1}\right)\,\left(1\,-\,\frac{y_4}{y_3}\right)\,\frac{y_4}{y_2}\,\mathcal{O}_{(2\,4\,3\,1)} \,+\, q_1\,\left(1\,-\,\frac{y_4}{y_1}\right)\,\frac{y_4}{y_3}\, \mathcal{O}_{(3\,4\,2\,1)}~,\\
     &\mathcal{O}_{(1\,3\,2\,4)} \,\otimes_q\, \mathcal{O}_{(1\,4\,3\,2)} \,=\, \frac{y_4}{y_2}\,\mathcal{O}_{(2\,4\,3\,1)}\, +\,\frac{y_4}{y_2}\,\mathcal{O}_{(3\,4\,1\,2)} \,-\,\frac{y_4}{y_2}\,\mathcal{O}_{(3\,4\,2\,1)}\,+\,q_2\, \frac{y_4}{y_2}\,\mathcal{O}_{(1\,3\,4\,2)}\,-\,q_2\,\frac{y_4}{y_2}\,\mathcal{O}_{(2\,3\,4\,1)} \\
      &\qquad \qquad \qquad\qquad \qquad -\,q_2\,\frac{y_4}{y_2}\,\mathcal{O}_{(3\,1\,4\,2)}\,+\,q_2\,\frac{y_4}{y_2}\,\mathcal{O}_{(3\,2\,4\,1)} \,+\,q_2\,q_3\,\frac{y_4}{y_2}\,-\,q_2\,q_3\,\frac{y_4}{y_2}\,\mathcal{O}_{(1\,3\,2\,4)}\\
      &\qquad \qquad \qquad\qquad \qquad -\,q_2\,q_3\,\frac{y_4}{y_2}\,\mathcal{O}_{(2\,1\,3\,4)}\,+\,q_2\,q_3\,\frac{y_4}{y_2}\,\mathcal{O}_{(2\,3\,1\,4)}\,+\,q_2\,q_3\,\frac{y_4}{y_2}\,\mathcal{O}_{(3\,1\,2\,4)}\\
      &\qquad \qquad \qquad\qquad \qquad-\,q_2\,q_3\,\frac{y_4}{y_2}\,\mathcal{O}_{(3\,2\,1\,4)}\,+\,\left(1\,-\,\frac{y_4}{y_2}\right)\,\mathcal{O}_{(1\,4\,3\,2)}~.
\end{align*}}

In the non-equivariant limit, these become:
{\small\begin{align*}
    &\mathcal{O}_{(1\,2\,4\,3)} \,\otimes_q\, \mathcal{O}_{(1\,3\,2\,4)}\, =\, \mathcal{O}_{(1\,3\,4\,2)}\, +\, \mathcal{O}_{(1\,4\,2\,3)} \,-\, \mathcal{O}_{(1\,4\,3\,2)}~,\\
      &\mathcal{O}_{(1\,2\,4\,3)} \,\otimes_q\, \mathcal{O}_{(3\,2\,4\,1)}\, =\, q_3\,\mathcal{O}_{(3\,2\,1\,4)}~,\\
      &\mathcal{O}_{(1\,2\,4\,3)} \,\otimes_q\, \mathcal{O}_{(2\,3\,4\,1)}\, =\,  q_3\,\mathcal{O}_{(2\,3\,1\,4)}~,\\
        &\mathcal{O}_{(2\,3\,4\,1)} \,\otimes_q\, \mathcal{O}_{(3\,2\,1\,4)} \,=\, \mathcal{O}_{(4\,3\,2\,1)} ~,\\
        &\mathcal{O}_{(1\,2\,4\,3)} \,\otimes_q\, \mathcal{O}_{(1\,2\,4\,3)}\,=\, \mathcal{O}_{(1\,3\,4\,2)}\, -\, q_3\,\mathcal{O}_{(1\,3\,2\,4)}\, +\,q_3~,\\
      &\mathcal{O}_{(1\,2\,4\,3)} \,\otimes_q\, \mathcal{O}_{(1\,3\,4\,2)} \,=\, \mathcal{O}_{(2\,3\,4\,1)}\, +\,q_3 \,\mathcal{O}_{(1\,3\,2\,4)} \,-\,q_3,\mathcal{O}_{(2\,3\,1\,4)}~,\\
      &\mathcal{O}_{(1\,3\,2\,4)} \,\otimes_q\, \mathcal{O}_{(1\,3\,2\,4)} \,=\,\mathcal{O}_{(1\,4\,2\,3)}\, +\,\mathcal{O}_{(2\,3\,1\,4)} \,-\,\mathcal{O}_{(2\,4\,1\,3)}\,+\,q_2\,-\,q_2\,\mathcal{O}_{(1\,2\,4\,3)} -\,q_2\,\mathcal{O}_{(2\,1\,3\,4)}\,+\,q_2\,\mathcal{O}_{(2\,1\,4\,3)} ~,\\
        &\mathcal{O}_{(1\,3\,2\,4)} \,\otimes_q\, \mathcal{O}_{(2\,1\,4\,3)} \,=\,\mathcal{O}_{(2\,3\,4\,1)}\, +\,\mathcal{O}_{(2\,4\,1\,3)} \,-\,\mathcal{O}_{(2\,4\,3\,1)}\,+\,\mathcal{O}_{(3\,1\,4\,2)} \,-\,\mathcal{O}_{(3\,2\,4\,1)} \,-\,\mathcal{O}_{(3\,4\,1\,2)}\\
    &\qquad \qquad \qquad\qquad \qquad +\,\mathcal{O}_{(3\,4\,2\,1)} +\,\mathcal{O}_{(4\,1\,2\,3)}\,-\,\mathcal{O}_{(4\,1\,3\,2)}\,-\,\mathcal{O}_{(4\,2\,1\,3)}\,+\,\mathcal{O}_{(4\,2\,3\,1)}\,+\,\mathcal{O}_{(4\,3\,1\,2)}\,-\,\mathcal{O}_{(4\,3\,2\,1)}~,\\
      &\mathcal{O}_{(4\,1\,2\,3)} \,\otimes_q\, \mathcal{O}_{(4\,2\,3\,1)} \,=\, q_1\, q_2 \,q_3 \,\mathcal{O}_{(3\,1\,2\,4)} ~,\\
     &\mathcal{O}_{(1\,3\,2\,4)} \,\otimes_q\, \mathcal{O}_{(1\,4\,3\,2)} \,=\,\mathcal{O}_{(2\,4\,3\,1)}\, +\,\mathcal{O}_{(3\,4\,1\,2)} \,-\,\mathcal{O}_{(3\,4\,2\,1)}\,+\,q_2\, \mathcal{O}_{(1\,3\,4\,2)}\,-\,q_2\,\mathcal{O}_{(2\,3\,4\,1)} \\
      &\qquad \qquad \qquad\qquad \qquad -\,q_2\,\mathcal{O}_{(3\,1\,4\,2)}\,+\,q_2\,\mathcal{O}_{(3\,2\,4\,1)} \,+\,q_2\,q_3\,-\,q_2\,q_3\,\mathcal{O}_{(1\,3\,2\,4)}\\
      &\qquad \qquad \qquad\qquad \qquad -\,q_2\,q_3\,\mathcal{O}_{(2\,1\,3\,4)}\,+\,q_2\,q_3\,\mathcal{O}_{(2\,3\,1\,4)}\,+\,q_2\,q_3\,\mathcal{O}_{(3\,1\,2\,4)}\,-\,q_2\,q_3\,\mathcal{O}_{(3\,2\,1\,4)}~.
\end{align*}}

%%%%%%%%%%%%%%%%%%%%%%%%%%%%%%%%%%%%%%%%%%%%%%
%%%%%%%%%%%%%%%%%%%%%%%%%%%%%%%%%%%%%%%%%%%%%%
\subsection{The incidence flag Fl\texorpdfstring{$(1,3;4)$}{134}}\label{app: QK Fl134}
In this subsection, we give the details of the quantum K-theory of the incidence flag Fl$(1,3;4)$. In this case, there are $12$ Schubert classes. In what follows, we use the ordering:
\begin{equation}
\begin{split}
    \{(1\,2\,3\,4)\,,\, (1\,2\,4\,3)\,,\, (1\,3\,4\,2)\,,\, (2\,1\,3\,4)\,,\, (2\,1\,4\,3)\,,\, (2\,3\,4\,1)\,,\\ (3\,1\,2\,4)\,, \,(3\,1\,4\,2)\,,\, (3\,2\,4\,1)\,,\, (4\,1\,2\,3)\,,\, (4\,1\,3\,2)\,,\, (4\,2\,3\,1)\}
\end{split}
\end{equation}
The corresponding $\ell(w)$ and $\dim(X_w)$ are given in \cite[Appendix B]{Closset:2026bnk}.

\medskip
\noindent
\textbf{The topological metric and the dual Schubert classes.}
Using the companion matrix algorithm discussed in section~\ref{sec:QK from Grobner}, we find the topological metric of QK$({\rm Fl}(1,3;4))$ is of the form:
{\begin{equation}
g_{w\,w'}(q) = \frac{1}{(1\,-\,q_1)\,(1\,-\,q_2)}\left(
\begin{array}{cccccccccccc}
 1 & 1 & 1 & 1 & 1 & 1 & 1 & 1 & 1 & 1 & 1 & 1 \\
 1 & 1 & 1 & 1 & 1 & q_2 & 1 & 1 & q_2 & 1 & 1 & q_2 \\
 1 & 1 & q_2 & 1 & 1 & q_2 & 1 & q_2 & q_2 & 1 & q_2 & q_2 \\
 1 & 1 & 1 & 1 & 1 & 1 & 1 & 1 & 1 & q_1 & q_1 & q_1 \\
 1 & 1 & 1 & 1 & 1 & q_2 & 1 & 1 & q_2 & q_1 & q_1 &q_1\,q_2 \\
 1 & q_2 & q_2 & 1 & q_2 & q_2 & 1 & q_2 & q_2 &q_1\,q_2 &q_1\,q_2 &q_1\,q_2 \\
 1 & 1 & 1 & 1 & 1 & 1 & q_1 & q_1 & q_1 & q_1 & q_1 & q_1 \\
 1 & 1 & q_2 & 1 & 1 & q_2 & q_1 &q_1\,q_2 &q_1\,q_2 & q_1 &q_1\,q_2 &q_1\,q_2 \\
 1 & q_2 & q_2 & 1 & q_2 & q_2 & q_1 &q_1\,q_2 &q_1\,q_2 &q_1\,q_2 &q_1\,q_2 &q_1\,q_2 \\
 1 & 1 & 1 & q_1 & q_1 &q_1\,q_2 & q_1 & q_1 &q_1\,q_2 & q_1 & q_1 &q_1\,q_2 \\
 1 & 1 & q_2 & q_1 & q_1 &q_1\,q_2 & q_1 &q_1\,q_2 &q_1\,q_2 & q_1 &q_1\,q_2 &q_1\,q_2 \\
 1 & q_2 & q_2 & q_1 &q_1\,q_2 &q_1\,q_2 & q_1 &q_1\,q_2 &q_1\,q_2 &q_1\,q_2 &q_1\,q_2 &
  q_1\,q_2 \\
\end{array}
\right)~,
\end{equation}}
and its inverse is of the form:
{\begin{equation}
g^{w\,w'}(q) =\left(
\begin{array}{cccccccccccc}
-\,q_1\,q_2 &q_1\,q_2 & 0 &q_1\,q_2 & 0 &-\,q_1 & 0 & 0 & 0 & -q_2 & 0 & 1 \\
q_1\,q_2 & 0 &-\,q_1 &-\,q_1\,q_2 & 0 & q_1 & 0 & 0 & 0 & 0 & 1 & -1 \\
 0 &-\,q_1 & q_1 & 0 & 0 & 0 & 0 & 0 & 0 & 1 & -1 & 0 \\
q_1\,q_2 &-\,q_1\,q_2 & 0 & 0 & 0 & 0 &-\,q_2 & 0 & 1 & q_2 & 0 & -1 \\
 0 & 0 & 0 & 0 & 0 & 0 & 0 & 1 & -1 & 0 & -1 & 1 \\
-\,q_1 & q_1 & 0 & 0 & 0 & 0 & 1 & -1 & 0 & -1 & 1 & 0 \\
 0 & 0 & 0 &-\,q_2 & 0 & 1 & q_2 & 0 & -1 & 0 & 0 & 0 \\
 0 & 0 & 0 & 0 & 1 & -1 & 0 & -1 & 1 & -1 & 1 & 0 \\
 0 & 0 & 0 & 1 & -1 & 0 & -1 & 1 & 0 & 1 & -1 & 0 \\
-\,q_2 & 0 & 1 & q_2 & 0 & -1 & 0 & -1 & 1 & 0 & 0 & 0 \\
 0 & 1 & -1 & 0 & -1 & 1 & 0 & 1 & -1 & 0 & 0 & 0 \\
 1 & -1 & 0 & -1 & 1 & 0 & 0 & 0 & 0 & 0 & 0 & 0 \\
\end{array}
\right)
\end{equation}}

From the inverse of the topological metric, one can work out the dual Schubert classes in the quantum K-theory ring as in \eqref{dual schubert}. More explicitly, we have:
{\small\begin{align*}
        &\CO^{\vee\,(1\,2\,3\,4)}\,=\,-\,q_1\,q_2\,+\,q_1\,q_2\,\CO_{(1\,2\,4\,3)}\,+\,q_1\,q_2\,\CO_{(2\,1\,3\,4)}\,-\,q_1\,\CO_{(2\,3\,4\,1)}\,-\,q_2\,\CO_{(4\,1\,2\,3)}\,+\,\CO_{(4\,2\,3\,1)}~,\\
        &\CO^{\vee\,(1\,2\,4\,3)}\,=\,q_1\,q_2\,-\,q_1\,\CO_{(1\,3\,4\,2)}\,-\,q_1\,q_2\,\CO_{(2\,1\,3\,4)}\,+\,q_1\,\CO_{(2\,3\,4\,1)}\,+\,\CO_{(4\,1\,3\,2)}\,-\,\CO_{(4\,2\,3\,1)}~,\\
        &\CO^{\vee\,(1\,3\,4\,2)}\,=\,-\,q_1\,\CO_{(1\,2\,4\,3)}\,+\,q_1\,\CO_{(1\,3\,4\,2)}\,+\,\CO_{(4\,1\,2\,3)}\,-\,\CO_{(4\,1\,3\,2)}~,\\
        &\CO^{\vee\,(2\,1\,3\,4)}\,=\,q_1\,q_2\,-\,q_1\,q_2\,\CO_{(1\,2\,4\,3)}\,-\,q_2\,\CO_{(3\,1\,2\,4)}\,+\,\CO_{(3\,2\,4\,1)}\,+\,q_2\,\CO_{(4\,1\,2\,3)}\,-\,\CO_{(4\,2\,3\,1)}~,\\
        &\CO^{\vee\,(2\,1\,4\,3)}\,=\,\CO_{(3\,1\,4\,2)}\,-\,\CO_{(3\,2\,4\,1)}\,-\,\CO_{(4\,1\,3\,2)}\,+\,\CO_{(4\,2\,3\,1)}~,\\
        &\CO^{\vee\,(2\,3\,4\,1)}\,=\,-\,q_1\,+\,q_1\,\CO_{(1\,2\,4\,3)}\,+\,\CO_{(3\,1\,2\,4)}\,-\,\CO_{(3\,1\,4\,2)}\,-\,\CO_{(4\,1\,2\,3)}\,+\,\CO_{(4\,1\,3\,2)}~,\\
        &\CO^{\vee\,(3\,1\,2\,4)}\,=\,-\,q_2\,\CO_{(2\,1\,3\,4)}\,+\,\CO_{(2\,3\,4\,1)}\,+\,q_2\,\CO_{(3\,1\,2\,4)}\,-\,\CO_{(3\,2\,4\,1)}~,\\
        &\CO^{\vee\,(3\,1\,4\,2)}\,=\,\CO_{(2\,1\,4\,3)}\,-\,\CO_{(2\,3\,4\,1)}\,-\,\CO_{(3\,1\,4\,2)}\,+\,\CO_{(3\,2\,4\,1)}\,-\,\CO_{(4\,1\,2\,3)}\,+\,\CO_{(4\,1\,3\,2)}~,\\
        &\CO^{\vee\,(3\,2\,4\,1)}\,=\,\CO_{(2\,1\,3\,4)}\,-\,\CO_{(2\,1\,4\,3)}\,-\,\CO_{(3\,1\,2\,4)}\,+\,\CO_{(3\,1\,4\,2)}\,+\,\CO_{(4\,1\,2\,3)}\,-\,\CO_{(4\,1\,3\,2)}~,\\
        &\CO^{\vee\,(4\,1\,2\,3)}\,=\,-\,q_2\,+\,\CO_{(1\,3\,4\,2)}\,+\,q_2\,\CO_{(2\,1\,3\,4)}\,-\,\CO_{(2\,3\,4\,1)}\,-\,\CO_{(3\,1\,4\,2)}\,+\,\CO_{(3\,2\,4\,1)}~,\\
        &\CO^{\vee\,(4\,1\,3\,2)}\,=\,\CO_{(1\,2\,4\,3)}\,-\,\CO_{(1\,3\,4\,2)}\,-\,\CO_{(2\,1\,4\,3)}\,+\,\CO_{(2\,3\,4\,1)}\,+\,\CO_{(3\,1\,4\,2)}\,-\,\CO_{(3\,2\,4\,1)}~,\\
        &\CO^{\vee\,(4\,2\,3\,1)}\,=\,1\,-\,\CO_{(1\,2\,4\,3)}\,-\,\CO_{(2\,1\,3\,4)}\,+\,\CO_{(2\,1\,4\,3)}~.
\end{align*}}

\begin{table}[t!]
    \centering
    \begin{tabular}{|c||c|c|c|c|c|c|}
    \hline
        $w$ &  $(4\,2\,3\,1)$ & $(4\,1\,3\,2)$ & $(4\,1\,2\,3)$ & $(3\,2\,4\,1)$ & $(3\,1\,4\,2)$ & $(3\,1\,2\,4)$\\
        \hline
         $w^{\rm opp}$& $(1\,2\,3\,4)$ & $(1\,2\,4\,3)$ & $(1\,3\,4\,2)$ & $(2\,1\,3\,4)$ & $(2\,1\,4\,3)$ & $(2\,3\,4\,1)$\\
         \hline
    \end{tabular}
    \caption{$w$ and $w^{\rm opp}$ for the incidence flag Fl$(1,3;4)$}
    \label{tab:wopp}
\end{table}

In the classical limit, these reduce to:
{\small\begin{align*}
        &\CO^{\vee\,(1\,2\,3\,4)}\,=\,\CO_{(4\,2\,3\,1)}~,\\
        &\CO^{\vee\,(1\,2\,4\,3)}\,=\,\CO_{(4\,1\,3\,2)}\,-\,\CO_{(4\,2\,3\,1)}~,\\
        &\CO^{\vee\,(1\,3\,4\,2)}\,=\,\CO_{(4\,1\,2\,3)}\,-\,\CO_{(4\,1\,3\,2)}~,\\
        &\CO^{\vee\,(2\,1\,3\,4)}\,=\,\CO_{(3\,2\,4\,1)}\,-\,\CO_{(4\,2\,3\,1)}~,\\
        &\CO^{\vee\,(2\,1\,4\,3)}\,=\,\CO_{(3\,1\,4\,2)}\,-\,\CO_{(3\,2\,4\,1)}\,-\,\CO_{(4\,1\,3\,2)}\,+\,\CO_{(4\,2\,3\,1)}~,\\
        &\CO^{\vee\,(2\,3\,4\,1)}\,=\,\CO_{(3\,1\,2\,4)}\,-\,\CO_{(3\,1\,4\,2)}\,-\,\CO_{(4\,1\,2\,3)}\,+\,\CO_{(4\,1\,3\,2)}~,\\
        &\CO^{\vee\,(3\,1\,2\,4)}\,=\,\CO_{(2\,3\,4\,1)}\,-\,\CO_{(3\,2\,4\,1)}~,\\
        &\CO^{\vee\,(3\,1\,4\,2)}\,=\,\CO_{(2\,1\,4\,3)}\,-\,\CO_{(2\,3\,4\,1)}\,-\,\CO_{(3\,1\,4\,2)}\,+\,\CO_{(3\,2\,4\,1)}\,-\,\CO_{(4\,1\,2\,3)}\,+\,\CO_{(4\,1\,3\,2)}~,\\
        &\CO^{\vee\,(3\,2\,4\,1)}\,=\,\CO_{(2\,1\,3\,4)}\,-\,\CO_{(2\,1\,4\,3)}\,-\,\CO_{(3\,1\,2\,4)}\,+\,\CO_{(3\,1\,4\,2)}\,+\,\CO_{(4\,1\,2\,3)}\,-\,\CO_{(4\,1\,3\,2)}~,\\
        &\CO^{\vee\,(4\,1\,2\,3)}\,=\,\CO_{(1\,3\,4\,2)}\,-\,\CO_{(2\,3\,4\,1)}\,-\,\CO_{(3\,1\,4\,2)}\,+\,\CO_{(3\,2\,4\,1)}~,\\
        &\CO^{\vee\,(4\,1\,3\,2)}\,=\,\CO_{(1\,2\,4\,3)}\,-\,\CO_{(1\,3\,4\,2)}\,-\,\CO_{(2\,1\,4\,3)}\,+\,\CO_{(2\,3\,4\,1)}\,+\,\CO_{(3\,1\,4\,2)}\,-\,\CO_{(3\,2\,4\,1)}~,\\
        &\CO^{\vee\,(4\,2\,3\,1)}\,=\,1\,-\,\CO_{(1\,2\,4\,3)}\,-\,\CO_{(2\,1\,3\,4)}\,+\,\CO_{(2\,1\,4\,3)}~.
\end{align*}}
These match with \cite[Equation~(4.14)]{xu2024quantum}. Keeping in mind that, in \cite{xu2024quantum}, Xu expands the dual Schubert classes in terms of the opposite Schubert classes $\CO^w$ rather than the Schubert classes $\CO_w$ as we do here. To match the two expressions, let us denote the opposite Schubert class of $\CO_w$ by $\CO^{w^{\rm opp}}$, where $w^{\rm opp}$ is the minimal representative of the coset of $w_0 w$ \footnote{We would like to thank Weihong Xu for pointing this out. See \cite[Section~10.8]{Anderson_Fulton_2023} for more details.} and ${(w^{\rm opp}})^{\rm opp} = w$. In table \ref{tab:wopp}, we list all the permutation pairs for Fl$(1,3;4)$.

\medskip
\noindent
\textbf{Equivariant ring relations.}
{\small
}

\medskip
\noindent
\textbf{The non-equivariant limit.}
{\small\begin{align*}
\mathcal{O}_{(1\,2\,4\,3)}\,\otimes_q\,\mathcal{O}_{(1\,2\,4\,3)}
&\,=\,\mathcal{O}_{(1\,3\,4\,2)}~,\\
\mathcal{O}_{(1\,2\,4\,3)}\,\otimes_q\,\mathcal{O}_{(2\,1\,3\,4)}
&\,=\,\mathcal{O}_{(2\,1\,4\,3)}~,\\
\mathcal{O}_{(1\,2\,4\,3)}\,\otimes_q\,\mathcal{O}_{(1\,3\,4\,2)}
&\,=\,\mathcal{O}_{(2\,3\,4\,1)}
\,+\,q_2\,\left(1\,-\,\mathcal{O}_{(2\,1\,3\,4)}\right)~,\\
\mathcal{O}_{(1\,2\,4\,3)}\,\otimes_q\,\mathcal{O}_{(2\,1\,4\,3)}
&\,=\,\mathcal{O}_{(2\,3\,4\,1)}\,-\,\mathcal{O}_{(3\,2\,4\,1)}
\,+\,\mathcal{O}_{(3\,1\,4\,2)}~,\\
\mathcal{O}_{(1\,2\,4\,3)}\,\otimes_q\,\mathcal{O}_{(3\,1\,2\,4)}
&\,=\,\mathcal{O}_{(3\,1\,4\,2)}\,-\,\mathcal{O}_{(4\,1\,3\,2)}
\,+\,\mathcal{O}_{(4\,1\,2\,3)}~,\\
\mathcal{O}_{(1\,2\,4\,3)}\,\otimes_q\,\mathcal{O}_{(2\,3\,4\,1)}
&\,=\,q_2\,\mathcal{O}_{(2\,1\,3\,4)}~,\\
\mathcal{O}_{(1\,2\,4\,3)}\,\otimes_q\,\mathcal{O}_{(3\,2\,4\,1)}
&\,=\,q_2\,\mathcal{O}_{(3\,1\,2\,4)}~,\\
\mathcal{O}_{(1\,2\,4\,3)}\,\otimes_q\,\mathcal{O}_{(4\,1\,3\,2)}
&\,=\,\mathcal{O}_{(4\,2\,3\,1)}~,\\
\mathcal{O}_{(1\,2\,4\,3)}\,\otimes_q\,\mathcal{O}_{(4\,1\,2\,3)}
&\,=\,\mathcal{O}_{(4\,1\,3\,2)}~,\\
\mathcal{O}_{(1\,2\,4\,3)}\,\otimes_q\,\mathcal{O}_{(3\,1\,4\,2)}
&\,=\,\mathcal{O}_{(3\,2\,4\,1)}~,\\
\mathcal{O}_{(1\,2\,4\,3)}\,\otimes_q\,\mathcal{O}_{(4\,2\,3\,1)}
&\,=\,q_2\,\mathcal{O}_{(4\,1\,2\,3)}
\,+\,q_1\,q_2\,\left(1\,-\,\mathcal{O}_{(1\,2\,4\,3)}\right)~,\\
\mathcal{O}_{(2\,1\,3\,4)}\,\otimes_q\,\mathcal{O}_{(2\,1\,3\,4)}
&\,=\,\mathcal{O}_{(3\,1\,2\,4)}~,\\
\mathcal{O}_{(2\,1\,3\,4)}\,\otimes_q\,\mathcal{O}_{(1\,3\,4\,2)}
&\,=\,\mathcal{O}_{(2\,3\,4\,1)}\,-\,\mathcal{O}_{(3\,2\,4\,1)}
\,+\,\mathcal{O}_{(3\,1\,4\,2)}~,\\
\mathcal{O}_{(2\,1\,3\,4)}\,\otimes_q\,\mathcal{O}_{(2\,1\,4\,3)}
&\,=\,\mathcal{O}_{(3\,1\,4\,2)}\,-\,\mathcal{O}_{(4\,1\,3\,2)}
\,+\,\mathcal{O}_{(4\,1\,2\,3)}~,\\
\mathcal{O}_{(2\,1\,3\,4)}\,\otimes_q\,\mathcal{O}_{(3\,1\,2\,4)}
&\,=\,\mathcal{O}_{(4\,1\,2\,3)}
\,+\,q_1\,\left(1\,-\,\mathcal{O}_{(1\,2\,4\,3)}\right)~,\\
\mathcal{O}_{(2\,1\,3\,4)}\,\otimes_q\,\mathcal{O}_{(2\,3\,4\,1)}
&\,=\,\mathcal{O}_{(3\,2\,4\,1)}~,\\
\mathcal{O}_{(2\,1\,3\,4)}\,\otimes_q\,\mathcal{O}_{(3\,2\,4\,1)}
&\,=\,\mathcal{O}_{(4\,2\,3\,1)}~,\\
\mathcal{O}_{(2\,1\,3\,4)}\,\otimes_q\,\mathcal{O}_{(4\,1\,3\,2)}
&\,=\,q_1\,\mathcal{O}_{(1\,3\,4\,2)}~,\\
\mathcal{O}_{(2\,1\,3\,4)}\,\otimes_q\,\mathcal{O}_{(4\,1\,2\,3)}
&\,=\,q_1\,\mathcal{O}_{(1\,2\,4\,3)}~,\\
\mathcal{O}_{(2\,1\,3\,4)}\,\otimes_q\,\mathcal{O}_{(3\,1\,4\,2)}
&\,=\,\mathcal{O}_{(4\,1\,3\,2)}~,\\
\mathcal{O}_{(2\,1\,3\,4)}\,\otimes_q\,\mathcal{O}_{(4\,2\,3\,1)}
&\,=\,q_1\,\mathcal{O}_{(2\,3\,4\,1)}
\,+\,q_1\,q_2\,\left(1\,-\,\mathcal{O}_{(2\,1\,3\,4)}\right)~,\\
\mathcal{O}_{(1\,3\,4\,2)}\,\otimes_q\,\mathcal{O}_{(1\,3\,4\,2)}
&\,=\,q_2\,\left(\mathcal{O}_{(1\,2\,4\,3)}
\,+\,\mathcal{O}_{(2\,1\,3\,4)}
\,-\,\mathcal{O}_{(2\,1\,4\,3)}\right)~,\\
\mathcal{O}_{(1\,3\,4\,2)}\,\otimes_q\,\mathcal{O}_{(2\,1\,4\,3)}
&\,=\,\mathcal{O}_{(3\,2\,4\,1)}
\,+\,q_2\,\left(\mathcal{O}_{(2\,1\,3\,4)}\,-\,\mathcal{O}_{(3\,1\,2\,4)}\right)~,\\
\mathcal{O}_{(1\,3\,4\,2)}\,\otimes_q\,\mathcal{O}_{(3\,1\,2\,4)}
&\,=\,\mathcal{O}_{(3\,2\,4\,1)}\,-\,\mathcal{O}_{(4\,2\,3\,1)}
\,+\,\mathcal{O}_{(4\,1\,3\,2)}~,\\
\mathcal{O}_{(1\,3\,4\,2)}\,\otimes_q\,\mathcal{O}_{(2\,3\,4\,1)}
&\,=\,q_2\,\mathcal{O}_{(2\,1\,4\,3)}~,\\
\mathcal{O}_{(1\,3\,4\,2)}\,\otimes_q\,\mathcal{O}_{(3\,2\,4\,1)}
&\,=\,q_2\,\left(\mathcal{O}_{(3\,1\,4\,2)}
\,-\,\mathcal{O}_{(4\,1\,3\,2)}
\,+\,\mathcal{O}_{(4\,1\,2\,3)}\right)~,\\
\mathcal{O}_{(1\,3\,4\,2)}\,\otimes_q\,\mathcal{O}_{(4\,1\,3\,2)}
&\,=\,q_2\,\mathcal{O}_{(4\,1\,2\,3)}
\,+\,q_1\,q_2\,\left(1\,-\,\mathcal{O}_{(1\,2\,4\,3)}\right)~,\\
\mathcal{O}_{(1\,3\,4\,2)}\,\otimes_q\,\mathcal{O}_{(4\,1\,2\,3)}
&\,=\,\mathcal{O}_{(4\,2\,3\,1)}~,\\
\mathcal{O}_{(1\,3\,4\,2)}\,\otimes_q\,\mathcal{O}_{(3\,1\,4\,2)}
&\,=\,q_2\,\mathcal{O}_{(3\,1\,2\,4)}~,\\
\mathcal{O}_{(1\,3\,4\,2)}\,\otimes_q\,\mathcal{O}_{(4\,2\,3\,1)}
&\,=\,q_2\,\mathcal{O}_{(4\,1\,3\,2)}
\,+\,q_1\,q_2\,\left(\mathcal{O}_{(1\,2\,4\,3)}\,-\,\mathcal{O}_{(1\,3\,4\,2)}\right)~,\\
\mathcal{O}_{(2\,1\,4\,3)}\,\otimes_q\,\mathcal{O}_{(3\,1\,2\,4)}
&\,=\,\mathcal{O}_{(4\,1\,3\,2)}
\,+\,q_1\,\left(\mathcal{O}_{(1\,2\,4\,3)}\,-\,\mathcal{O}_{(1\,3\,4\,2)}\right)~,\\
\mathcal{O}_{(2\,1\,4\,3)}\,\otimes_q\,\mathcal{O}_{(2\,1\,4\,3)}
&\,=\,\mathcal{O}_{(3\,2\,4\,1)}
\,+\,\mathcal{O}_{(4\,1\,3\,2)}
\,-\,\mathcal{O}_{(4\,2\,3\,1)}~,\\
\mathcal{O}_{(2\,1\,4\,3)}\,\otimes_q\,\mathcal{O}_{(2\,3\,4\,1)}
&\,=\,q_2\,\mathcal{O}_{(3\,1\,2\,4)}~,\\
\mathcal{O}_{(2\,1\,4\,3)}\,\otimes_q\,\mathcal{O}_{(3\,2\,4\,1)}
&\,=\,q_2\,\mathcal{O}_{(4\,1\,2\,3)}
\,+\,q_1\,q_2\,\left(1\,-\,\mathcal{O}_{(1\,2\,4\,3)}\right)~,\\
\mathcal{O}_{(2\,1\,4\,3)}\,\otimes_q\,\mathcal{O}_{(4\,1\,3\,2)}
&\,=\,q_1\,\mathcal{O}_{(2\,3\,4\,1)}
\,+\,q_1\,q_2\,\left(1\,-\,\mathcal{O}_{(2\,1\,3\,4)}\right)~,\\
\mathcal{O}_{(2\,1\,4\,3)}\,\otimes_q\,\mathcal{O}_{(4\,1\,2\,3)}
&\,=\,q_1\,\mathcal{O}_{(1\,3\,4\,2)}~,\\
\mathcal{O}_{(2\,1\,4\,3)}\,\otimes_q\,\mathcal{O}_{(3\,1\,4\,2)}
&\,=\,\mathcal{O}_{(4\,2\,3\,1)}~,\\
\mathcal{O}_{(2\,1\,4\,3)}\,\otimes_q\,\mathcal{O}_{(4\,2\,3\,1)}
&\,=\,q_1\,q_2\,\left(\mathcal{O}_{(1\,2\,4\,3)}
\,+\,\mathcal{O}_{(2\,1\,3\,4)}
\,-\,\mathcal{O}_{(2\,1\,4\,3)}\right)~,\\
\mathcal{O}_{(3\,1\,2\,4)}\,\otimes_q\,\mathcal{O}_{(3\,1\,2\,4)}
&\,=\,q_1\,\left(\mathcal{O}_{(1\,2\,4\,3)}
\,+\,\mathcal{O}_{(2\,1\,3\,4)}
\,-\,\mathcal{O}_{(2\,1\,4\,3)}\right)~,\\
\mathcal{O}_{(3\,1\,2\,4)}\,\otimes_q\,\mathcal{O}_{(2\,3\,4\,1)}
&\,=\,\mathcal{O}_{(4\,2\,3\,1)}~,\\
\mathcal{O}_{(3\,1\,2\,4)}\,\otimes_q\,\mathcal{O}_{(3\,2\,4\,1)}
&\,=\,q_1\,\mathcal{O}_{(2\,3\,4\,1)}
\,+\,q_1\,q_2\,\left(1\,-\,\mathcal{O}_{(2\,1\,3\,4)}\right)~,\\
\mathcal{O}_{(3\,1\,2\,4)}\,\otimes_q\,\mathcal{O}_{(4\,1\,3\,2)}
&\,=\,q_1\,\left(\mathcal{O}_{(2\,3\,4\,1)}
\,-\,\mathcal{O}_{(3\,2\,4\,1)}
\,+\,\mathcal{O}_{(3\,1\,4\,2)}\right)~,\\
\mathcal{O}_{(3\,1\,2\,4)}\,\otimes_q\,\mathcal{O}_{(4\,1\,2\,3)}
&\,=\,q_1\,\mathcal{O}_{(2\,1\,4\,3)}~,\\
\mathcal{O}_{(3\,1\,2\,4)}\,\otimes_q\,\mathcal{O}_{(3\,1\,4\,2)}
&\,=\,q_1\,\mathcal{O}_{(1\,3\,4\,2)}~,\\
\mathcal{O}_{(3\,1\,2\,4)}\,\otimes_q\,\mathcal{O}_{(4\,2\,3\,1)}
&\,=\,q_1\,\mathcal{O}_{(3\,2\,4\,1)}
\,+\,q_1\,q_2\,\left(\mathcal{O}_{(2\,1\,3\,4)}\,-\,\mathcal{O}_{(3\,1\,2\,4)}\right)~,\\
\mathcal{O}_{(2\,3\,4\,1)}\,\otimes_q\,\mathcal{O}_{(2\,3\,4\,1)}
&\,=\,q_2\,\mathcal{O}_{(3\,1\,4\,2)}~,\\
\mathcal{O}_{(2\,3\,4\,1)}\,\otimes_q\,\mathcal{O}_{(3\,2\,4\,1)}
&\,=\,q_2\,\mathcal{O}_{(4\,1\,3\,2)}~,\\
\mathcal{O}_{(2\,3\,4\,1)}\,\otimes_q\,\mathcal{O}_{(4\,1\,3\,2)}
&\,=\,q_1\,q_2\,\mathcal{O}_{(1\,2\,4\,3)}~,\\
\mathcal{O}_{(2\,3\,4\,1)}\,\otimes_q\,\mathcal{O}_{(4\,1\,2\,3)}
&\,=\,q_1\,q_2~,\\
\mathcal{O}_{(2\,3\,4\,1)}\,\otimes_q\,\mathcal{O}_{(3\,1\,4\,2)}
&\,=\,q_2\,\mathcal{O}_{(4\,1\,2\,3)}~,\\
\mathcal{O}_{(2\,3\,4\,1)}\,\otimes_q\,\mathcal{O}_{(4\,2\,3\,1)}
&\,=\,q_1\,q_2\,\mathcal{O}_{(1\,3\,4\,2)}~,\\
\mathcal{O}_{(3\,2\,4\,1)}\,\otimes_q\,\mathcal{O}_{(3\,2\,4\,1)}
&\,=\,q_1\,q_2\,\mathcal{O}_{(1\,3\,4\,2)}~,\\
\mathcal{O}_{(3\,2\,4\,1)}\,\otimes_q\,\mathcal{O}_{(4\,1\,3\,2)}
&\,=\,q_1\,q_2\,\mathcal{O}_{(2\,1\,4\,3)}~,\\
\mathcal{O}_{(3\,2\,4\,1)}\,\otimes_q\,\mathcal{O}_{(4\,1\,2\,3)}
&\,=\,q_1\,q_2\,\mathcal{O}_{(2\,1\,3\,4)}~,\\
\mathcal{O}_{(3\,2\,4\,1)}\,\otimes_q\,\mathcal{O}_{(3\,1\,4\,2)}
&\,=\,q_1\,q_2\,\mathcal{O}_{(1\,2\,4\,3)}~,\\
\mathcal{O}_{(3\,2\,4\,1)}\,\otimes_q\,\mathcal{O}_{(4\,2\,3\,1)}
&\,=\,q_1\,q_2\,\left(\mathcal{O}_{(2\,3\,4\,1)}
\,+\,\mathcal{O}_{(3\,1\,4\,2)}
\,-\,\mathcal{O}_{(3\,2\,4\,1)}\right)~,\\
\mathcal{O}_{(4\,1\,3\,2)}\,\otimes_q\,\mathcal{O}_{(4\,1\,2\,3)}
&\,=\,q_1\,\mathcal{O}_{(3\,2\,4\,1)}~,\\
\mathcal{O}_{(4\,1\,3\,2)}\,\otimes_q\,\mathcal{O}_{(4\,1\,3\,2)}
&\,=\,q_1\,q_2\,\mathcal{O}_{(3\,1\,2\,4)}~,\\
\mathcal{O}_{(4\,1\,3\,2)}\,\otimes_q\,\mathcal{O}_{(3\,1\,4\,2)}
&\,=\,q_1\,q_2\,\mathcal{O}_{(2\,1\,3\,4)}~,\\
\mathcal{O}_{(4\,1\,3\,2)}\,\otimes_q\,\mathcal{O}_{(4\,2\,3\,1)}
&\,=\,q_1\,q_2\,\left(\mathcal{O}_{(3\,1\,4\,2)}
\,-\,\mathcal{O}_{(4\,1\,3\,2)}\,+\, \mathcal{O}_{(4\,1\,2\,3)}\right)~,\\
\mathcal{O}_{(4\,1\,2\,3)}\,\otimes_q\,\mathcal{O}_{(3\,1\,4\,2)}
&\,=\,q_1\,\mathcal{O}_{(2\,3\,4\,1)}~,\\
\mathcal{O}_{(4\,1\,2\,3)}\,\otimes_q\,\mathcal{O}_{(4\,1\,2\,3)}
&\,=\,q_1\,\mathcal{O}_{(3\,1\,4\,2)}~,\\
\mathcal{O}_{(4\,1\,2\,3)}\,\otimes_q\,\mathcal{O}_{(4\,2\,3\,1)}
&\,=\,q_1\,q_2\,\mathcal{O}_{(3\,1\,2\,4)}~,\\
\mathcal{O}_{(3\,1\,4\,2)}\,\otimes_q\,\mathcal{O}_{(3\,1\,4\,2)}
&\,=\,q_1\,q_2~,\\
\mathcal{O}_{(3\,1\,4\,2)}\,\otimes_q\,\mathcal{O}_{(4\,2\,3\,1)}
&\,=\,q_1\,q_2\,\mathcal{O}_{(2\,1\,4\,3)}~,\\
\mathcal{O}_{(4\,2\,3\,1)}\,\otimes_q\,\mathcal{O}_{(4\,2\,3\,1)}
&\,=\,q_1\,q_2\,\left(\mathcal{O}_{(4\,1\,3\,2)}
\,+\,\mathcal{O}_{(3\,2\,4\,1)}
\,-\,\mathcal{O}_{(4\,2\,3\,1)}\right)~.
\end{align*}}

\medskip
\noindent
\textbf{Comparison with the literature.} The equivariant relations above involving either the Schubert class $\CO_{(2\,1\,3\,4)}$ or $\CO_{(1\,2\,4\,3)}$ match with those given in \cite[Theorem~4.5]{xu2024quantum}. As for the non-equivariant relations, they match those obtained by \cite[Theorem~5.1]{xu2024quantum}.

\subsection{The partial flag Fl\texorpdfstring{$(1,2;4)$}{124}}\label{app: QK Fl124}
In this subsection, we list all the ring relations -- in both, equivariant and non-equivariant forms -- for the partial flag Fl$(1,2;4)$. In this case, we have $12$ Schubert classes; therefore, there are $67$ independent non-trivial relations. We order these classes as follows:
\begin{equation}
    \begin{split}
         \{&(1\, 2\, 3\, 4)\,,\, (1\, 3\, 2\, 4)\,,\, (1\, 4\, 2\, 3)\,,\, (2\, 1\, 3\, 4)\,,\, (2\, 3\, 1\, 
  4)\,,\, (2\, 4\, 1\, 3)\,,\\&(3\, 1\, 2\, 4)\,,\, (3\, 2\, 1\, 4)\,,\, (3\, 4\, 1\, 2)\,,\, (4\, 1\, 
  2\, 3)\,,\, (4\, 2\, 1\, 3)\,,\, (4\, 3\, 1\, 2)\}~.
    \end{split}
\end{equation}
The corresponding $\ell(w)$ and $\dim(X_w)$ are given in \cite[Appendix A]{Closset:2026bnk}.

\medskip
\noindent
\textbf{The topological metric and the dual Schubert classes.} Before listing the quantum ring relations, let us exhibit the non-equivariant form of the topological metric of this model. 
% \OK{This is not the right place!!}
% The topological metric in the non-equivariant limit is
{\begin{equation*}
    g_{w\,w'} (q)\,=\, \frac{1}{(1\,-\,q_1)\,(1\,-\,q_2)}\,\left(
\begin{array}{cccccccccccc}
 1 & 1 & 1 & 1 & 1 & 1 & 1 & 1 & 1 & 1 & 1 & 1 \\
 1 & 1 & 1 & 1 & 1 & 1 & 1 & 1 & q_2 & 1 & 1 & q_2 \\
 1 & 1 & 1 & 1 & q_2 & q_2 & 1 & q_2 & q_2 & 1 & q_2 & q_2 \\
 1 & 1 & 1 & 1 & 1 & 1 & 1 & 1 & 1 & q_1 & q_1 & q_1 \\
 1 & 1 & q_2 & 1 & 1 & q_2 & 1 & 1 & q_2 & q_1\, q_2 & q_1 \,q_2 & q_1 \,q_2 \\
 1 & 1 & q_2 & 1 & q_2 & q_2 & 1 & q_2 & q_2 & q_1\, q_2 & q_1 \,q_2 & q_1\, q_2 \\
 1 & 1 & 1 & 1 & 1 & 1 & q_1 & q_1 & q_1\, q_2 & q_1 & q_1 & q_1\, q_2 \\
 1 & 1 & q_2 & 1 & 1 & q_2 & q_1 & q_1 & q_1\, q_2 & q_1\, q_2 & q_1\, q_2 & q_1 \,q_2 \\
 1 & q_2 & q_2 & 1 & q_2 & q_2 & q_1 \,q_2 & q_1 \,q_2 & q_1 \,q_2^2 & q_1\, q_2 & q_1\, q_2 & q_1\, q_2^2 \\
 1 & 1 & 1 & q_1 & q_1 q_2 & q_1\, q_2 & q_1 & q_1\, q_2 & q_1\, q_2 & q_1 & q_1\, q_2 & q_1\, q_2 \\
 1 & 1 & q_2 & q_1 & q_1 \,q_2 & q_1\, q_2 & q_1 & q_1\, q_2 & q_1\, q_2 & q_1 \,q_2 & q_1\, q_2 & q_1\, q_2 \\
 1 & q_2 & q_2 & q_1 & q_1 \,q_2 & q_1\, q_2 & q_1 \,q_2 & q_1 \,q_2 & q_1 \,q_2^2 & q_1 \,q_2 & q_1\, q_2 & q_1 \,q_2^2 \\
\end{array}
\right)~.
\end{equation*}}
This is obtained from the companion matrix method that we reviewed in subsection~\ref{QK from ComMatrix} in the main text. The inverse of the topological metric is of the form:
{\begin{equation*}
    g^{w\,w'}(q)\,=\, \left(
\begin{array}{cccccccccccc}
 0 & -\,q_1 \,q_2 & q_1 \,q_2 & 0 & 0 & 0 & q_1 \,q_2 & 0 & -\,q_1 & -\,q_2 & 0 & 1 \\
 -\,q_1 \,q_2 & q_1 \,q_2 & 0 & q_1\, q_2 & 0 & -\,q_1 & -\,q_1\, q_2 & 0 & q_1 & 0 & 1 & -1 \\
 q_1\, q_2 & 0 & -\,q_1 & -\,q_1\, q_2 & 0 & q_1 & 0 & 0 & 0 & 1 & -1 & 0 \\
 0 & q_1\, q_2 & -\,q_1\, q_2 & 0 & 0 & 0 & -\,q_2 & 0 & 1 & q_2 & 0 & -1 \\
 0 & 0 & 0 & 0 & -\,q_1 & q_1 & 0 & 1 & -1 & 0 & -1 & 1 \\
 0 & -\,q_1 & q_1 & 0 & q_1 & -\,q_1 & 1 & -1 & 0 & -1 & 1 & 0 \\
 q_1\, q_2 & -\,q_1\, q_2 & 0 & -\,q_2 & 0 & 1 & q_2 & 0 & -1 & 0 & -1 & 1 \\
 0 & 0 & 0 & 0 & 1 & -1 & 0 & -1 & 1 & 0 & 1 & -1 \\
 -\,q_1 & q_1 & 0 & 1 & -1 & 0 & -1 & 1 & 0 & 0 & 0 & 0 \\
 -\,q_2 & 0 & 1 & q_2 & 0 & -1 & 0 & 0 & 0 & -1 & 1 & 0 \\
 0 & 1 & -1 & 0 & -1 & 1 & -1 & 1 & 0 & 1 & -1 & 0 \\
 1 & -1 & 0 & -1 & 1 & 0 & 1 & -1 & 0 & 0 & 0 & 0 \\
\end{array}
\right)~.
\end{equation*}}

Using the inverse of the metric and the definition \eqref{dual schubert}, one can work out the explicit form of the dual Schubert classes in the quantum K-theory of Fl$(1,2;4)$. They are given by:
{\small\begin{align*}
        &\CO^{\vee\,(1\,2\,3\,4)}\,=\,-\,q_1\,q_2\,\CO_{(1\,3\,2\,4)}\,+\,q_1\,q_2\,\CO_{(1\,4\,2\,3)}\,+\,q_1\,q_2\,\CO_{(3\,1\,2\,4)}\,-\,q_1\,\CO_{(3\,4\,1\,2)}\,-\,q_2\,\CO_{(4\,1\,2\,3)}\,+\,\CO_{(4\,3\,1\,2)}~,\\
        &\CO^{\vee\,(1\,3\,2\,4)}\,=\,-\,q_1\,q_2\,+\,q_1\,q_2\,\CO_{(1\,3\,2\,4)}\,+\,q_1\,q_2\,\CO_{(2\,1\,3\,4)}\,-\,q_1\,\CO_{(2\,4\,1\,3)}\,-\,q_1\,q_2\,\CO_{(3\,1\,2\,4)}\,+\,q_1\,\CO_{(3\,4\,1\,2)}\\
        &\qquad \qquad ~+\,\CO_{(4\,2\,1\,3)}\,-\,\CO_{(4\,3\,1\,2)}~,\\
        &\CO^{\vee\,(1\,4\,2\,3)}\,=\,q_1\,q_2\,-\,q_1\,\CO_{(1\,4\,2\,3)}\,-\,q_1\,q_2\,\CO_{(2\,1\,3\,4)}\,+\,q_1\,\CO_{(2\,4\,1\,3)}\,+\,\CO_{(4\,1\,2\,3)}\,-\,\CO_{(4\,2\,1\,3)}~,\\
        &\CO^{\vee\,(2\,1\,3\,4)}\,=\,q_1\,q_2\,\CO_{(1\,3\,2\,4)}\,-\,q_1\,q_2\,\CO_{(1\,4\,2\,3)}\,-\,q_2\,\CO_{(3\,1\,2\,4)}\,+\,\CO_{(3\,4\,1\,2)}\,+\,q_2\,\CO_{(4\,1\,2\,3)}\,-\,\CO_{(4\,3\,1\,2)}~,\\
        &\CO^{\vee\,(2\,3\,1\,4)}\,=\,-\,q_1\,\CO_{(2\,3\,1\,4)}\,+\,q_1\,\CO_{(2\,4\,1\,3)}\,+\,\CO_{(3\,2\,1\,4)}\,-\,\CO_{(3\,4\,1\,2)}\,-\,\CO_{(4\,2\,1\,3)}\,+\,\CO_{(4\,3\,1\,2)}~,\\
        &\CO^{\vee\,(2\,4\,1\,3)}\,=\,-\,q_1\,\CO_{(1\,3\,2\,4)}\,+\,q_1\,\CO_{(1\,4\,2\,3)}\,+\,q_1\,\CO_{(2\,3\,1\,4)}\,-\,q_1\,\CO_{(2\,4\,1\,3)}\,+\,\CO_{(3\,1\,2\,4)}\,-\,\CO_{(3\,2\,1\,4)}\\&\qquad \qquad ~-\,\CO_{(4\,1\,2\,3)}\,+\,\CO_{(4\,2\,1\,3)}~,\\
        &\CO^{\vee\,(3\,1\,2\,4)}\,=\,q_1\,q_2\,-\,q_1\,q_2\,\CO_{(1\,3\,2\,4)}\,-\,q_2\,\CO_{(2\,1\,3\,4)}\,+\,\CO_{(2\,4\,1\,3)}\,+\,q_2\,\CO_{(3\,1\,2\,4)}\,-\,\CO_{(3\,4\,1\,2)}\,-\,\CO_{(4\,2\,1\,3)}\,\\
        &\qquad\qquad~  +\,\CO_{(4\,3\,1\,2)}~,\\
        &\CO^{\vee\,(3\,2\,1\,4)}\,=\,\CO_{(2\,3\,1\,4)}\,-\,\CO_{(2\,4\,1\,3)}\,-\,\CO_{(3\,2\,1\,4)}\,+\,\CO_{(3\,4\,1\,2)}\,+\,\CO_{(4\,2\,1\,3)}\,-\,\CO_{(4\,3\,1\,2)}~,\\
        &\CO^{\vee\,(3\,4\,1\,2)}\,=\,-\,q_1\,+\,q_1\,\CO_{(1\,3\,2\,4)}\,+\,\CO_{(2\,1\,3\,4)}\,-\,\CO_{(2\,3\,1\,4)}\,-\,\CO_{(3\,1\,2\,4)}\,+\,\CO_{(3\,2\,1\,4)}~,\\
        &\CO^{\vee\,(4\,1\,2\,3)}\,=\,-\,q_2\,+\,\CO_{(1\,4\,2\,3)}\,+\,q_2\,\CO_{(2\,1\,3\,4)}\,-\,\CO_{(2\,4\,1\,3)}\,-\,\CO_{(4\,1\,2\,3)}\,+\,\CO_{(4\,2\,1\,3)}~,\\
        &\CO^{\vee\,(4\,2\,1\,3)}\,=\,\CO_{(1\,3\,2\,4)}\,-\,\CO_{(1\,4\,2\,3)}\,-\,\CO_{(2\,3\,1\,4)}\,+\,\CO_{(2\,4\,1\,3)}\,-\,\CO_{(3\,1\,2\,4)}\,+\,\CO_{(3\,2\,1\,4)}\,+\,\CO_{(4\,1\,2\,3)}\,-\,\CO_{(4\,2\,1\,3)}~,\\
        &\CO^{\vee\,(4\,3\,1\,2)}\,=\,1\,-\,\CO_{(1\,3\,2\,4)}\,-\,\CO_{(2\,1\,3\,4)}\,+\,\CO_{(2\,3\,1\,4)}\,+\,\CO_{(3\,1\,2\,4)}\,-\,\CO_{(3\,2\,1\,4)}~.
    \end{align*}}

\medskip
\noindent
\textbf{The equivariant quantum relations.}
{\small
}

\medskip
\noindent
\textbf{The non-equivariant limit.}
{\small\begin{align*}
&\CO_{(1\,3\,2\,4)}\,\otimes_q\,\CO_{(1\,3\,2\,4)}\,=\,\CO_{(1\,4\,2\,3)}\,+\,\CO_{(2\,3\,1\,4)}\,-\,\CO_{(2\,4\,1\,3)}~,\\
&\CO_{(1\,3\,2\,4)}\,\otimes_q\, \CO_{(1\,4\,2\,3)}\,=\,-\,q_2\, \CO_{(2\,1\,3\,4)}\,+\,\CO_{(2\,4\,1\,3)}\,+\,q_2~,\\
&\CO_{(1\,3\,2\,4)}\otimes_q\CO_{(4\,1\,2\,3)}=\CO_{(4\,2\,1\,3)}~,\\
&\CO_{(1\,4\,2\,3)}\,\otimes_q\,\CO_{(1\,4\,2\,3)}\,=\,q_2\, \CO_{(1\,3\,2\,4)}\,-\,q_2 \,\CO_{(3\,1\,2\,4)}\,+\,\CO_{(3\,4\,1\,2)}~,\\
&\CO_{(1\,3\,2\,4)}\,\otimes_q\,
\CO_{(2\,1\,3\,4)}\,=\,\CO_{(2\,3\,1\,4)}\,+\,\CO_{(3\,1\,2\,4)}\,-\,\CO_{(3\,2\,1\,4)}~,\\
 &\CO_{(1\,4\,2\,3)}\,\otimes_q\,
\CO_{(2\,1\,3\,4)}\,=\,\CO_{(2\,4\,1\,3)}\,+\,\CO_{(4\,1\,2\,3)}\,-\,\CO_{(4\,2\,1\,3)}~,\\
  &\CO_{(2\,1\,3\,4)}\,\otimes_q\,\CO_{(2\,1\,3\,4)}\,=\,-\,q_1 \,\CO_{(1\,3\,2\,4)}\,+\,\CO_{(3\,1\,2\,4)}\,+\,q_1~,\\
  &\CO_{(1\,3\,2\,4)}\,\otimes_q\, \CO_{(2\,3\,1\,4)}\,=\,\CO_{(2\,4\,1\,3)}~,\\
  &\CO_{(1\,4\,2\,3)}\,\otimes_q\, \CO_{(2\,3\,1\,4)}\,=\,q_2\, \CO_{(2\,1\,3\,4)}~,\\
&\CO_{(1\,4\,2\,3)}\otimes_q\CO_{(4\,1\,2\,3)}=\CO_{(4\,3\,1\,2)}~,\\
  &\CO_{(2\,1\,3\,4)}\,\otimes_q\, \CO_{(2\,3\,1\,4)}\,=\,\CO_{(3\,2\,1\,4)}~,\\
&\CO_{(2\,1\,3\,4)}\otimes_q\CO_{(3\,4\,1\,2)}=\CO_{(4\,3\,1\,2)}~,\\
&\CO_{(2\,3\,1\,4)}\otimes_q\CO_{(3\,4\,1\,2)}=q_2\CO_{(4\,1\,2\,3)}~,\\
  &\CO_{(2\,3\,1\,4)}\,\otimes_q\,\CO_{(2\,3\,1\,4)}\,=\,
  \CO_{(3\,4\,1\,2)}~,\\
  &\CO_{(1\,3\,2\,4)}\,\otimes_q\,
   \CO_{(2\,4\,1\,3)}\,=\,q_2\, \CO_{(2\,1\,3\,4)}\,-\,q_2 \,\CO_{(3\,1\,2\,4)}\,+\,\CO_{(3\,4\,1\,2)}~,\\
&\CO_{(2\,4\,1\,3)}\otimes_q\CO_{(3\,4\,1\,2)}=q_2\CO_{(4\,2\,1\,3)}~,\\
   &\CO_{(1\,4\,2\,3)}\,\otimes_q\, \CO_{(2\,4\,1\,3)}\,=\,q_2\, \CO_{(2\,3\,1\,4)}\,+\,q_2\, \CO_{(3\,1\,2\,4)}\,-\,q_2\,
   \CO_{(3\,2\,1\,4)}~,\\
   &\CO_{(2\,1\,3\,4)}\,\otimes_q\,
\CO_{(2\,4\,1\,3)}\,=\,\CO_{(3\,4\,1\,2)}\,+\,\CO_{(4\,2\,1\,3)}\,-\,\CO_{(4\,3\,1\,2)}~,\\
   &\CO_{(2\,3\,1\,4)} \,\otimes_q\,\CO_{(2\,4\,1\,3)}\,=\,q_2\, \CO_{(3\,1\,2\,4)}~,\\
&\CO_{(2\,4\,1\,3)}\,\otimes_q\,\CO_{(2\,4\,1\,3)}\,=\,q_2\, \CO_{(3\,2\,1\,4)}\,+\,q_2\, \CO_{(4\,1\,2\,3)}\,-\,q_2\,
   \CO_{(4\,2\,1\,3)}~,\\
   &\CO_{(1\,3\,2\,4)}\,\otimes_q\,
\CO_{(3\,1\,2\,4)}\,=\,\CO_{(3\,2\,1\,4)}\,+\,\CO_{(4\,1\,2\,3)}\,-\,\CO_{(4\,2\,1\,3)}~,\\
   &\CO_{(1\,4\,2\,3)}\,\otimes_q\,
\CO_{(3\,1\,2\,4)}\,=\,\CO_{(3\,4\,1\,2)}\,+\,\CO_{(4\,2\,1\,3)}\,-\,\CO_{(4\,3\,1\,2)}~,\\
   &\CO_{(2\,1\,3\,4)}\,\otimes_q\,
   \CO_{(3\,1\,2\,4)}\,=\,q_1\, \CO_{(1\,3\,2\,4)}\,-\,q_1 \,\CO_{(1\,4\,2\,3)}\,+\,\CO_{(4\,1\,2\,3)}~,\\
   &\CO_{(2\,3\,1\,4)} \,\otimes_q\,\CO_{(3\,1\,2\,4)}\,=\,\CO_{(4\,2\,1\,3)}~,\\
   &\CO_{(2\,4\,1\,3)} \,\otimes_q\,\CO_{(3\,1\,2\,4)}\,=\,-\,q_1\, q_2\, \CO_{(1\,3\,2\,4)}\,+\,\CO_{(4\,3\,1\,2)}\,+\,q_1\,
   q_2~,\\
   &\CO_{(3\,1\,2\,4)}\,\otimes_q\,\CO_{(3\,1\,2\,4)}\,=\,q_1 \,\CO_{(1\,4\,2\,3)}\,+\,q_1\, \CO_{(2\,3\,1\,4)}\,-\,q_1\,
   \CO_{(2\,4\,1\,3)}~,\\
   &\CO_{(1\,3\,2\,4)} \,\otimes_q\,\CO_{(3\,2\,1\,4)}\,=\,\CO_{(3\,4\,1\,2)}\,+\,\CO_{(4\,2\,1\,3)}\,-\,\CO_{(4\,3\,1\,2)}~,\\
   &\CO_{(1\,4\,2\,3)} \,\otimes_q\,\CO_{(3\,2\,1\,4)}\,=\,-\,q_1\, q_2\, \CO_{(1\,3\,2\,4)}\,+\,q_2 \,\CO_{(3\,1\,2\,4)}\,+\,q_1 \,q_2~,\\
   &\CO_{(2\,1\,3\,4)}\,\otimes_q\,
   \CO_{(3\,2\,1\,4)}\,=\,q_1\, \CO_{(2\,3\,1\,4)}\,-\,q_1\, \CO_{(2\,4\,1\,3)}\,+\,\CO_{(4\,2\,1\,3)}~,\\
   &\CO_{(2\,3\,1\,4)}\,\otimes_q\, \CO_{(3\,2\,1\,4)}\,=\,\CO_{(4\,3\,1\,2)}~,\\
   &\CO_{(2\,4\,1\,3)}\,\otimes_q\, \CO_{(3\,2\,1\,4)}\,=\,q_1\, q_2 \,\CO_{(1\,3\,2\,4)}\,-\,q_1\, q_2\, \CO_{(1\,4\,2\,3)}\,+\,q_2\,
   \CO_{(4\,1\,2\,3)}~,\\
   &\CO_{(3\,1\,2\,4)}\,\otimes_q\,\CO_{(3\,2\,1\,4)}\,=\,-\,q_1 \,q_2\, \CO_{(2\,1\,3\,4)}\,+\,q_1\, \CO_{(2\,4\,1\,3)}\,+\,q_1 \,q_2~,\\
&\CO_{(3\,2\,1\,4)}\,\otimes_q\,\CO_{(3\,2\,1\,4)}\,=\,q_1\, q_2\, \CO_{(1\,3\,2\,4)}\,-\,q_1\, q_2 \,\CO_{(3\,1\,2\,4)}\,+\,q_1\,
   \CO_{(3\,4\,1\,2)}~,\\
  &\CO_{(1\,3\,2\,4)}\,\otimes_q\, \CO_{(3\,4\,1\,2)}\,=\,q_2\, \CO_{(3\,1\,2\,4)}~,\\
  &\CO_{(1\,4\,2\,3)}\,\otimes_q\, \CO_{(3\,4\,1\,2)}\,=\,q_2 \,\CO_{(3\,2\,1\,4)}~,\\
  &\CO_{(2\,1\,3\,4)}\,\otimes_q\, \CO_{(4\,1\,2\,3)}\,=\,q_1\, \CO_{(1\,4\,2\,3)}~,\\
  &\CO_{(2\,3\,1\,4)}\,\otimes_q\, \CO_{(4\,1\,2\,3)}\,=\,q_1\, q_2~,\\
  &\CO_{(2\,4\,1\,3)}\,\otimes_q\, \CO_{(4\,1\,2\,3)}\,=\,q_1 \,q_2\, \CO_{(1\,3\,2\,4)}~,\\
  &\CO_{(3\,1\,2\,4)}\,\otimes_q\, \CO_{(4\,1\,2\,3)}\,=\,q_1 \CO_{(2\,4\,1\,3)}~,\\
  &\CO_{(3\,2\,1\,4)}\,\otimes_q\, \CO_{(4\,1\,2\,3)}\,=\,q_1\, q_2\, \CO_{(2\,1\,3\,4)}~,\\
  &\CO_{(3\,4\,1\,2)}\,\otimes_q \,\CO_{(4\,1\,2\,3)}\,=\,q_1 \,q_2\, \CO_{(2\,3\,1\,4)}~,\\
  &\mathcal{O}_{(4\,1\,2\,3)}\,\otimes_q\, \mathcal{O}_{(4\,1\,2\,3)}\,=\, q_1 \,\mathcal{O}_{(3\,4\,1\,2)}~,\\
  &\mathcal{O}_{(1\,3\,2\,4)}\,\otimes_q\, \mathcal{O}_{(4\,2\,1\,3)}\,
=\, q_1\, q_2\, - \,q_1\, q_2\, \mathcal{O}_{(1\,3\,2\,4)}\, +\, \mathcal{O}_{(4\,3\,1\,2)}~,\\
&\mathcal{O}_{(1\,4\,2\,3)}\,\otimes_q\, \mathcal{O}_{(4\,2\,1\,3)}
\,=\, q_1\, q_2\, \mathcal{O}_{(1\,3\,2\,4)} \,- \,q_1 \,q_2\, \mathcal{O}_{(1\,4\,2\,3)} \,+\, q_2 \,\mathcal{O}_{(4\,1\,2\,3)}
~,\\
&\mathcal{O}_{(2\,1\,3\,4)}\,\otimes_q\, \mathcal{O}_{(4\,2\,1\,3)}
\,= \,q_1\, q_2\, -\, q_1\, q_2\, \mathcal{O}_{(2\,1\,3\,4)} \,+\, q_1\, \mathcal{O}_{(2\,4\,1\,3)}
~,\\
&\mathcal{O}_{(2\,3\,1\,4)}\,\otimes_q\, \mathcal{O}_{(4\,2\,1\,3)}
\,=\, q_1\, q_2\, \mathcal{O}_{(1\,3\,2\,4)}~,\\
&\mathcal{O}_{(2\,4\,1\,3)}\,\otimes_q\, \mathcal{O}_{(4\,2\,1\,3)}\,
=\, q_1\, q_2\, \mathcal{O}_{(1\,4\,2\,3)}\, +\, q_1 \,q_2\, \mathcal{O}_{(2\,3\,1\,4)}\, - \,q_1\, q_2\, \mathcal{O}_{(2\,4\,1\,3)}
~,\\
&\mathcal{O}_{(3\,1\,2\,4)}\,\otimes_q\, \mathcal{O}_{(4\,2\,1\,3)}\,
=\, q_1\, q_2\, \mathcal{O}_{(2\,1\,3\,4)}\, -\, q_1 \,q_2\, \mathcal{O}_{(3\,1\,2\,4)}\, +\, q_1 \mathcal{O}_{(3\,4\,1\,2)}~,\\
&\CO_{(3\,1\,2\,4)}\otimes_q\CO_{(3\,4\,1\,2)}=q_1q_2\CO_{(1\,3\,2\,4)}~,\\
&\CO_{(3\,2\,1\,4)}\otimes_q\CO_{(3\,4\,1\,2)}=q_1q_2\CO_{(1\,4\,2\,3)}~,\\
&\CO_{(3\,4\,1\,2)}\otimes_q\CO_{(3\,4\,1\,2)}=q_1q_2^2~,\\
&\mathcal{O}_{(3\,2\,1\,4)}\,\otimes_q\, \mathcal{O}_{(4\,2\,1\,3)}\,
=\, q_1\, q_2\, \mathcal{O}_{(2\,3\,1\,4)}\, +\, q_1\, q_2\, \mathcal{O}_{(3\,1\,2\,4)}\, -\, q_1\, q_2\, \mathcal{O}_{(3\,2\,1\,4)}~,\\
&\mathcal{O}_{(3\,4\,1\,2)}\,\otimes_q\, \mathcal{O}_{(4\,2\,1\,3)}
\,= \,q_1 \,q_2\, \mathcal{O}_{(2\,4\,1\,3)}~,\\
&\mathcal{O}_{(4\,1\,2\,3)}\,\otimes_q\, \mathcal{O}_{(4\,2\,1\,3)}
\,= \,q_1 \,q_2\, \mathcal{O}_{(3\,1\,2\,4)}~,\\
&\mathcal{O}_{(4\,2\,1\,3)}\,\otimes_q\,\mathcal{O}_{(4\,2\,1\,3)}\,
= \,q_1\, q_2\, \mathcal{O}_{(3\,2\,1\,4)} \,+\, q_1\, q_2\, \mathcal{O}_{(4\,1\,2\,3)} \,-\, q_1\, q_2\, \mathcal{O}_{(4\,2\,1\,3)}~,\\
&\mathcal{O}_{(1\,3\,2\,4)}\,\otimes_q\, \mathcal{O}_{(4\,3\,1\,2)}
\,=\, q_1\ q_2\, \mathcal{O}_{(1\,3\,2\,4)}\, -\, q_1\, q_2\, \mathcal{O}_{(1\,4\,2\,3)} \,+\, q_2\, \mathcal{O}_{(4\,1\,2\,3)}~,\\
&\mathcal{O}_{(1\,4\,2\,3)} \,\otimes_q\,\mathcal{O}_{(4\,3\,1\,2)}
\,=\, q_1\, q_2\, \mathcal{O}_{(2\,3\,1\,4)} \,-\, q_1 \,q_2\, \mathcal{O}_{(2\,4\,1\,3)} \,+ \,q_2 \,\mathcal{O}_{(4\,2\,1\,3)}~,\\
&\mathcal{O}_{(2\,1\,3\,4)} \,\otimes_q\,\mathcal{O}_{(4\,3\,1\,2)}\,
=\, q_1\, q_2\, \mathcal{O}_{(1\,3\,2\,4)} \,-\, q_1 \,q_2\, \mathcal{O}_{(3\,1\,2\,4)}\, +\, q_1\, \mathcal{O}_{(3\,4\,1\,2)}~,\\
&\mathcal{O}_{(2\,3\,1\,4)}\,\otimes_q\, \mathcal{O}_{(4\,3\,1\,2)}\,
= \,q_1\, q_2\, \mathcal{O}_{(1\,4\,2\,3)}~,\\
&\mathcal{O}_{(2\,4\,1\,3)}\,\otimes_q\, \mathcal{O}_{(4\,3\,1\,2)}\,
= \,q_1\, q_2^2\, - \,q_1 \,q_2^2\, \mathcal{O}_{(2\,1\,3\,4)} \,+\, q_1 \,q_2\, \mathcal{O}_{(2\,4\,1\,3)}~,\\
&\mathcal{O}_{(3\,1\,2\,4)} \,\otimes_q\, \mathcal{O}_{(4\,3\,1\,2)}\,
=\, q_1 \,q_2 \,\mathcal{O}_{(2\,3\,1\,4)} \,+\, q_1\, q_2\, \mathcal{O}_{(3\,1\,2\,4)} \,-\, q_1 \,q_2\, \mathcal{O}_{(3\,2\,1\,4)}~,\\
&\mathcal{O}_{(3\,2\,1\,4)}\,\otimes_q\,  \mathcal{O}_{(4\,3\,1\,2)}\,
=\, q_1\, q_2\, \mathcal{O}_{(2\,4\,1\,3)} \,+\, q_1\, q_2\, \mathcal{O}_{(4\,1\,2\,3)}\, - \,q_1 \,q_2 \,\mathcal{O}_{(4\,2\,1\,3)}~,\\
&\mathcal{O}_{(3\,4\,1\,2)} \,\otimes_q\, \mathcal{O}_{(4\,3\,1\,2)}
\,=\, q_1 \,q_2^2\, \mathcal{O}_{(2\,1\,3\,4)}~,\\
&\mathcal{O}_{(4\,1\,2\,3)} \,\otimes_q\, \mathcal{O}_{(4\,3\,1\,2)}
\,=\, q_1 \,q_2 \,\mathcal{O}_{(3\,2\,1\,4)}~,\\
&\mathcal{O}_{(4\,2\,1\,3)} \,\otimes_q\, \mathcal{O}_{(4\,3\,1\,2)}
\,=\, q_1 \,q_2\, \mathcal{O}_{(3\,4\,1\,2)} \,+\, q_1 \,q_2\, \mathcal{O}_{(4\,2\,1\,3)} \,-\, q_1\, q_2\, \mathcal{O}_{(4\,3\,1\,2)}
~,\\
&\mathcal{O}_{(4\,3\,1\,2)}\,\otimes_q\, \mathcal{O}_{(4\,3\,1\,2)}\,
= \,q_1^2\, q_2^2\, - \,q_1^2\, q_2^2 \,\mathcal{O}_{(1\,3\,2\,4)}\, + \,q_1\, q_2^2 \,\mathcal{O}_{(3\,1\,2\,4)}
~.
\end{align*}}

%%%%%%%%%%%%%%%%%%%%%%%%%
\bibliography{flagGLSM} \addcontentsline{toc}{section}{References}

@article{Closset:2023bdr,
    author = "Closset, Cyril and Khlaif, Osama",
    title = "{Grothendieck lines in 3d $ \mathcal{N} $ = 2 SQCD and the quantum K-theory of the Grassmannian}",
    eprint = "2309.06980",
    archivePrefix = "arXiv",
    primaryClass = "hep-th",
    doi = "10.1007/JHEP12(2023)082",
    journal = "JHEP",
    volume = "12",
    pages = "082",
    year = "2023"
}

@article{Gu:2025tda,
    author = "Gu, W. and Mihalcea, L. and Sharpe, E. and Xu, W. and Zhang, H. and Zou, H.",
    title = "{Schubert defects in Lagrangian Grassmannians}",
 JOURNAL = {JHEP},
  FJOURNAL = {Journal of High Energy Physics},
      YEAR = {2025},
    NUMBER = {6},
     PAGES = {148},
DOI = {10.1007/jhep06(2025)148},
    eprint = "2502.04438",
    archivePrefix = "arXiv",
    primaryClass = "hep-th",
}

@article{Gu:2023tcv,
    author = "Gu, Wei and Mihalcea, Leonardo and Sharpe, Eric and Xu, Weihong and Zhang, Hao and Zou, Hao",
    title = "{Quantum K theory rings of partial flag manifolds}",
    eprint = "2306.11094",
    archivePrefix = "arXiv",
    primaryClass = "hep-th",
    doi = "10.1016/j.geomphys.2024.105127",
    journal = "J. Geom. Phys.",
    volume = "198",
    pages = "105127",
    year = "2024"
}

@book{Anderson_Fulton_2023, 
title={Equivariant Cohomology in Algebraic Geometry}, 
author={Anderson, David and Fulton, William}, 
publisher={Cambridge University Press}, 
year={2023}, 
place={Cambridge}, 
series={Cambridge Studies in Advanced Mathematics}, 
collection={Cambridge Studies in Advanced Mathematics},
isbn={9781009349963},
url={https://books.google.com/books?id=yjbcEAAAQBAJ},

}

@article{LamShimozono,
 ISSN = {00029939, 10886826},
 URL = {http://www.jstor.org/stable/23807903},
    AUTHOR = {Lam, Thomas and Shimozono, Mark},
     TITLE = {Quantum double {S}chubert polynomials represent {S}chubert
              classes},
   JOURNAL = {Proc. Amer. Math. Soc.},
  FJOURNAL = {Proceedings of the American Mathematical Society},
    VOLUME = {142},
      YEAR = {2014},
 number = {3},
 pages = {835--850},
 publisher = {American Mathematical Society},
DOI = {10.1090/S0002-9939-2013-11831-9},
eprint="1108.4958",
archivePrefix="arXiv",
primaryClass="math.CO",

}

@article{Astashkevich:1993ks,
    author = "Astashkevich, Alexander and Sadov, Vladimir",
    title = "{Quantum cohomology of partial flag manifolds $F_{n_1 ... n_k}$}",
    eprint = "hep-th/9401103",
    archivePrefix = "arXiv",
    reportNumber = "HUTP-93-A040",
    doi = "10.1007/BF02099147",
    journal = "Commun. Math. Phys.",
    volume = "170",
    pages = "503--528",
    year = "1995"
}

@article{Gu:2022yvj,
    title={{Quantum K theory of Grassmannians, Wilson line operators and Schur bundles}},
  author={Gu, Wei and Mihalcea, Leonardo and Sharpe, Eric and Zou, Hao},
  JOURNAL={Forum Math. Sigma},
  volume={13},
  pages={e140},
  year={2025},
DOI={10.1017/fms.2025.10088}, 
  organization={Cambridge University Press},
    eprint = "2208.01091",
    archivePrefix = "arXiv",
    primaryClass = "math.AG",
}

@article{Donagi:2007hi,
    author = "Donagi, Ron and Sharpe, Eric",
    title = "{GLSM's for partial flag manifolds}",
    eprint = "0704.1761",
    archivePrefix = "arXiv",
    primaryClass = "hep-th",
    doi = "10.1016/j.geomphys.2008.07.010",
    journal = "J. Geom. Phys.",
    volume = "58",
    pages = "1662--1692",
    year = "2008"
}

@incollection{brion-lec,
    AUTHOR = {Brion, Michel},
     TITLE = {Lectures on the geometry of flag varieties},
 BOOKTITLE = {Topics in cohomological studies of algebraic varieties},
    SERIES = {Trends Math.},
     PAGES = {33--85},
 PUBLISHER = {Birkh\"auser, Basel},
      YEAR = {2005},
      ISBN = {3-7643-7214-1; 978-3-7643-7214-9},
       DOI = {10.1007/3-7643-7342-3\_2},
       URL = {https://doi-org.ezproxy.lib.vt.edu/10.1007/3-7643-7342-3_2},
eprint = "math/0410240",
archivePrefix = "arXiv",
primaryClass = "math.AG",
}

@incollection{Witten:1993xi,
title="{The Verlinde algebra and the cohomology of the Grassmannian}",
author={E.~Witten},
BOOKTITLE = {Geometry, topology, \& physics},
    SERIES = {Conf. Proc. Lecture Notes Geom. Topology},
    VOLUME = {IV},
     PAGES = {357--422},
 PUBLISHER = {Int. Press, Cambridge, MA},
      YEAR = {1995},
ISBN = {1-57146-024-1},
eprint="hep-th/9312104",
archivePrefix="arXiv",
primaryClass = "hep-th",
}

@article{Kapustin:2013hpk,
author = "Kapustin, Anton and Willett, Brian",
    title = "{Wilson loops in supersymmetric Chern-Simons-matter theories and duality}",
    eprint = "1302.2164",
    archivePrefix = "arXiv",
    primaryClass = "hep-th",
    month = "2",
    year = "2013"
}

@article {ahkmox,
    AUTHOR = {Amini, Kamyar and Huq-Kuruvilla, Irit and Mihalcea, Leonardo
              C. and Orr, Daniel and Xu, Weihong},
     TITLE = {Toda-type presentations for the quantum {K} theory of partial flag varieties},
   JOURNAL = {SIGMA Symmetry Integrability Geom. Methods Appl.},
  FJOURNAL = {SIGMA. Symmetry, Integrability and Geometry. Methods and
              Applications},
    VOLUME = {21},
      YEAR = {2025},
     PAGES = {Paper No. 098},
      ISSN = {1815-0659},
eprint="2504.07412",
    archivePrefix="arXiv",
    primaryClass = "math.AG",
   MRCLASS = {14M15 (05E05 14N35 37K10)},
  MRNUMBER = {4991109},
       DOI = {10.3842/SIGMA.2025.098},
       URL = {https://doi-org.ezproxy.lib.vt.edu/10.3842/SIGMA.2025.098},
}

@article{Gu:2023fpw,
    author = "Gu, Wei and Mihalcea, Leonardo C. and Sharpe, Eric and Xu, Weihong and Zhang, Hao and Zou, Hao",
    title = "{Quantum K Whitney relations for partial flag varieties}",
    eprint = "2310.03826",
    archivePrefix = "arXiv",
    primaryClass = "math.AG",
    month = "10",
    year = "2023"
}

@article{Closset:2023vos,
    author = "Closset, Cyril and Khlaif, Osama",
    title = "{Twisted indices, Bethe ideals and 3d $ \mathcal{N} $ = 2 infrared dualities}",
    eprint = "2301.10753",
    archivePrefix = "arXiv",
    primaryClass = "hep-th",
    doi = "10.1007/JHEP05(2023)148",
    journal = "JHEP",
    volume = "05",
    pages = "148",
    year = "2023"
}

@article{Closset:2023jiq,
    author = "Closset, Cyril and Khlaif, Osama",
    title = "{On the Witten index of 3d $\mathcal{N}=2$ unitary SQCD with general CS levels}",
    eprint = "2305.00534",
    archivePrefix = "arXiv",
    primaryClass = "hep-th",
    doi = "10.21468/SciPostPhys.15.3.085",
    journal = "SciPost Phys.",
    volume = "15",
    number = "3",
    pages = "085",
    year = "2023"
}

@article{Closset:2023izb,
    author = "Closset, Cyril and Khlaif, Osama",
    title = "{New results on 3d {\ensuremath{\mathscr{N}}}=2 SQCD and its 3d GLSM interpretation}",
    eprint = "2312.05076",
    archivePrefix = "arXiv",
    primaryClass = "hep-th",
    doi = "10.1142/S0217751X24460114",
    journal = "Int. J. Mod. Phys. A",
    volume = "39",
    number = "33",
    pages = "2446011",
    year = "2024"
}

@article{Huq-Kuruvilla:2025nlf,
    author = "Huq-Kuruvilla, I. and Mihalcea, L. and Sharpe, E. and Zhang, H.",
    title = "{Quantum K-theory levels in physics and math}",
    eprint = "2507.00116",
    archivePrefix = "arXiv",
    primaryClass = "hep-th",
    month = "6",
    year = "2025"
}

@article{Jockers:2019lwe,
    author = "Jockers, Hans and Mayr, Peter and Ninad, Urmi and Tabler, Alexander",
    title = "{Wilson loop algebras and quantum K-theory for Grassmannians}",
    eprint = "1911.13286",
    archivePrefix = "arXiv",
    primaryClass = "hep-th",
    reportNumber = "BONN-TH-2019-08, LMU-ASC 53/19",
    doi = "10.1007/JHEP10(2020)036",
    journal = "JHEP",
    volume = "10",
    pages = "036",
    year = "2020"
}

@article{Closset:2026bnk,
    author = "Closset, Cyril and Gu, Wei and Khlaif, Osama and Sharpe, Eric and Zhang, Hao and Zou, Hao",
    title = "{Schubert line defects in 3d GLSMs. Part II. Partial flag manifolds and parabolic quantum polynomials}",
    eprint = "2601.18881",
    archivePrefix = "arXiv",
    primaryClass = "hep-th",
    doi = "10.1007/JHEP04(2026)075",
    journal = "JHEP",
    volume = "04",
    pages = "075",
    year = "2026"
}

@article{Closset:2025akk,
    author = "Closset, Cyril and Gu, Wei and Khlaif, Osama and Sharpe, Eric and Zhang, Hao and Zou, Hao",
    title = "{Schubert line defects in 3d GLSMs. Part I. Complete flag manifolds and quantum Grothendieck polynomials}",
    eprint = "2512.19802",
    archivePrefix = "arXiv",
    primaryClass = "hep-th",
    doi = "10.1007/JHEP04(2026)074",
    journal = "JHEP",
    volume = "04",
    pages = "074",
    year = "2026"
}

@article{Bullimore:2014awa,
    author = "Bullimore, Mathew and Kim, Hee-Cheol and Koroteev, Peter",
    title = "{Defects and quantum Seiberg-Witten geometry}",
    eprint = "1412.6081",
    archivePrefix = "arXiv",
    primaryClass = "hep-th",
    doi = "10.1007/JHEP05(2015)095",
    journal = "JHEP",
    volume = "05",
    pages = "095",
    year = "2015"
}

@article{Jockers:2018sfl,
    author = "Jockers, Hans and Mayr, Peter",
    title = "{A 3d gauge theory/quantum K-theory correspondence}",
    eprint = "1808.02040",
    archivePrefix = "arXiv",
    primaryClass = "hep-th",
    reportNumber = "BONN-TH-2018-12, LMU-ASC 47/18",
    doi = "10.4310/ATMP.2020.v24.n2.a4",
    journal = "Adv. Theor. Math. Phys.",
    volume = "24",
    number = "2",
    pages = "327--457",
    year = "2020"
}

@article{Jockers:2019wjh,
    author = "Jockers, Hans and Mayr, Peter",
    title = "{Quantum K-theory of Calabi-Yau manifolds}",
    eprint = "1905.03548",
    archivePrefix = "arXiv",
    primaryClass = "hep-th",
    reportNumber = "BONN-TH-2019-03, LMU-ASC 19/19",
    doi = "10.1007/JHEP11(2019)011",
    journal = "JHEP",
    volume = "11",
    pages = "011",
    year = "2019"
}

@article{Jockers:2021omw,
    author = "Jockers, Hans and Mayr, Peter and Ninad, Urmi and Tabler, Alexander",
    title = "{BPS indices, modularity and perturbations in quantum K-theory}",
    eprint = "2106.07670",
    archivePrefix = "arXiv",
    primaryClass = "hep-th",
    reportNumber = "BONN-TH-2021-03, LMU-ASC 14/21, MITP/21-027",
    doi = "10.1007/JHEP02(2022)044",
    journal = "JHEP",
    volume = "02",
    pages = "044",
    year = "2022"
}

@article{Koroteev:2017nab,
    author = "Koroteev, Peter and Pushkar, Petr P. and Smirnov, Andrey V. and Zeitlin, Anton M.",
    title = "{Quantum K-theory of quiver varieties and many-body systems}",
    eprint = "1705.10419",
    archivePrefix = "arXiv",
    primaryClass = "math.AG",
    doi = "10.1007/s00029-021-00698-3",
    journal = "Selecta Math.",
    volume = "27",
    number = "5",
    pages = "87",
    year = "2021"
}

@article{Huq-Kuruvilla:2024tsg,
    author = "Huq-Kuruvilla, Irit",
    title = "{Quantum K-rings of partial flag varieties, Coulomb branches, and the Bethe ansatz}",
    eprint = "2409.15575",
    archivePrefix = "arXiv",
    primaryClass = "math.AG",
    month = "9",
    year = "2024"
}

@article{Sharpe:2024ujm,
    author = "Sharpe, Eric and Zhang, Hao",
    title = "{Decomposition squared}",
    eprint = "2405.12269",
    archivePrefix = "arXiv",
    primaryClass = "hep-th",
    doi = "10.1007/JHEP10(2024)168",
    journal = "JHEP",
    volume = "10",
    pages = "168",
    year = "2024"
}

@article{Gu:2021yek,
    author = "Gu, Wei and Pei, Du and Zhang, Ming",
    title = "{On phases of 3d N=2 Chern-Simons-matter theories}",
    eprint = "2105.02247",
    archivePrefix = "arXiv",
    primaryClass = "hep-th",
    doi = "10.1016/j.nuclphysb.2021.115604",
    journal = "Nucl. Phys. B",
    volume = "973",
    pages = "115604",
    year = "2021"
}

@Inbook{Gu:2021beo,
    author="Gu, Wei",
    title="Vacuum Structures Revisited",
    bookTitle="2021-2022 MATRIX Annals",
    year="2024",
    eprint = "2110.13156",
    archivePrefix = "arXiv",
    primaryClass = "hep-th",
    publisher="Springer Nature Switzerland",
    address="Cham",
    pages="835--854"
}

@article{Ueda:2019qhg,
    author = "Ueda, Kazushi and Yoshida, Yutaka",
    title = "{3d $ \mathcal{N} $ = 2 Chern-Simons-matter theory, Bethe ansatz, and quantum $K$-theory of Grassmannians}",
    eprint = "1912.03792",
    archivePrefix = "arXiv",
    primaryClass = "hep-th",
    reportNumber = "IPMU19-0176",
    doi = "10.1007/JHEP08(2020)157",
    journal = "JHEP",
    volume = "08",
    pages = "157",
    year = "2020"
}

@article{Gu:2020zpg,
    author = "Gu, Wei and Mihalcea, Leonardo and Sharpe, Eric and Zou, Hao",
    title = "{Quantum K theory of symplectic Grassmannians}",
    eprint = "2008.04909",
    archivePrefix = "arXiv",
    primaryClass = "hep-th",
    doi = "10.1016/j.geomphys.2022.104548",
    journal = "J. Geom. Phys.",
    volume = "177",
    pages = "104548",
    year = "2022"
}

@article{Gu:2025abc,
author = "Gu, Wei and Mihalcea, Leonardo C. and Sharpe, E. and Xu, Weihong and Zhang, Hao and Zou, Hao",
title = "{A Nakayama result for the quantum K theory of homogeneous spaces}",
eprint="2507.15183",
archivePrefix="arXiv",
primaryClass="math.AG",
doi = {10.46298/epiga.2025.17016},
journal = {Épijournal de Géométrie Algébrique},
volume = {9},
year = {2025},
pages = {no. 25},
url = {https://epiga.episciences.org/17016},    
issn       = {2491-6765},
eid        = 25,
month      = {Dec}
}

@article {MR4881172,
    AUTHOR = {Maeno, Toshiaki and Naito, Satoshi and Sagaki, Daisuke},
     TITLE = {A presentation of the torus-equivariant quantum {$K$}-theory
              ring of flag manifolds of type {$A$}, {P}art {I}: {T}he
              defining ideal},
   JOURNAL = {J. Lond. Math. Soc. (2)},
  FJOURNAL = {Journal of the London Mathematical Society. Second Series},
    VOLUME = {111},
      YEAR = {2025},
    NUMBER = {3},
     PAGES = {Paper No. e70095, 43},
      ISSN = {0024-6107,1469-7750},
      eprint="2302.09485",
archivePrefix="arXiv",
primaryClass="math.QA",
   MRCLASS = {14M15 (05E30 14N15 14N35 20C08)},
  MRNUMBER = {4881172},
       DOI = {10.1112/jlms.70095},
       URL = {https://doi.org/10.1112/jlms.70095},
}

@article {MNS23,
    AUTHOR = {Maeno, Toshiaki and Naito, Satoshi and Sagaki, Daisuke},
     TITLE = {A presentation of the torus-equivariant quantum {$K$}-theory
              ring of flag manifolds of type {$A$}, {P}art {II}: quantum
              double {G}rothendieck polynomials},
   JOURNAL = {Forum Math. Sigma},
  FJOURNAL = {Forum of Mathematics. Sigma},
    VOLUME = {13},
      YEAR = {2025},
     PAGES = {Paper No. e19, 26},
      ISSN = {2050-5094},
eprint="2305.17685",
archivePrefix="arXiv",
primaryClass="math.QA",
   MRCLASS = {14N15 (05E05 05E14 14M15 14N35)},
  MRNUMBER = {4856932},
       DOI = {10.1017/fms.2024.147},
       URL = {https://doi-org.ezproxy.lib.vt.edu/10.1017/fms.2024.147},
}

@article{Closset:2016arn,
    author = "Closset, Cyril and Kim, Heeyeon",
    title = "{Comments on twisted indices in 3d supersymmetric gauge theories}",
    eprint = "1605.06531",
    archivePrefix = "arXiv",
    primaryClass = "hep-th",
    doi = "10.1007/JHEP08(2016)059",
    journal = "JHEP",
    volume = "08",
    pages = "059",
    year = "2016"
}

@article{Nekrasov:2009uh,
    author = "Nekrasov, Nikita A. and Shatashvili, Samson L.",
    editor = "Baulieu, L. and de Boer, J. and Douglas, M. R. and Rabinovici, E. and Vanhove, P. and Windey, P.",
    title = "{Supersymmetric vacua and Bethe ansatz}",
    eprint = "0901.4744",
    archivePrefix = "arXiv",
    primaryClass = "hep-th",
    reportNumber = "IHES-P-09-09, TCD-MATH-09-04, HMI-09-01, NSF-KITP-09-11",
    doi = "10.1016/j.nuclphysbps.2009.07.047",
    journal = "Nucl. Phys. B Proc. Suppl.",
    volume = "192-193",
    pages = "91--112",
    year = "2009"
}

@article{Closset:2015rna,
    author = "Closset, Cyril and Cremonesi, Stefano and Park, Daniel S.",
    title = "{The equivariant A-twist and gauged linear sigma models on the two-sphere}",
    eprint = "1504.06308",
    archivePrefix = "arXiv",
    primaryClass = "hep-th",
    doi = "10.1007/JHEP06(2015)076",
    journal = "JHEP",
    volume = "06",
    pages = "076",
    year = "2015"
}

@article{Bullimore:2018jlp,
    author = "Bullimore, Mathew and Ferrari, Andrea and Kim, Heeyeon",
    title = "{Twisted indices of 3d $ \mathcal{N} $ = 4 gauge theories and enumerative geometry of quasi-maps}",
    eprint = "1812.05567",
    archivePrefix = "arXiv",
    primaryClass = "hep-th",
    doi = "10.1007/JHEP07(2019)014",
    journal = "JHEP",
    volume = "07",
    pages = "014",
    year = "2019"
}

@article{Bullimore:2019qnt,
    author = "Bullimore, Mathew and Ferrari, Andrea E. V. and Kim, Heeyeon",
    title = "{The 3d twisted index and wall-crossing}",
    eprint = "1912.09591",
    archivePrefix = "arXiv",
    primaryClass = "hep-th",
    doi = "10.21468/SciPostPhys.12.6.186",
    journal = "SciPost Phys.",
    volume = "12",
    number = "6",
    pages = "186",
    year = "2022"
}

@article{Bullimore:2020nhv,
    author = "Bullimore, Mathew and Ferrari, Andrea E. V. and Kim, Heeyeon and Xu, Guangyu",
    title = "{The twisted index and topological saddles}",
    eprint = "2007.11603",
    archivePrefix = "arXiv",
    primaryClass = "hep-th",
    doi = "10.1007/JHEP05(2022)116",
    journal = "JHEP",
    volume = "05",
    pages = "116",
    year = "2022"
}

@article{Dedushenko:2023qjq,
    author = "Dedushenko, Mykola and Nekrasov, Nikita",
    title = "{Interfaces and quantum algebras, II: Cigar partition function}",
    eprint = "2306.16434",
    archivePrefix = "arXiv",
    primaryClass = "hep-th",
    month = "6",
    year = "2023"
}

@article{Dedushenko:2021mds,
    author = "Dedushenko, Mykola and Nekrasov, Nikita",
    title = "{Interfaces and quantum algebras, I: Stable envelopes}",
    eprint = "2109.10941",
    archivePrefix = "arXiv",
    primaryClass = "hep-th",
    doi = "10.1016/j.geomphys.2023.104991",
    journal = "J. Geom. Phys.",
    volume = "194",
    pages = "104991",
    year = "2023"
}

@article {buch-lmi,
    AUTHOR = {Buch, Anders S. and Mihalcea, Leonardo C.},
     TITLE = {Quantum {$K$}-theory of {G}rassmannians},
   JOURNAL = {Duke Math. J.},
  FJOURNAL = {Duke Mathematical Journal},
    VOLUME = {156},
      YEAR = {2011},
    NUMBER = {3},
     PAGES = {501--538},
      ISSN = {0012-7094,1547-7398},
eprint="0810.0981",
archivePrefix="arXiv",
primaryClass="math.AG",
   MRCLASS = {14N35 (14C35 14M15 19E08)},
  MRNUMBER = {2772069},
MRREVIEWER = {Hsian-Hua\ Tseng},
       DOI = {10.1215/00127094-2010-218},
       URL = {https://doi-org.ezproxy.lib.vt.edu/10.1215/00127094-2010-218},
}

@article{xu2024quantum,
	author = {Xu, Weihong},
	date = {2024/03/19},
 eprint="math/2112.13036",
   archivePrefix="arXiv",
   primaryClass = "math.AG",
	doi = {10.1007/s40879-024-00738-0},
	id = {Xu2024},
	isbn = {2199-6768},
	journal = {European Journal of Mathematics},
	number = {2},
	pages = {22},
	title = "{Quantum K-theory of incidence varieties}",
	url = {https://doi.org/10.1007/s40879-024-00738-0},
	volume = {10},
	year = {2024}}

@article{Nekrasov:2014xaa,
    author = "Nekrasov, Nikita A. and Shatashvili, Samson L.",
    title = "{Bethe/Gauge correspondence on curved spaces}",
    eprint = "1405.6046",
    archivePrefix = "arXiv",
    primaryClass = "hep-th",
    doi = "10.1007/JHEP01(2015)100",
    journal = "JHEP",
    volume = "01",
    pages = "100",
    year = "2015"
}

@article{Closset:2017zgf,
    author = "Closset, Cyril and Kim, Heeyeon and Willett, Brian",
    title = "{Supersymmetric partition functions and the three-dimensional A-twist}",
    eprint = "1701.03171",
    archivePrefix = "arXiv",
    primaryClass = "hep-th",
    reportNumber = "CERN-TH-2017-006",
    doi = "10.1007/JHEP03(2017)074",
    journal = "JHEP",
    volume = "03",
    pages = "074",
    year = "2017"
}

@article{Witten:1981nf,
    author = "Witten, Edward",
    title = "{Dynamical Breaking of Supersymmetry}",
    reportNumber = "Print-81-0317 (PRINCETON)",
    doi = "10.1016/0550-3213(81)90006-7",
    journal = "Nucl. Phys. B",
    volume = "188",
    pages = "513",
    year = "1981"
}

@article{Witten:1993yc,
    author = "Witten, Edward",
    editor = "Greene, B. and Yau, Shing-Tung",
    title = "{Phases of N=2 theories in two-dimensions}",
    eprint = "hep-th/9301042",
    archivePrefix = "arXiv",
    reportNumber = "IASSNS-HEP-93-3",
    doi = "10.1016/0550-3213(93)90033-L",
    journal = "Nucl. Phys. B",
    volume = "403",
    pages = "159--222",
    year = "1993"
}

@misc{samuel2026triplequantumschubertcalculus,
      title={Triple Quantum Schubert Calculus}, 
      author={Matthew J. Samuel},
      year={2026},
      eprint={2610.31533},
      archivePrefix={arXiv},
      primaryClass={math.CO},
      url={https://arxiv.org/abs/2610.31533}, 
}

@misc{schubmult,
	author = {Matthew J. Samuel},
	title = {{\texttt{schubmult}: a {P}ython package for multiplying {S}chubert polynomials}},
	howpublished = {Available from the {P}ython {P}ackage {I}ndex, \url{https://pypi.org/project/schubmult/}, with sources at \url{https://github.com/matthematics/schubmult}},
	year = {2026}
}

@article{Witten:1999ds,
    author = "Witten, Edward",
    editor = "Shifman, Mikhail A.",
    title = "{Supersymmetric index of three-dimensional gauge theory}",
    eprint = "hep-th/9903005",
    archivePrefix = "arXiv",
    reportNumber = "IASSNS-HEP-99-20",
    doi = "10.1142/9789812793850_0013",
    pages = "156--184",
    month = "2",
    year = "1999"
}

@misc{Singularm,
 title = {Singular.m},
 author = {M.~Kauers and V.~Levandovskyy},
 howpublished = {\url{https://www3.risc.jku.at/research/combinat/software/Singular}},
}

@misc{DGPS,
 title = {{\sc Singular} {4-3-0} --- {A} computer algebra system for polynomial computations},
 author = {Decker, Wolfram and Greuel, Gert-Martin and Pfister, Gerhard and Sch\"onemann, Hans},
 year = {2022},
 howpublished = {\url{http://www.singular.uni-kl.de}},
}

@article {bertram1997quantumschubertcalculus,
    AUTHOR = {Bertram, Aaron},
     TITLE = {Quantum {S}chubert calculus},
   JOURNAL = {Adv. Math.},
  FJOURNAL = {Advances in Mathematics},
    VOLUME = {128},
      YEAR = {1997},
    NUMBER = {2},
     PAGES = {289--305},
      ISSN = {0001-8708,1090-2082},
   MRCLASS = {14M15 (14N10)},
  MRNUMBER = {1454400},
MRREVIEWER = {Sara\ C.\ Billey},
       DOI = {10.1006/aima.1997.1627},
       URL = {https://doi.org/10.1006/aima.1997.1627},
}

@article {buch2018,
    AUTHOR = {Buch, Anders S. and Chaput, Pierre-Emmanuel and Mihalcea,
              Leonardo C. and Perrin, Nicolas},
     TITLE = {A {C}hevalley formula for the equivariant quantum {$K$}-theory
              of cominuscule varieties},
   JOURNAL = {Algebr. Geom.},
  FJOURNAL = {Algebraic Geometry},
    VOLUME = {5},
      YEAR = {2018},
    NUMBER = {5},
     PAGES = {568--595},
      ISSN = {2313-1691,2214-2584},
   MRCLASS = {14N35 (14C35 14M15 14N15 19L47)},
  MRNUMBER = {3847206},
MRREVIEWER = {Feng\ Qu},
       DOI = {10.14231/ag-2018-015},
       URL = {https://doi.org/10.14231/ag-2018-015},
}

@phdthesis{Khlaif:2025ccg,
    author = "Khlaif, Osama",
    title = "{Novel Aspects of 3D $\mathcal{N}= 2$ Chern{\textendash}Simons{\textendash}Matter theories}",
    school = "Birmingham U.",
    month = "4",
    year = "2025",
    URL = {https://etheses.bham.ac.uk//id/eprint/16524/}
}

@misc{MihalceaLecture,
  author       = {L.~C.~Mihalcea},
  title        = {Notes on Quantum K Theory of Flag Manifolds},
  howpublished = {\url{https://personal.math.vt.edu/lmihalce/QKlectures(MSJ23).pdf}},
  year         = {2024}
}

@book{fulton2013representation,
  title={Representation Theory: A First Course},
  author={Fulton, W. and Harris, J.},
  isbn={9781461209799},
  series={Graduate Texts in Mathematics},
  url={https://books.google.com/books?id=6TwmBQAAQBAJ},
  year={2013},
  publisher={Springer New York}
}

@article{Ohmori:2018qza,
    author = "Ohmori, Kantaro and Seiberg, Nathan and Shao, Shu-Heng",
    title = "{Sigma Models on Flags}",
    eprint = "1809.10604",
    archivePrefix = "arXiv",
    primaryClass = "hep-th",
    doi = "10.21468/SciPostPhys.6.2.017",
    journal = "SciPost Phys.",
    volume = "6",
    number = "2",
    pages = "017",
    year = "2019"
}

@article{Gu:2024mqk,
    author = "Gu, Wei and Guo, Jirui and Mihalcea, Leonardo and Wen, Yaoxiong and Yan, Xiaohan",
    title = "{A correspondence between the quantum K theory and quantum cohomology of Grassmannians}",
    eprint = "2406.13739",
    archivePrefix = "arXiv",
    primaryClass = "hep-th",
    reportNumber = "MPIM-Bonn-2024",
    doi = "10.1016/j.geomphys.2025.105437",
    journal = "J. Geom. Phys.",
    volume = "210",
    pages = "105437",
    year = "2025"
}

@article{2024arXiv240702703S,
       author = {{Summers}, Kevin},
        title = "{A dual basis for the equivariant quantum $K$-theory of cominuscule varieties}",
      journal = {arXiv e-prints},
         year = 2024,
        month = jul,
          eid = {arXiv:2407.02703},
        pages = {arXiv:2407.02703},
          doi = {10.48550/arXiv.2407.02703},
archivePrefix = {arXiv},
       eprint = {2407.02703},
 primaryClass = {math.AG},
       adsurl = {https://ui.adsabs.harvard.edu/abs/2024arXiv240702703S}
}

@article{10.1093/imrn/rnaa108,
    author = {Anderson, David and Chen, Linda and Tseng, Hsian-Hua},
    title = {On the Finiteness of Quantum K-Theory of a Homogeneous Space},
    journal = {International Mathematics Research Notices},
    volume = {2022},
    number = {2},
    pages = {1313-1349},
    year = {2022},
    month = {01},
    issn = {1073-7928},
    doi = {10.1093/imrn/rnaa108},
    url = {https://doi.org/10.1093/imrn/rnaa108},
    archivePrefix = {arXiv},
       eprint = {1804.04579},
 primaryClass = {math.AG}
}

@article{kouno2023quantumktheorychevalleyformulas,
      title={Quantum K-theory Chevalley formulas in the parabolic case}, 
      author={Takafumi Kouno and Cristian Lenart and Satoshi Naito and Daisuke Sagaki and with an Appendix by Takafumi Kouno and Cristian Lenart and Satoshi Naito and Daisuke Sagaki and Weihong Xu},
      year={2023},
      eprint={2109.11596},
      archivePrefix={arXiv},
      primaryClass={math.CO},
      url={https://arxiv.org/abs/2109.11596}, 
}
\bibliographystyle{JHEP}

\end{document}